\documentclass [aps,letterpaper,twocolumn,superscriptaddress,amsmath,amssymb,floatfix] {revtex4}

\usepackage{graphicx}
\usepackage{amsmath}
\makeatletter

\pdfpageheight\paperheight
\pdfpagewidth\paperwidth

\@ifundefined{textcolor}{}
{%
 \definecolor{BLACK}{gray}{0}
 \definecolor{WHITE}{gray}{1}
 \definecolor{RED}{rgb}{1,0,0}
 \definecolor{GREEN}{rgb}{0,1,0}
 \definecolor{BLUE}{rgb}{0,0,1}
 \definecolor{CYAN}{cmyk}{1,0,0,0}
 \definecolor{MAGENTA}{cmyk}{0,1,0,0}
 \definecolor{YELLOW}{cmyk}{0,0,1,0}
}

\usepackage{xcolor}\usepackage{soul}

\newcommand{\bra}[1]{\ensuremath{\left\langle#1\right|}}
\newcommand{\ket}[1]{\ensuremath{\left|#1\right\rangle}}

\definecolor{blue}{rgb}{0,0,1}
\definecolor{red}{rgb}{1,0,0}
\definecolor{green}{rgb}{0,1,0}

\begin{document}

\title{Engineering superconducting qubits to reduce quasiparticles and charge noise}

\newcommand{\SIQSE}{\affiliation{1}{Shenzhen Institute for Quantum Science and Engineering, Southern University of Science and Technology, Shenzhen, Guangdong, China}}
\newcommand{\IQA}{\affiliation{2}{International Quantum Academy, Shenzhen, Guangdong, China}}
\newcommand{\GDKL}{\affiliation{3}{Guangdong Provincial Key Laboratory of Quantum Science and Engineering, Southern University of Science and Technology, Shenzhen, Guangdong, China}}
\newcommand{\DPHY}{\affiliation{4}{Department of Physics, Southern University of Science and Technology, Shenzhen, Guangdong, China}}
\newcommand{\FJ}{\affiliation{5}{JARA Institute for Quantum Information (PGI-11), Forschungszentrum J\"ulich, 52425 J\"ulich, Germany}}
\newcommand{\TII}{\affiliation{6}{Quantum Research Centre, Technology Innovation Institute, Abu Dhabi, UAE}}
\newcommand{\SEU}{\affiliation{7}{State Key Laboratory of Millimeter Waves, School of Information Science and Engineering, Southeast University, Nanjing, China}}

\author{Xianchuang Pan}
\thanks{These authors have contributed equally to this work.}
\affiliation{\SIQSE}
\affiliation{\IQA}
\affiliation{\GDKL}

\author{Yuxuan Zhou}
\thanks{These authors have contributed equally to this work.}
\affiliation{\SIQSE}
\affiliation{\IQA}
\affiliation{\GDKL}
\affiliation{\DPHY}

\author{Haolan Yuan}
\affiliation{\SIQSE}
\affiliation{\IQA}
\affiliation{\GDKL}
\affiliation{\DPHY}

\author{Lifu Nie}
\affiliation{\SIQSE}
\affiliation{\IQA}
\affiliation{\GDKL}

\author{Weiwei Wei}
\affiliation{\SIQSE}
\affiliation{\IQA}
\affiliation{\GDKL}

\author{Libo Zhang}
\affiliation{\SIQSE}
\affiliation{\IQA}
\affiliation{\GDKL}

\author{Jian Li}
\affiliation{\SIQSE}
\affiliation{\IQA}
\affiliation{\GDKL}

\author{Song Liu}
\affiliation{\SIQSE}
\affiliation{\IQA}
\affiliation{\GDKL}

\author{Zhi Hao Jiang}
\affiliation{\SEU}

\author{Gianluigi Catelani}
\email{g.catelani@fz-juelich.de}
\affiliation{\FJ}
\affiliation{\TII}

\author{Ling Hu}
\email{hul@sustech.edu.cn}
\affiliation{\SIQSE}
\affiliation{\IQA}
\affiliation{\GDKL}

\author{Fei Yan}
\email{yanf7@sustech.edu.cn}
\affiliation{\SIQSE}
\affiliation{\IQA}
\affiliation{\GDKL}

\author{Dapeng Yu}
\affiliation{\SIQSE}
\affiliation{\IQA}
\affiliation{\GDKL}
\affiliation{\DPHY}


\begin{abstract}
Identifying, quantifying, and suppressing decoherence mechanisms in qubits are important steps towards the goal of engineering a quantum computer or simulator. Superconducting circuits offer flexibility in qubit design; however, their performance is adversely affected by quasiparticles (broken Cooper pairs). Developing a quasiparticle mitigation strategy compatible with scalable, high-coherence devices is therefore highly desirable.
Here we experimentally demonstrate how to control quasiparticle generation by downsizing the qubit, capping it with a metallic cover, and equipping it with suitable quasiparticle traps. Using a flip-chip design, we shape the electromagnetic environment of the qubit above the superconducting gap, inhibiting quasiparticle poisoning. Our findings support the hypothesis that quasiparticle generation is dominated by the breaking of Cooper pairs at the junction, as a result of photon absorption by the antenna-like qubit structure. We achieve record low charge-parity switching rate ($<1$~Hz). Our aluminium devices also display improved stability with respect to discrete charging events.
\end{abstract}

\maketitle

\section*{Introduction}
Quantum computers and simulators are highly anticipated transformative technologies, and superconducting quantum circuits based on Josephson junctions are a leading candidate for their realization.
The proper functioning of superconducting circuits requires a pristine environment to protect the collective behaviour of Cooper pairs. An acknowledged potential danger is quasiparticle poisoning, that is, the presence of broken pairs ubiquitously seen in superconducting devices; these quasiparticles can be a significant source of decoherence in qubits based on Josephson junctions~\cite{SPLN,MQE}.
Moreover, recent studies with superconducting circuits~\cite{Cardani2020,Wilen2021,GOOburst,karatsu2019mitigation} have shown that energy deposited in the substrate may cause quasiparticle generation not only locally (i.e., in a single qubit) but also across multiple qubits within a short time, leading to correlated errors that can impede quantum error correction. Therefore, a deeper understanding of the generation mechanisms of quasiparticles, from ionizing radiation~\cite{MITrad} to stray photons~\cite{Houzet2019}, is imperative.

Quasiparticles have been intensely investigated over the last decade~\cite{Catelani2011l,Lenander2011,Paik2011,Pop2014}.
Puzzlingly, experiments conducted using various devices unanimously suggest a much higher number of quasiparticles at experimental temperatures (typically $\sim 10\,$mK) than expected in thermal equilibrium, a phenomenon that is not yet fully understood.
The number of quasiparticles is determined by the balance between generation, i.e., the breaking of a Cooper pair into two quasiparticles, and recombination, the reverse process. While recombination is determined by material properties that can be difficult to modify, generation can be controlled to some extent, for example, by using phonon traps~\cite{Henriques2019}. For small superconducting islands, schemes to pump out quasiparticles have been developed~\cite{Gustavsson2016} and protection by a ground plane has been shown to reduce quasiparticle generation~\cite{mannila_superconductor_2021}. Alternatively, trapping quasiparticles in normal-metal islands, so that they cannot tunnel through Josephson junctions, can also protect qubits~\cite{Riwar2016}. 
However, a mitigation method that is compatible with state-of-the-art quantum processor designs~\cite{arute_quantum_2019} has not yet been demonstrated.

\begin{figure*}[ht]
\includegraphics[]{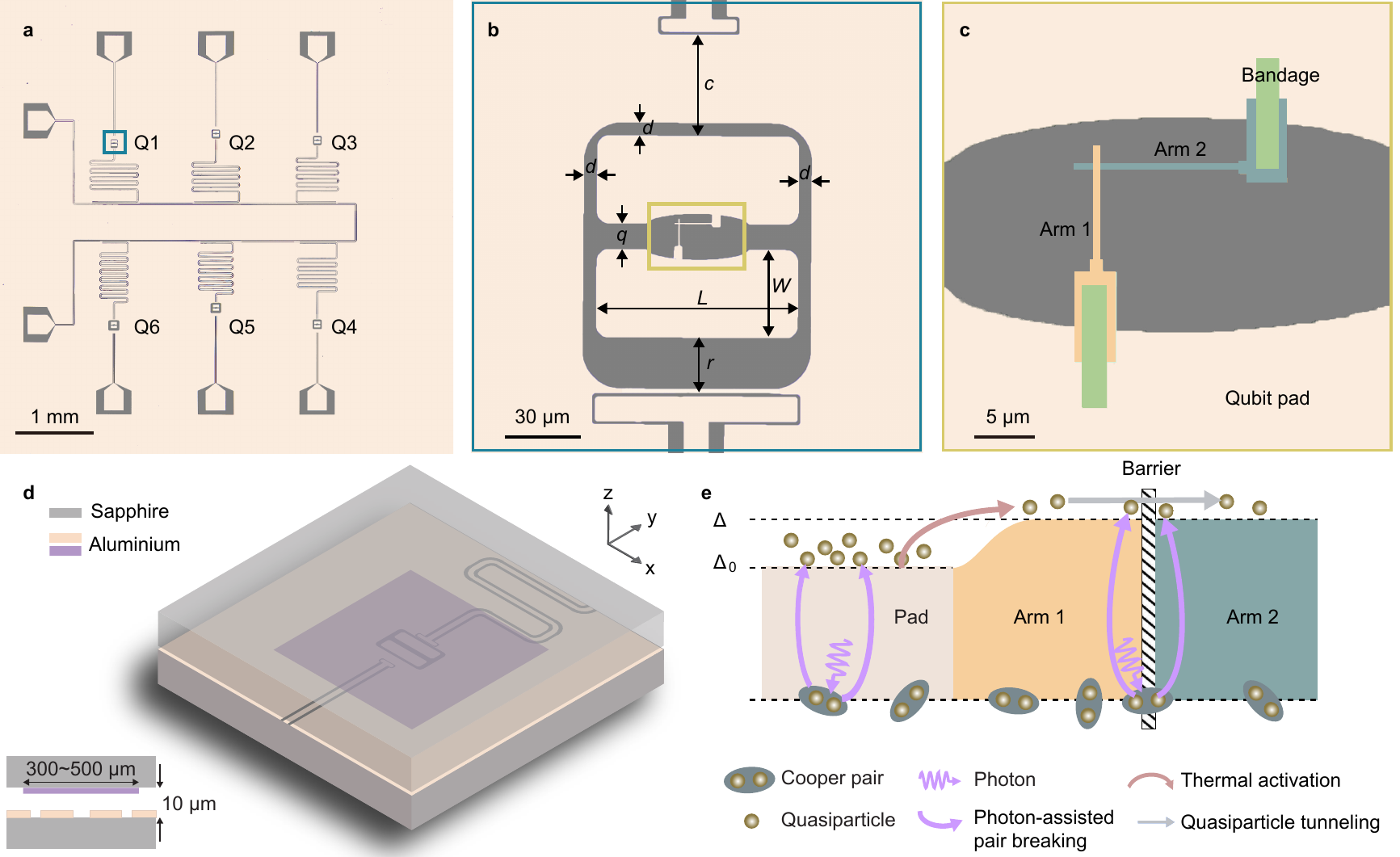}
\caption{\textbf{Device layout and photon-assisted quasiparticle generation.}
\textbf{a}, Optical micrograph of a planar-design 6~mm$\times$6~mm sample chip. The light area indicates the base aluminium layer; the dark area indicates the exposed sapphire substrate. Each qubit (Q1--Q6) has a dedicated charge drive line (straight transmission line) and a dedicated readout resonator (meandering transmission line). The six resonators share a common feed line for a multiplexed readout.
\textbf{b}, Close-up view of Q1 (blue rectangle in \textbf{a}), showing two aluminium pads (length $L$ and width $W$) floating inside an aperture.
The pad-to-resonator distance is $r$; the pad-to-ground distance on the other three sides is $d$; the pad-to-drive-line distance is $c$; the pad-to-pad distance is $q$.
In the shown case, $L=80~\mu$m, $W=35~\mu$m, $r=23~\mu$m, $d=5~\mu$m, $c=41~\mu$m, $q=10~\mu$m.
The qubits shown in \textbf{a} have varying pad-to-ground distances ($d=5-30~\mu$m).
\textbf{c}, Close-up view of the junction area (yellow rectangle in \textbf{b}).
Two layers (colour-coded) of aluminium film are deposited on the base layer to form the junction, i.e., the region where the two thin strips (Arm1 and Arm2) overlap. An additional aluminium layer ($10~\mu$m$\times2~\mu$m rectangular patches) forms the bandage structures that are used to improve the galvanic contact~\cite{Dunsworth2017}.
The film thicknesses are 100~nm for the pads, 30~nm for Arm1, 40~nm for Arm2, and 200~nm for the bandages.
\textbf{d}, Schematic of a vertically integrated device with the qubit, resonator and drive line patterned on the bottom die and a square-shaped aluminium pad (purple) on the top die floating above the qubit. The inset on the bottom left shows a cross-sectional view of the device.
\textbf{e}, Schematic illustrating the quasiparticle processes. $\Delta_0$ and $\Delta$ are the superconducting energy gaps of the pads and the junction leads, respectively. 
}
\label{fig:Fig1}
\end{figure*}

Here we experimentally investigate how variations in our superconducting qubit design affect the rate of quasiparticle generation and show that the main quasiparticle source is local: quasiparticles originate from the breaking of Cooper pairs at the Josephson junction via the absorption of stray photons.
This corroborates the conjecture that photons with energy greater than twice the superconducting energy gap and whose absorption is mediated by the antenna-like structure of the qubit are responsible for the observed excess quasiparticles~\cite{Houzet2019,Serniak2019,Rafferty2021}.
Leveraging the design flexibility of superconducting circuits, particularly flip-chip technology, we demonstrate convenient control of the antenna mode and hence the quasiparticle generation, achieving an exceedingly low charge-parity switching rate ($\Gamma_P\lesssim 1$~Hz) in our aluminium qubits.
The charge offset stability is also improved, and the occurrence rate of strong charge jumps (jump amplitudes greater than $0.1~e$, where $e$ is an electron charge) is on the order of 0.01~mHz.
In addition, the measured temperature dependence of the charge-parity switching rates is consistent with quasiparticles being thermally excited out of the capacitor pads, which act as superconducting traps~\cite{Riwar2019}, and into the junction leads.


\section*{Results}
\noindent\textbf{Devices and quasiparticle generation mechanism.} Our devices have two different types of architectures, planar and vertically integrated~\cite{rosenberg_3d_2017,foxen_qubit_2017,satzinger_simple_2019}, both consisting of aluminium on c-plane sapphire substrates. 
A planar sample consists of six qubits separated by at least 1.3~mm (Fig.~\ref{fig:Fig1}a), with each qubit (transition frequency between the ground and excited state $\omega_{\rm ge}$) coupled to a local resonator (frequency $\omega_{\rm r}$) for dispersive readout~\cite{Blais2004} and to a dedicated control line (feeding both direct current and radio frequency signals).
As shown in Fig.~\ref{fig:Fig1}b, the qubits share a floating transmon design~\cite{Koch2007}, two rectangular-shaped capacitor pads (charging energy $E_{\rm C}=e^2/C_\Sigma$, where $C_\Sigma$ is the total capacitance) shunting a Josephson junction (Josephson energy $E_{\rm J}$).
We fabricated Manhattan-style Josephson junctions using two aluminium leads, Arm1 and Arm2 in Fig.~\ref{fig:Fig1}c, that extend from the pads and overlap each other, separated by an aluminium oxide barrier.
In addition, we fabricated a second type of device using flip-chip technology to cover the qubit structure with a floating aluminium cap separated by $10~\mu$m (Fig.~\ref{fig:Fig1}d).
In both the capped and uncapped devices, we added variations in the circuit design across the different qubits to investigate the generation mechanism of the nonequilibrium quasiparticles.
We explored an extended parameter regime of the $E_{\rm J}/E_{\rm C}$ ratio (2--30) between the transmon and the Cooper pair box or charge qubit~\cite{Nakamura1999,Duty2004,Astafiev2004}, which retains sensitivity to charge fluctuations and quasiparticle tunnelling.
The samples were packaged in aluminium and copper boxes that were thermally anchored to the mixing chamber stage ($<$10~mK) of a dilution refrigerator.
Similar to other studies~\cite{Serniak2019,Kurter2021,Gordon2021}, we find that careful shielding and filtering are important to reduce quasiparticles.
See Supplementary Note 1$\sim$2~\cite{supplement} for more information concerning the device and experimental setup.

The aluminium film is thinner at the junction leads (30--40~nm) than at the pads (100~nm); therefore, the superconducting gap frequency near the junction ($f^*=2\Delta/h\approx 2\times217~\mu {\rm eV}/h=105~$GHz) is higher than that in the pads ($f_0=2\Delta_0/h \approx 2\times180~\mu{\rm eV}/h=87~{\rm GHz}$)~\cite{Chubov1969,Court2007}.
Accordingly, we hypothesize the following scenario for the generation and tunnelling processes of nonequilibrium quasiparticles (Fig.~\ref{fig:Fig1}e). Photons or phonons with energy greater than twice the superconducting energy gap can break a Cooper pair and create two quasiparticles. Such bulk generation is more likely in the pads than in the arms, because of the much larger area and volume of the pads.
However, these nonequilibrium quasiparticles may not directly contribute to tunnelling across the junction, unless, for example, they are excited by phonons to overcome the gap difference between the thinner junction arms and the pads.
Conversely, the coherent tunnelling of a Cooper pair across the junction can accompany a photon absorption event, as a form of photon-assisted tunnelling~\cite{Nakamura1998} that breaks the pair and creates one quasiparticle on each side of the barrier.
Therefore, by measuring the quasiparticle tunnelling rate, one can infer the absorption efficiency of sub-terahertz ($\sim$100~GHz) photons.

The qubit Hamiltonian can be expressed in the form of a generalized Cooper-pair box~\cite{Serniak2019}:
\begin{equation}\label{equ:Hamiltonian_qubit}
       \hat{H}_{\rm q} = 4 E_{\rm C}\left(\hat{n}-n_{\rm g}+\frac{P-1}{4}\right)^2-E_{\rm J}\cos\hat{\phi} \,,
\end{equation}
where $\hat{n}$ is the number of Cooper pairs that have traversed the junction and $\hat{\phi}$ is the superconducting phase difference across the junction. $n_{\rm g}$ indicates the offset charge in units of $2e$ and the Hamiltonian has a $2e$-periodicity.
$P$ is the charge parity of the circuit, where $P=1$ corresponds to even parity and $P=-1$ corresponds to odd parity.
The Hamiltonian implies that a change in the charge parity of the junction electrodes is equivalent to a shift of $1e$ in the offset charge.
Compared to the usual transmon Hamiltonian~\cite{Koch2007}, in which the parity is conventionally fixed to be even, the system described by Eq.~\eqref{equ:Hamiltonian_qubit} has twice as many eigenstates, one for each parity.

\begin{figure}[t]
\includegraphics{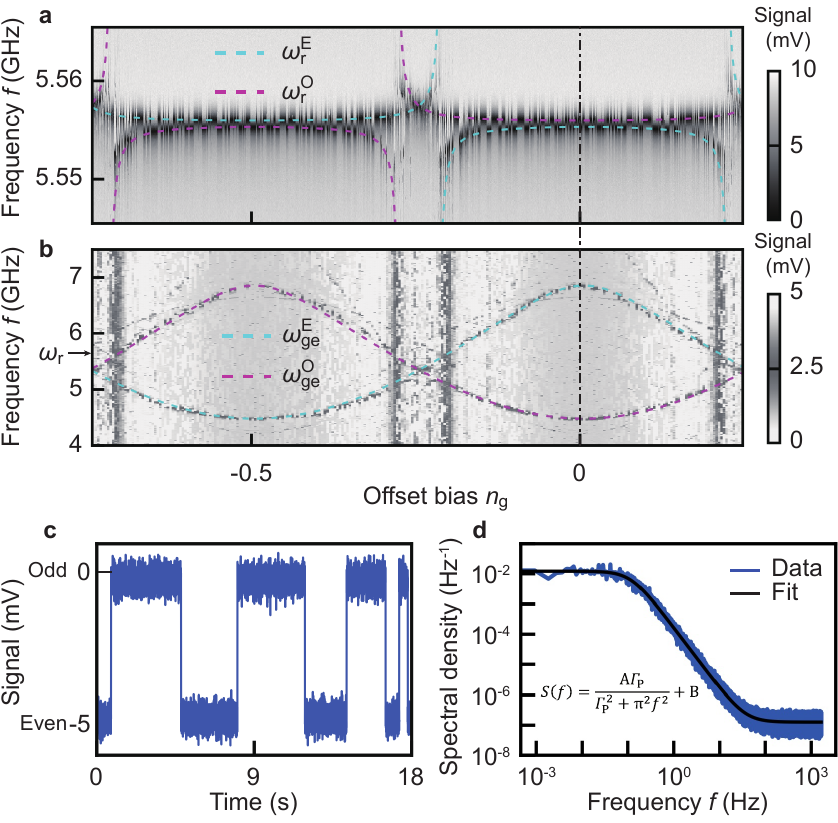}
\caption{\textbf{Spectroscopy and charge-parity detection.}
\textbf{a}, Resonator ($\omega_{\rm r}$) and \textbf{b}, qubit spectroscopy as a function of the offset charge bias $n_{\rm g}$, showing $2e$ periodicity and a shift of $1e$ between even and odd parity.
The dashed lines are the identified resonator frequency $\omega_{r}^{\rm E(O)}$ and the g-e transition frequency of the qubit $\omega_{ge}^{\rm E(O)}$ (the superscript letter E and O indicate even and odd parity, respectively) from fitting to the Jaynes-Cumming model. 
The resonator spectrum was acquired by a network analyzer at a rate of 0.2~s per offset bias or vertical linecut, while the qubit spectrum was obtained from pulsed measurements with each data point taking approximately 0.1~s. 
See Supplementary Note 3~\cite{supplement} for the other identified transitions.
\textbf{c}, Example of the time evolution (time interval: 0.3~ms; total length 18~s) of the charge parity measured at $n_{\rm g}=0$ showing random telegraph behaviour between even ($P=1$) and odd ($P=-1$) parity.
\textbf{d}, Power spectrum of the charge-parity fluctuations obtained from 1200 repetitions of the measurement in \textbf{c}. The inset shows the Lorentzian fitting function, where $\Gamma_P$ (the average switching rate), A and B are fitting parameters. In the illustrated case, the extracted charge-parity lifetime $T_P =1/\Gamma_P=2.7~$s.
The white noise (offset term $B$) is due to the sampling noise.
}
\label{fig:Fig2}
\end{figure}

\begin{figure*}[t]
\includegraphics{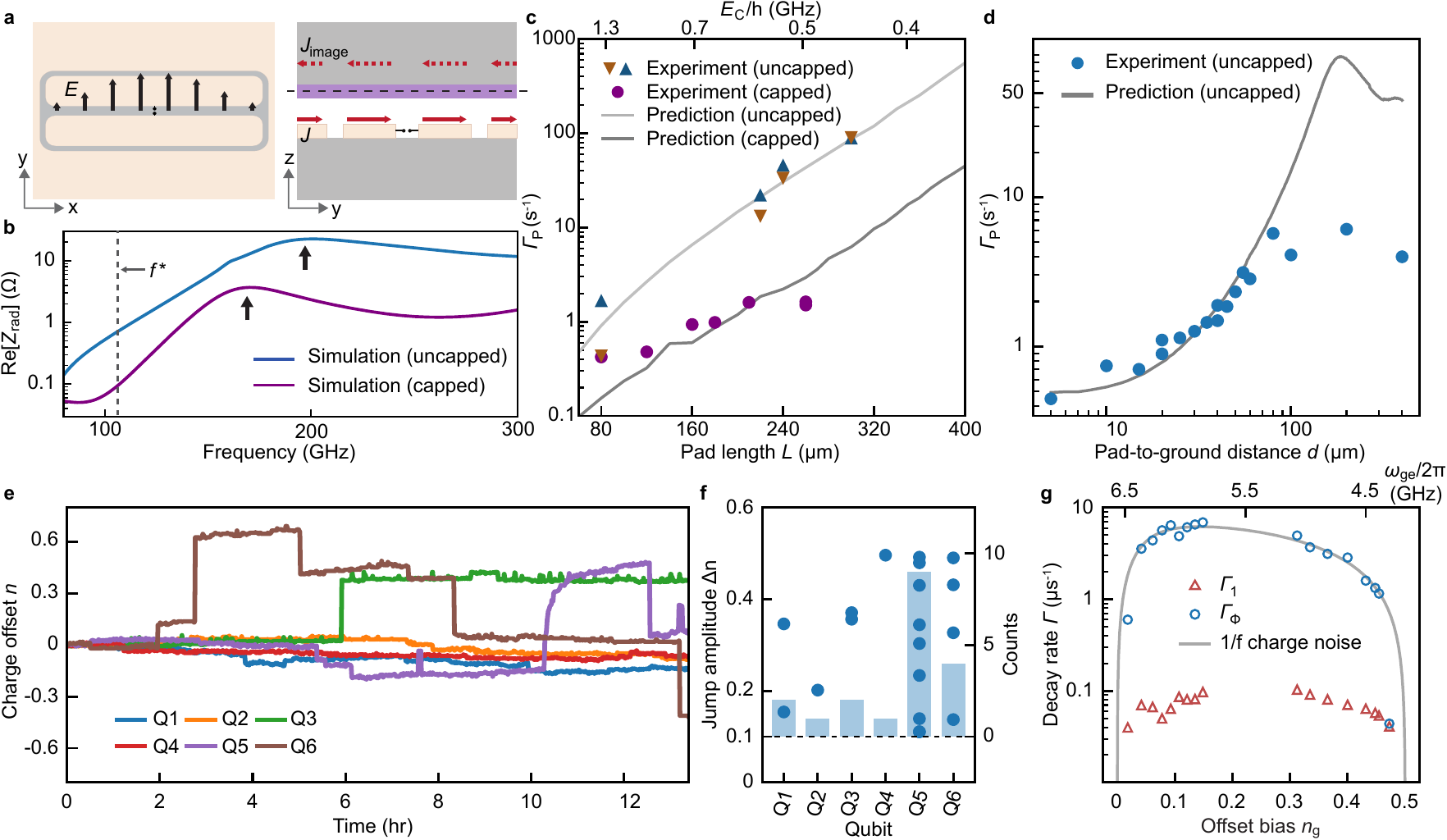}
\caption{\textbf{Effect of circuit geometry on parity switching, charge offset stability, and coherence.}
\textbf{a}, Top view of the electric field ($E$, arrows) of the fundamental radiation mode formed by the floating transmon structure (left).
Mode current ($J$, solid arrows) and its image ($J_{\rm image}$, dashed arrows) on the two sides of the aluminium cap indicated by the dashed line (right). 
\textbf{b}, Real part of the simulated input impedance $Z_{\rm rad}$ of a typical qubit ($L$=260~$\mu$m, $W$=35~$\mu$m) with and without a cap.
The arrows indicate the peaks corresponding to the fundamental mode.
The dashed line indicates the superconducting gap frequency at 105~GHz.
\textbf{c}, Measured parity switching rates for capped (dots) and uncapped (triangles) qubits with varying pad size plotted as a function of the pad length.
In the uncapped case, the pad width is also varying, but the effect is much smaller.
The solid lines indicate finite-element simulation predictions.
The top axis indicates the charging energy $E_{\rm C}$ that corresponds to the different pad lengths in the capped case.
\textbf{d}, Measured (dots) and simulated (line) parity switching rate of qubits with varying pad-to-ground distance $d$.
\textbf{e}, Offset charge drift for six qubits on a single chip (Fig.~\ref{fig:Fig1}a) with the same pad size but varying pad-to-ground distance ($d=5-30~\mu$m), simultaneously monitored over a 13-hour period.
\textbf{f}, Amplitudes (dots, left axis) and total counts (bars, right axis) of all offset charge jumps ($|\Delta q|$) greater than $0.1\rm e$ identified in the data in \textbf{c} and the extended data~\cite{supplement} during a total of 40~h of monitoring.
\textbf{g}, Measured relaxation ($\Gamma_{1}$) and pure-dephasing ($\Gamma_{\rm \phi}$) rates of Q1 as a function of $n_{\rm g}$ sampled over half a period. The top axis indicates the corresponding qubit frequency.
The solid line is fit to the $1/f$ charge noise model.
It is difficult to characterize the qubit around $n_{\rm g}=0.25$ because of the stronger sensitivity to charge noise around this bias and because of its adjacency to the resonator at 5.55~GHz.
}
\label{fig:Fig3}
\end{figure*}

\noindent\textbf{Measuring the charge-parity switch rate.}
Both the single-tone resonator and two-tone qubit spectroscopy (Fig.~\ref{fig:Fig2}a and b, respectively) exhibit random switching between the two spectral curves corresponding to the different parities.
In the displayed case, the qubit is in the charge regime ($E_{\rm J}/E_{\rm C}\sim3$), leading to distinct qubit transition frequencies for the two parities, $\omega_{\rm ge}^{\rm E}/2\pi=6.850~$GHz and $\omega_{\rm ge}^{\rm O}/2\pi=4.478$~GHz at $n_{\rm g}=0$, and hence different dispersive shifts of the resonator frequency.
By fitting both spectra to Eq.~(\ref{equ:Hamiltonian_qubit}) and the Jaynes-Cummings model~\cite{Blais2004}, we can extract the actual device parameters: $E_{\rm J}/h=4.6$~GHz, $E_{\rm C}/h=1.4$~GHz, and the qubit--resonator coupling $g/h=24$~MHz.

To track how the charge parity evolves over time, we repeatedly measured the transmitted signal. With our approach, events faster than the sampling rate of 3.3~kHz may be missed. However, if the parity switching is a random process without strong correlation, the only missed events would be those with parity consecutively switching for an even number of times during a short time interval. Such events are not the same type of random telegraph processes studied in this work. Indeed, a typical trace is shown in Fig.~\ref{fig:Fig2}c, displaying a telegraph signal that randomly switches between even and odd parity every few seconds. We acquired a few hundred such traces and computed their power spectral density (Fig.~\ref{fig:Fig2}d); the Lorentzian spectral shape is consistent with a random telegraph process. The spectral width is proportional to the average switching rate $\Gamma_P$; in this case, $T_P=1/\Gamma_P=$2.7~s, which is a state-of-the-art result for superconducting qubits.
Previously reported $T_P$ values range from 1~ms~\cite{Riste2013} to approximately 10 ms under similar shielding and filtering conditions~\cite{Serniak2019}, and have recently been prolonged to 100~ms level by creating a light-tight environment~\cite{Gordon2021}. Our result -- more than one order of magnitude better without sophisticated shielding -- implies the significant influence of the device geometry.
This influence, as we discuss next, is much stronger than the variation in switching rates (about a factor of 2) between devices with identical design measured in the same cooldown, and between different cooldowns for a given device (see Supplementary Fig.~5~\cite{supplement} for details).

\noindent\textbf{Effect of circuit geometry on parity switching.}
We performed a parametric study of the dependence of the charge-parity switching rate on the geometry to investigate the origin of the nonequilibrium quasiparticles in our devices.
The qubit structure, typically a few hundred microns in size, can be a good antenna~\cite{Rafferty2021}, channelling stray photons at a few hundred gigahertz to the junction.
The entire structure can be regarded as a pair of folded slots~\cite{supplement,Kraus1997} (Fig.~\ref{fig:Fig3}a, left) that support multiple resonant modes, determined primarily by the length of the long edges $L$ of the metallic pads. For the fundamental mode, $L=\lambda/2$, $\lambda$ being the effective mode wavelength obtainable from the real part of the input impedance $Z_{\rm rad}$ calculated via finite-element electromagnetic field simulations (Fig.~\ref{fig:Fig3}b).
For a capped qubit with the same geometry, the radiator mode frequency is slightly redshifted because of the additional capacitance between the metallic cap and the qubit. 
Moreover, because the cap behaves as a floating ground plane located in close proximity to the radiator, the induced currents also contribute to the radiated field; the virtual currents located on the other side of the cap are out of phase with respect to the currents on the qubit (Fig.~\ref{fig:Fig3}a, right), therefore cancelling each other and leading to near-zero radiated power. This explains the much smaller radiation impedance when the same qubit is capped (Fig.~\ref{fig:Fig3}b).
The real part of the input impedance is directly related to the power transfer efficiency of the pair-breaking photons~\cite{Rafferty2021}:
\begin{equation}
e_{\rm c}(f) = \frac{4{\rm Re}[Z_{\rm rad}] {\rm Re}[Z_{\rm J}]}{|Z_{\rm rad}+Z_{\rm J}|^2} \; ,
\label{equ:ec}
\end{equation}
where $Z_{\rm J}$ is the junction impedance.

Figure \ref{fig:Fig3}c shows the parity switching rate $\Gamma_P$ for qubits with different pad lengths, exhibiting a monotonic increase with the pad length $L$ by two orders of magnitude for the uncapped qubits.
With larger $L$, the frequency of the fundamental mode is reduced from 500~GHz ($L=80~\mu$m) to 160~GHz ($L=300~\mu$m), approaching twice the superconducting gap frequency $f^*=2\Delta/h$.
Consequently, the power transfer efficiency at this critical frequency is enhanced.
Note that for $L=240~\mu$m, $\Gamma_P\approx30~{\rm s^{-1}}$, in agreement with the switching rate measured for a device having similar $E_{\rm J}/E_{\rm C}$ ratio (20-30)~\cite{Serniak2019}.
Assuming the linear relation $\Gamma_P = \gamma\, e_{\rm c}(f^*)$, where $\gamma$ is a constant indicating the conversion efficiency between the incoming photon flux (excluding the geometry-dependent factor $e_{\rm c}$) and the observed parity switching events, we find that $\gamma=3\times10^5~\rm{s^{-1}}$ gives the best agreement between the experiments and simulation predictions. This value of $\gamma$ is used in all cases, since they share a nominally identical setup.
For the capped qubits with a similar size, $\Gamma_P$ is approximately an order of magnitude lower, consistent with the predictions.

The above result supports the hypothesis that pair-breaking photons absorbed by the antenna mode are responsible for the excessive quasiparticles in our devices; it also validates our method of protecting qubits from stray photons via capping, which is predicted to be effective across different regimes, from $E_{\rm C}>1$~GHz (charge qubits) to $E_{\rm C}<0.3$~GHz (transmon qubits).
The parity switching time $T_P$ for the low energy states of transmon qubits, while not detectable because of their insensitivity to charge, is estimated to be 10--100~ms (see Fig.~\ref{fig:Fig3}c).
Since $T_P$ can set an upper limit on the qubit coherence times~\cite{Catelani2014}, it is important to prevent nonequilibrium quasiparticles from compromising the qubit performance, especially with the state-of-the-art coherence time of transmon qubits approaching the millisecond mark~\cite{place_new_2021, wang2021transmon}.

We also investigated the dependence of the charge-parity switching rate on the pad-to-ground distance $d$ (the aperture size, see Fig.~\ref{fig:Fig1}b).
The experimental result again showed good agreement with predictions using the same method and parameters as before (Fig.~\ref{fig:Fig3}d).
Enlarging the gap between the pads and the ground plane increases the effective wavelength $\lambda$ of the fundamental mode, giving rise to larger $\Gamma_P$ values~\cite{supplement}.

In addition to parity switching, we observed significantly improved offset charge stability.
The charge offset for all six qubits on a single chip was monitored continuously and simultaneously via repeated spectral scans  (Fig.~\ref{fig:Fig3}e).
The occurrence of discrete charge jumps (jump amplitude $>0.1e$) was approximately 1--9 times over 40 hours (Fig.~\ref{fig:Fig3}f), corresponding to an offset charge jump rate of 0.007--0.07~mHz; the accumulated offset deviation during the long-term drift was within $1e$, a significantly less volatile result than those previously observed (jump rate $\sim 1.35\,$mHz)~\cite{Wilen2021}.
This difference can be explained by the reduced scattering cross section inside the substrate of our devices, which have smaller capacitor pads and a floating design~\cite{supplement}.
However, differences in the fabrication materials may be another important factor, as suggested recently in~\cite{Tennant2021}, where an average jump rate of $\sim0.7\,$mHz was measured.
Note that we do not observe simultaneous jumps between different qubits. Considering the relatively small footprint of our qubits and their separation (1.3~mm), this is consistent with the previous observation of charge jumps being correlated for qubits separated by up to 0.64~mm but not if separated by 3~mm~\cite{Wilen2021}.

We also measured the coherence properties of the qubits.
As shown in Fig.~\ref{fig:Fig3}g, the energy relaxation rates $\Gamma_1$ of the charge qubit (Q1) measured at different offset biases (different qubit frequencies) are relatively uniform, with slight variations between $10~\mu$s and $25~\mu$s, comparable to that of our regular transmon qubits fabricated using similar processes.
The spin-echo pure-dephasing rates $\Gamma_{\rm \phi}$ are in good agreement with a low-frequency, presumably $1/f$-type, charge-noise model.
The extracted noise spectral density is approximately $(2.0\times10^{-3}~e)^2$ at 1~Hz, in line with charge noise observed in other experiments~\cite{Verbrugh1995, Zimmerli1992Noise}.

\begin{figure}[ht]
\includegraphics{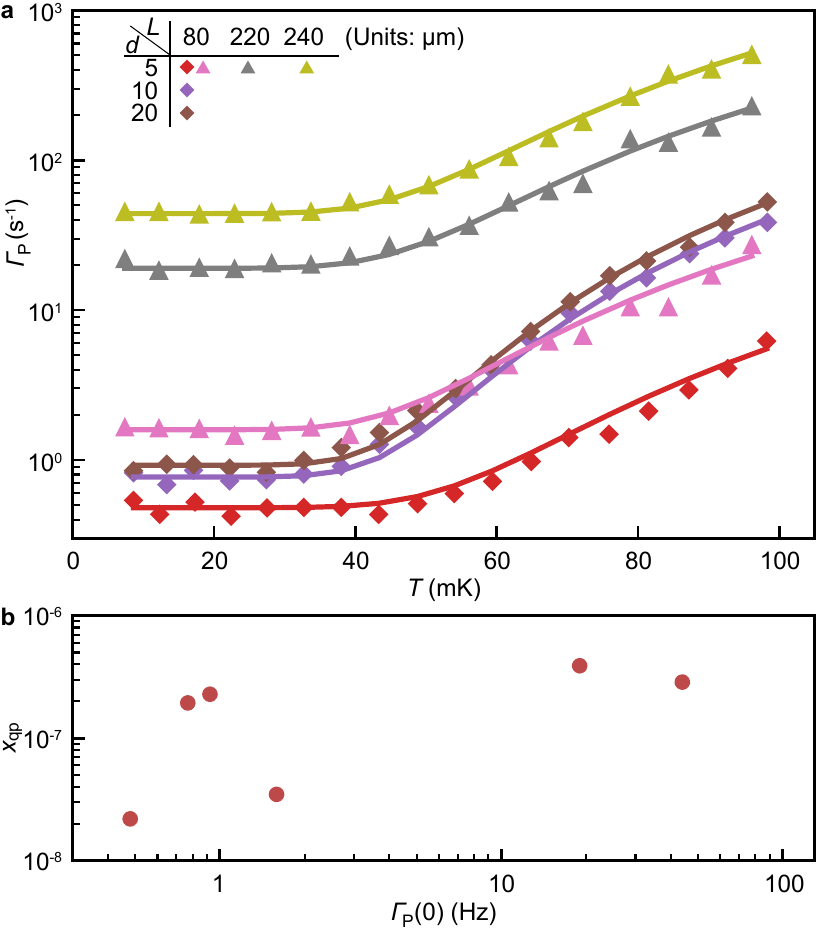} 
\caption{\textbf{Effect of temperature on parity switching}. 
\textbf{a}, Parity switching rates measured at $n_{\rm g}=0$ as a function of the nominal mixing chamber temperature $T$ for several uncapped qubits with different $L$ and $d$, whose combinations are colour-coded. Symbols identify different samples.
The solid lines are fit to Eq.~\eqref{equ:gammap_thermal}.
All curves share the same value for the superconducting gap in the pads ($2\Delta_0/h = 87\,$GHz) and in the junction leads ($2\Delta/h\simeq 99-105\,$GHz~\cite{supplement}) on a given chip. The only remaining free parameter for a single qubit is the normalized quasiparticle density $x_{\rm qp}$. 
\textbf{b}, Extracted quasiparticle density $x_{\rm qp}$ plotted versus the low-temperature parity switching rate $\Gamma_P(0)$.}
\label{fig:Fig4}
\end{figure}

\noindent\textbf{Effect of temperature on parity switching.} 
Next, we investigated how temperature affects the parity switching rate. After intentionally heating the mixing chamber, we measured the temperature dependence of $\Gamma_P$ at $n_{\rm g}=0$ for qubits with different $L$ and $d$ (Fig.~\ref{fig:Fig4}a).
At low temperatures, $\Gamma_P(T)$ did not display any clear temperature dependence and fluctuated around a qubit-specific average value $\Gamma_P(0)$ (up to these fluctuations, $\Gamma_P(0)$ agrees with the data in Figs.~\ref{fig:Fig3}c and d); $\Gamma_P(T)$ then started to increase with the temperature near $40-60\,$mK.

There is an energy difference of approximately $30~\mu$eV, equivalent to $350$~mK, between the superconducting gaps in the pads ($\Delta_0$) and in the junction strips ($\Delta$) because of the unequal aluminium film thicknesses (Fig.~\ref{fig:Fig1}e).
Therefore, at temperatures well below this value the pads act as quasiparticle traps~\cite{Riwar2019}. As the temperature increases, the quasiparticles can be thermally excited from the pads to the strips and hence reach the junction, which could explain the increase in $\Gamma_P(T)$ with temperature starting near 50~mK.


To quantify the above consideration, we can relate the parity switching rate to the normalized density of the quasiparticles in the pads $x_{\rm qp}\ll1$ (see Supplementary Note 8~\cite{supplement} for details):
\begin{align}\label{equ:gammap_thermal}
       \Gamma_P(T) & =\Gamma_P(0)\\  + & \frac{16E_{\rm J}}{\Delta}c_0^2\frac{\epsilon_0}{h}e^{\frac{-(\Delta-\Delta_0)}{k_{\rm B}T}}x_{\rm qp}\sqrt{\frac{\Delta_0}{2\pi k_{\rm B}T}}F\left(\frac{\epsilon_0}{2k_{\rm B}T},\frac{k_{\rm B}T}{2\Delta}\right)\!, \notag
\end{align}
where $\Gamma_P(0)$ accounts for all possible temperature-independent contributions, $\epsilon_0$ is the energy difference between $\ket{\rm g^E}$ and $\ket{\rm g^O}$, $c_0=|\bra{\rm g^E} \cos(\hat{\frac{\phi}{2}}) \ket{\rm g^O}|$ is the tunnelling matrix element between them, and the function $F(x,y)={\rm cosh}(x)[K_1(x)-xyK_0(x)]$ (with $K_i$ being the modified Bessel function of the second kind).
For each qubit, $\epsilon_0$ and $c_0$ can be evaluated using $E_{\rm J}$ and $E_{\rm C}$ values obtained from previous measurements.
We assume the same $\Delta$ for all qubits on a given chip and perform simultaneous fit the data in Fig.~\ref{fig:Fig4}a to Eq.~\eqref{equ:gammap_thermal}. We obtain $f^*=2\Delta/h$ between 99~GHz and 105~GHz (depending on the chip), which is consistent with a 30~nm aluminium thin film~\cite{Court2007}.
The quasiparticle density $x_{\rm qp}$ is the only adjustable parameter available to fit the temperature effect for a given qubit and Fig.~\ref{fig:Fig4}b shows the extracted $x_\mathrm{qp}$ as function of $\Gamma_P(0)$. Qualitatively, the increase in density with the parity switching rate indicates that pair breaking at the junction (followed by diffusion to and trapping in the pads) is a significant source of quasiparticles, with the steady-state nonequilibrium density determined by the balance between generation and recombination. Therefore, although our measurements of $\Gamma_P$ cannot directly distinguish between parity switching due to photon assisted tunneling or due to quasiparticles generated by other mechanisms, such as high-energy impacts of elevated temperature, the study of the temperature dependence enable us to conclude that these mechanisms, if present, are not the main sources of parity switching at low temperature (see Supplementary Note 8~\cite{supplement} for more details).

\section*{Discussion}
Our characterization of several tens of superconducting qubits with extended parameter regimes indicates that pair breaking at the junction by stray photons of sufficient energy is the main mechanism responsible for parity switching as well as the generation of excess quasiparticles. This mechanism is local and affects each qubit independently. 
Because the parity switching rate $\Gamma_P$ can be reduced by one to two orders of magnitude with simple engineering of the device geometry, e.g., using the capping technique and reducing the qubit footprint, novel miniaturized qubit designs~\cite{zhao_merged-element_2020,mamin_merged-element_2021,wang2021hexagonal} may be advantageous in this regard. Meanwhile, further studies are required to evaluate the quasiparticle trapping effectiveness of the lower-gap capacitor pads and to optimize their design. 

Finally, note that in our experiment we found no evidence of correlated increases in the parity switching rates of different qubits during simultaneous parity monitoring (similar to the simultaneous charge offset monitoring shown in Fig.~\ref{fig:Fig3}e). This is in contrast to the correlated frequency shifts measured in resonators~\cite{Cardani2020} or the correlated increase of the relaxation errors in qubits~\cite{GOOburst}. This suggests that our floating design with a ground plane (similar to the phonon traps of Ref.~\cite{Henriques2019}) and superconducting traps may also effectively suppress correlated errors, as suggested in Ref.~\cite{Martinis2021}.

\section*{Methods}

{\large Device fabrication and measurement setup}

The devices are made in a two-step process on c-plane sapphire wafers. In the first step, a layer of 100-nm-thick aluminum is deposited on the sapphire substrate at a growth rate of 1~nm/s with a base pressure of $10^{-10}$ Torr. The base metal is patterned with photolithography and subsequent dry etching using $\rm BCl_3/Cl_2$. In the second step, the Josephson junctions are made in the Manhattan style using the double-angle evaporation to form the $\rm Al/AlO_x/Al$ stack. The thickness of the first and second aluminium film is about 30~nm and 40~nm respectively. After ion milling, a final 200~nm-thick aluminum layer is deposited for making the bandage. For the flip-chip sample, the fabrication processes are identical. The two single-sided sapphire dies are bonded together using four rectangular spacers (2~mm$\times$2~mm$\times10~\mu$m) made of SU-8 photoresist at the corners.

The samples are mounted inside a BlueFors LD400 dilution refrigerator at a nominal base temperature less than 10~mK. In our standard setup, the sample is protected by a aluminum or copper holder box, a $\mu$-metal shield, a few layers of copper and aluminum shields, and an outer $\mu$-metal shield. Infrared filters are used in all control and readout lines. These measures help block stray photons from reaching the sample via open space and cables. More details can be found in Supplementary Note 2.

{\large Charge parity monitor}

For qubits with small $E_{\rm J}/E_{\rm C}$ ratio, the difference between the g-e transition frequencies of even and odd parity at $n_{\rm g}=0$ is relatively large (typically a few GHz). This leads to very different resonator frequencies due to strong qubit-resonator coupling. Utilizing the difference in resonator response, we send a probe tone at the resonator frequency of certain parity to detect the parity.
The measurement is done either with pulsed signals generated from an AWG and collected by a digitizer, or with continuous signals using a network analyzer.
In the pulsed case, the probe pulse is typically $10~\mu$s long and repeated every 0.3~ms. The single-shot result -- $99.14\%$ fidelity for parity classification -- is smoothed by taking a moving average to remove noise from thermal and measurement-induced excitation (see Supplementary Note 4 for more details).

For qubits with larger $E_{\rm J}/E_{\rm C}$ ratio (20$\sim$30), the frequency discrepancy between different parities becomes small (0.1$\sim$1~MHz) leading to unnoticeable difference in resonator frequency. Instead of direct dispersive readout, we use the Ramsey-type parity monitor as introduced in Ref.~\cite{Riste2013} for parity detection.
In the Ramsey experiment, we set the carrier frequency of the $\pi/2$-pulses at $\omega_{\rm drive}=(\omega_{\rm ge}^E+\omega_{\rm ge}^{\rm O})/2$ and the free-evolution time between the pulses to $\tau=\pi/2(\omega_{\rm ge}^E-\omega_{\rm ge}^{\rm O})$. Under such a pulse sequence, the qubit evolves to the excited (ground) state for even (odd) parity, allowing us to differentiate parity state.
The sequence is typically repeated every 0.1~ms.

\section*{Data Availability}
Source data are provided with this paper. All other data that support the plots within this paper and other findings of this study are available from the corresponding author upon reasonable request.

\bibliographystyle{naturemag}
\bibliography{main}

\begin{thebibliography}{10}
\expandafter\ifx\csname url\endcsname\relax
  \def\url#1{\texttt{#1}}\fi
\expandafter\ifx\csname urlprefix\endcsname\relax\def\urlprefix{URL }\fi
\providecommand{\bibinfo}[2]{#2}
\providecommand{\eprint}[2][]{\url{#2}}

\bibitem{SPLN}
\bibinfo{author}{Glazman, L.~I.} \& \bibinfo{author}{Catelani, G.}
\newblock \bibinfo{title}{{Bogoliubov Quasiparticles in Superconducting
  Qubits}}.
\newblock \emph{\bibinfo{journal}{SciPost Phys. Lect. Notes}}
  \textbf{\bibinfo{volume}{31}} (\bibinfo{year}{2021}).

\bibitem{MQE}
\bibinfo{author}{Catelani, G.} \& \bibinfo{author}{Pekola, J.~P.}
\newblock \bibinfo{title}{Using materials for quasiparticle engineering}.
\newblock \emph{\bibinfo{journal}{Mater. Quantum Technol.}}
  \textbf{\bibinfo{volume}{2}}, \bibinfo{pages}{013001} (\bibinfo{year}{2022}).

\bibitem{Cardani2020}
\bibinfo{author}{Cardani, L.} \emph{et~al.}
\newblock \bibinfo{title}{Reducing the impact of radioactivity on quantum
  circuits in a deep-underground facility}.
\newblock \emph{\bibinfo{journal}{Nat. Commun.}} \textbf{\bibinfo{volume}{12}},
  \bibinfo{pages}{2733} (\bibinfo{year}{2021}).

\bibitem{Wilen2021}
\bibinfo{author}{Wilen, C.~D.} \emph{et~al.}
\newblock \bibinfo{title}{{Correlated charge noise and relaxation errors in
  superconducting qubits}}.
\newblock \emph{\bibinfo{journal}{Nature}} \textbf{\bibinfo{volume}{594}},
  \bibinfo{pages}{369--373} (\bibinfo{year}{2021}).

\bibitem{GOOburst}
\bibinfo{author}{McEwen, M.} \emph{et~al.}
\newblock \bibinfo{title}{Resolving catastrophic error bursts from cosmic rays
  in large arrays of superconducting qubits}.
\newblock \emph{\bibinfo{journal}{Nat. Phys.}} \textbf{\bibinfo{volume}{18}},
  \bibinfo{pages}{107} (\bibinfo{year}{2022}).

\bibitem{karatsu2019mitigation}
\bibinfo{author}{Karatsu, K.} \emph{et~al.}
\newblock \bibinfo{title}{Mitigation of cosmic ray effect on microwave kinetic
  inductance detector arrays}.
\newblock \emph{\bibinfo{journal}{Applied Physics Letters}}
  \textbf{\bibinfo{volume}{114}}, \bibinfo{pages}{032601}
  (\bibinfo{year}{2019}).

\bibitem{MITrad}
\bibinfo{author}{Veps\"al\"ainen, A.~P.} \emph{et~al.}
\newblock \bibinfo{title}{Impact of ionizing radiation on superconducting qubit
  coherence}.
\newblock \emph{\bibinfo{journal}{Nature}} \textbf{\bibinfo{volume}{584}},
  \bibinfo{pages}{551} (\bibinfo{year}{2020}).

\bibitem{Houzet2019}
\bibinfo{author}{Houzet, M.}, \bibinfo{author}{Serniak, K.},
  \bibinfo{author}{Catelani, G.}, \bibinfo{author}{Devoret, M.} \&
  \bibinfo{author}{Glazman, L.~I.}
\newblock \bibinfo{title}{Photon-assisted charge-parity jumps in a
  superconducting qubit}.
\newblock \emph{\bibinfo{journal}{Phys. Rev. Lett.}}
  \textbf{\bibinfo{volume}{123}}, \bibinfo{pages}{107704}
  (\bibinfo{year}{2019}).

\bibitem{Catelani2011l}
\bibinfo{author}{Catelani, G.} \emph{et~al.}
\newblock \bibinfo{title}{Quasiparticle relaxation of superconducting qubits in
  the presence of flux}.
\newblock \emph{\bibinfo{journal}{Phys. Rev. Lett.}}
  \textbf{\bibinfo{volume}{106}}, \bibinfo{pages}{077002}
  (\bibinfo{year}{2011}).

\bibitem{Lenander2011}
\bibinfo{author}{Lenander, M.} \emph{et~al.}
\newblock \bibinfo{title}{Measurement of energy decay in superconducting qubits
  from nonequilibrium quasiparticles}.
\newblock \emph{\bibinfo{journal}{Phys. Rev. B}} \textbf{\bibinfo{volume}{84}},
  \bibinfo{pages}{024501} (\bibinfo{year}{2011}).

\bibitem{Paik2011}
\bibinfo{author}{Paik, H.} \emph{et~al.}
\newblock \bibinfo{title}{Observation of high coherence in josephson junction
  qubits measured in a three-dimensional circuit qed architecture}.
\newblock \emph{\bibinfo{journal}{Phys. Rev. Lett.}}
  \textbf{\bibinfo{volume}{107}}, \bibinfo{pages}{240501}
  (\bibinfo{year}{2011}).

\bibitem{Pop2014}
\bibinfo{author}{Pop, I.~M.} \emph{et~al.}
\newblock \bibinfo{title}{{Coherent suppression of electromagnetic dissipation
  due to superconducting quasiparticles}}.
\newblock \emph{\bibinfo{journal}{Nature}} \textbf{\bibinfo{volume}{508}},
  \bibinfo{pages}{369--372} (\bibinfo{year}{2014}).

\bibitem{Henriques2019}
\bibinfo{author}{Henriques, F.} \emph{et~al.}
\newblock \bibinfo{title}{Phonon traps reduce the quasiparticle density in
  superconducting circuits}.
\newblock \emph{\bibinfo{journal}{Appl. Phys. Lett.}}
  \textbf{\bibinfo{volume}{115}}, \bibinfo{pages}{212601}
  (\bibinfo{year}{2019}).

\bibitem{Gustavsson2016}
\bibinfo{author}{Gustavsson, S.} \emph{et~al.}
\newblock \bibinfo{title}{Suppressing relaxation in superconducting qubits by
  quasiparticle pumping}.
\newblock \emph{\bibinfo{journal}{Science}} \textbf{\bibinfo{volume}{354}},
  \bibinfo{pages}{1573--1577} (\bibinfo{year}{2016}).

\bibitem{mannila_superconductor_2021}
\bibinfo{author}{Mannila, E.~T.} \emph{et~al.}
\newblock \bibinfo{title}{A superconductor free of quasiparticles for seconds}.
\newblock \emph{\bibinfo{journal}{Nat. Phys.}} \textbf{\bibinfo{volume}{18}},
  \bibinfo{pages}{145} (\bibinfo{year}{2022}).

\bibitem{Riwar2016}
\bibinfo{author}{Riwar, R.-P.} \emph{et~al.}
\newblock \bibinfo{title}{Normal-metal quasiparticle traps for superconducting
  qubits}.
\newblock \emph{\bibinfo{journal}{Phys. Rev. B}} \textbf{\bibinfo{volume}{94}},
  \bibinfo{pages}{104516} (\bibinfo{year}{2016}).

\bibitem{arute_quantum_2019}
\bibinfo{author}{Arute, F.} \emph{et~al.}
\newblock \bibinfo{title}{Quantum supremacy using a programmable
  superconducting processor}.
\newblock \emph{\bibinfo{journal}{Nature}} \textbf{\bibinfo{volume}{574}},
  \bibinfo{pages}{505--510} (\bibinfo{year}{2019}).

\bibitem{Dunsworth2017}
\bibinfo{author}{Dunsworth, A.} \emph{et~al.}
\newblock \bibinfo{title}{{Characterization and reduction of capacitive loss
  induced by sub-micron Josephson junction fabrication in superconducting
  qubits}}.
\newblock \emph{\bibinfo{journal}{Appl. Phys. Lett.}}
  \textbf{\bibinfo{volume}{111}}, \bibinfo{pages}{022601}
  (\bibinfo{year}{2017}).

\bibitem{Serniak2019}
\bibinfo{author}{Serniak, K.} \emph{et~al.}
\newblock \bibinfo{title}{{Direct Dispersive Monitoring of Charge Parity in
  Offset-Charge-Sensitive Transmons}}.
\newblock \emph{\bibinfo{journal}{Phys. Rev. Applied}}
  \textbf{\bibinfo{volume}{12}}, \bibinfo{pages}{014052}
  (\bibinfo{year}{2019}).

\bibitem{Rafferty2021}
\bibinfo{author}{Rafferty, O.} \emph{et~al.}
\newblock \bibinfo{title}{{Spurious Antenna Modes of the Transmon Qubit}}
  (\bibinfo{year}{2021}).
\newblock \urlprefix\url{http://arxiv.org/abs/2103.06803}.
\newblock \eprint{2103.06803}.

\bibitem{Riwar2019}
\bibinfo{author}{Riwar, R.-P.} \& \bibinfo{author}{Catelani, G.}
\newblock \bibinfo{title}{Efficient quasiparticle traps with low dissipation
  through gap engineering}.
\newblock \emph{\bibinfo{journal}{Phys. Rev. B}}
  \textbf{\bibinfo{volume}{100}}, \bibinfo{pages}{144514}
  (\bibinfo{year}{2019}).

\bibitem{rosenberg_3d_2017}
\bibinfo{author}{Rosenberg, D.} \emph{et~al.}
\newblock \bibinfo{title}{{3D} integrated superconducting qubits}.
\newblock \emph{\bibinfo{journal}{npj Quantum Inf.}}
  \textbf{\bibinfo{volume}{3}}, \bibinfo{pages}{42} (\bibinfo{year}{2017}).

\bibitem{foxen_qubit_2017}
\bibinfo{author}{Foxen, B.} \emph{et~al.}
\newblock \bibinfo{title}{Qubit compatible superconducting interconnects}.
\newblock \emph{\bibinfo{journal}{Quantum Sci. Technol.}}
  \textbf{\bibinfo{volume}{3}}, \bibinfo{pages}{014005} (\bibinfo{year}{2017}).

\bibitem{satzinger_simple_2019}
\bibinfo{author}{Satzinger, K.~J.} \emph{et~al.}
\newblock \bibinfo{title}{Simple non-galvanic flip-chip integration method for
  hybrid quantum systems}.
\newblock \emph{\bibinfo{journal}{Appl. Phys. Lett.}}
  \textbf{\bibinfo{volume}{114}}, \bibinfo{pages}{173501}
  (\bibinfo{year}{2019}).

\bibitem{Blais2004}
\bibinfo{author}{Blais, A.}, \bibinfo{author}{Huang, R.-S.},
  \bibinfo{author}{Wallraff, A.}, \bibinfo{author}{Girvin, S.~M.} \&
  \bibinfo{author}{Schoelkopf, R.~J.}
\newblock \bibinfo{title}{Cavity quantum electrodynamics for superconducting
  electrical circuits: An architecture for quantum computation}.
\newblock \emph{\bibinfo{journal}{Phys. Rev. A}} \textbf{\bibinfo{volume}{69}},
  \bibinfo{pages}{062320} (\bibinfo{year}{2004}).

\bibitem{Koch2007}
\bibinfo{author}{Koch, J.} \emph{et~al.}
\newblock \bibinfo{title}{Charge-insensitive qubit design derived from the
  cooper pair box}.
\newblock \emph{\bibinfo{journal}{Phys. Rev. A}} \textbf{\bibinfo{volume}{76}},
  \bibinfo{pages}{042319} (\bibinfo{year}{2007}).

\bibitem{Nakamura1999}
\bibinfo{author}{Nakamura, Y.}, \bibinfo{author}{Pashkin, Y.} \&
  \bibinfo{author}{Tsai, J.}
\newblock \bibinfo{title}{Coherent control of macroscopic quantum states in a
  single-cooper-pair box}.
\newblock \emph{\bibinfo{journal}{Nature}} \textbf{\bibinfo{volume}{398}},
  \bibinfo{pages}{786--788} (\bibinfo{year}{1999}).

\bibitem{Duty2004}
\bibinfo{author}{Duty, T.}, \bibinfo{author}{Gunnarsson, D.},
  \bibinfo{author}{Bladh, K.} \& \bibinfo{author}{Delsing, P.}
\newblock \bibinfo{title}{Coherent dynamics of a josephson charge qubit}.
\newblock \emph{\bibinfo{journal}{Phys. Rev. B}} \textbf{\bibinfo{volume}{69}},
  \bibinfo{pages}{140503} (\bibinfo{year}{2004}).

\bibitem{Astafiev2004}
\bibinfo{author}{Astafiev, O.}, \bibinfo{author}{Pashkin, Y.~A.},
  \bibinfo{author}{Yamamoto, T.}, \bibinfo{author}{Nakamura, Y.} \&
  \bibinfo{author}{Tsai, J.~S.}
\newblock \bibinfo{title}{Single-shot measurement of the josephson charge
  qubit}.
\newblock \emph{\bibinfo{journal}{Phys. Rev. B}} \textbf{\bibinfo{volume}{69}},
  \bibinfo{pages}{180507} (\bibinfo{year}{2004}).

\bibitem{Kurter2021}
\bibinfo{author}{Kurter, C.} \emph{et~al.}
\newblock \bibinfo{title}{{Quasiparticle tunneling as a probe of Josephson
  junction quality and capacitor material in superconducting qubits}}.
\newblock \emph{\bibinfo{journal}{npj Quantum Inf.}}
  \textbf{\bibinfo{volume}{8}}, \bibinfo{pages}{31} (\bibinfo{year}{2022}).

\bibitem{Gordon2021}
\bibinfo{author}{Gordon, R.~T.} \emph{et~al.}
\newblock \bibinfo{title}{{Environmental Radiation Impact on Lifetimes and
  Quasiparticle Tunneling Rates of Fixed-Frequency Transmon Qubits}}.
\newblock \emph{\bibinfo{journal}{Appl. Phys. Lett.}}
  \textbf{\bibinfo{volume}{120}}, \bibinfo{pages}{074002}
  (\bibinfo{year}{2022}).

\bibitem{supplement}
\emph{\bibinfo{journal}{Supplementary Material}} .

\bibitem{Chubov1969}
\bibinfo{author}{Chubov, P.~N.}, \bibinfo{author}{Eremenko, V.~V.} \&
  \bibinfo{author}{Pilipenko, Y.~A.}
\newblock \bibinfo{title}{Dependence of the critical temperature and energy gap
  on the thickness of superconducting aluminum films}.
\newblock \emph{\bibinfo{journal}{Sov. Phys. JETP}}
  \textbf{\bibinfo{volume}{28}}, \bibinfo{pages}{389} (\bibinfo{year}{1969}).

\bibitem{Court2007}
\bibinfo{author}{Court, N.~A.}, \bibinfo{author}{Ferguson, A.~J.} \&
  \bibinfo{author}{Clark, R.~G.}
\newblock \bibinfo{title}{Energy gap measurement of nanostructured aluminium
  thin films for single cooper-pair devices}.
\newblock \emph{\bibinfo{journal}{Supercond. Sci. Technol.}}
  \textbf{\bibinfo{volume}{21}}, \bibinfo{pages}{015013}
  (\bibinfo{year}{2007}).

\bibitem{Nakamura1998}
\bibinfo{author}{Nakamura, Y.} \& \bibinfo{author}{Tsai, J.}
\newblock \bibinfo{title}{Photon-assisted cooper-pair tunneling in a
  superconducting single-electron transistor}.
\newblock \emph{\bibinfo{journal}{Solid-State Electronics}}
  \textbf{\bibinfo{volume}{42}}, \bibinfo{pages}{1471--1475}
  (\bibinfo{year}{1998}).

\bibitem{Riste2013}
\bibinfo{author}{Rist{\`{e}}, D.} \emph{et~al.}
\newblock \bibinfo{title}{{Millisecond charge-parity fluctuations and induced
  decoherence in a superconducting transmon qubit}}.
\newblock \emph{\bibinfo{journal}{Nat. Commun.}} \textbf{\bibinfo{volume}{4}},
  \bibinfo{pages}{1913} (\bibinfo{year}{2013}).

\bibitem{Kraus1997}
\bibinfo{author}{Kraus, J.~D.}
\newblock \emph{\bibinfo{title}{Antennas}} (\bibinfo{publisher}{McGraw-Hill
  Science}, \bibinfo{year}{2001}).

\bibitem{Catelani2014}
\bibinfo{author}{Catelani, G.}
\newblock \bibinfo{title}{Parity switching and decoherence by quasiparticles in
  single-junction transmons}.
\newblock \emph{\bibinfo{journal}{Phys. Rev. B}} \textbf{\bibinfo{volume}{89}},
  \bibinfo{pages}{094522} (\bibinfo{year}{2014}).

\bibitem{place_new_2021}
\bibinfo{author}{Place, A. P.~M.} \emph{et~al.}
\newblock \bibinfo{title}{New material platform for superconducting transmon
  qubits with coherence times exceeding 0.3 milliseconds}.
\newblock \emph{\bibinfo{journal}{Nat. Commun.}} \textbf{\bibinfo{volume}{12}},
  \bibinfo{pages}{1779} (\bibinfo{year}{2021}).

\bibitem{wang2021transmon}
\bibinfo{author}{Wang, C.} \emph{et~al.}
\newblock \bibinfo{title}{Transmon qubit with relaxation time exceeding 0.5
  milliseconds}.
\newblock \emph{\bibinfo{journal}{npj Quantum Inf.}}
  \textbf{\bibinfo{volume}{8}}, \bibinfo{pages}{3} (\bibinfo{year}{2022}).

\bibitem{Tennant2021}
\bibinfo{author}{Tennant, D.~M.} \emph{et~al.}
\newblock \bibinfo{title}{Low-frequency correlated charge-noise measurements
  across multiple energy transitions in a tantalum transmon}.
\newblock \emph{\bibinfo{journal}{PRX Quantum}} \textbf{\bibinfo{volume}{3}},
  \bibinfo{pages}{030307} (\bibinfo{year}{2022}).

\bibitem{Verbrugh1995}
\bibinfo{author}{Verbrugh, S.~M.}, \bibinfo{author}{Benhamadi, M.~L.},
  \bibinfo{author}{Visscher, E.~H.} \& \bibinfo{author}{Mooij, J.~E.}
\newblock \bibinfo{title}{Optimization of island size in single electron
  tunneling devices: Experiment and theory}.
\newblock \emph{\bibinfo{journal}{Journal of Applied Physics}}
  \textbf{\bibinfo{volume}{78}}, \bibinfo{pages}{2830--2836}
  (\bibinfo{year}{1995}).

\bibitem{Zimmerli1992Noise}
\bibinfo{author}{Zimmerli, G.}, \bibinfo{author}{Eiles, T.~M.},
  \bibinfo{author}{Kautz, R.~L.} \& \bibinfo{author}{Martinis, J.~M.}
\newblock \bibinfo{title}{Noise in the coulomb blockade electrometer}.
\newblock \emph{\bibinfo{journal}{Appl. Phys. Lett.}}
  \textbf{\bibinfo{volume}{61}}, \bibinfo{pages}{237--239}
  (\bibinfo{year}{1992}).

\bibitem{zhao_merged-element_2020}
\bibinfo{author}{Zhao, R.} \emph{et~al.}
\newblock \bibinfo{title}{Merged-{Element} {Transmon}}.
\newblock \emph{\bibinfo{journal}{Phys. Rev. Applied}}
  \textbf{\bibinfo{volume}{14}}, \bibinfo{pages}{064006}
  (\bibinfo{year}{2020}).
\newblock \bibinfo{note}{Publisher: American Physical Society}.

\bibitem{mamin_merged-element_2021}
\bibinfo{author}{Mamin, H.} \emph{et~al.}
\newblock \bibinfo{title}{Merged-{Element} {Transmons}: {Design} and {Qubit}
  {Performance}}.
\newblock \emph{\bibinfo{journal}{Phys. Rev. Applied}}
  \textbf{\bibinfo{volume}{16}}, \bibinfo{pages}{024023}
  (\bibinfo{year}{2021}).

\bibitem{wang2021hexagonal}
\bibinfo{author}{Wang, J.~I.} \emph{et~al.}
\newblock \bibinfo{title}{Hexagonal boron nitride (hbn) as a low-loss
  dielectric for superconducting quantum circuits and qubits}.
\newblock \emph{\bibinfo{journal}{Nat. Mater.}} \textbf{\bibinfo{volume}{21}},
  \bibinfo{pages}{398} (\bibinfo{year}{2022}).

\bibitem{Martinis2021}
\bibinfo{author}{Martinis, J.~M.}
\newblock \bibinfo{title}{Saving superconducting quantum processors from decay
  and correlated errors generated by gamma and cosmic rays}.
\newblock \emph{\bibinfo{journal}{npj Quantum Inf.}}
  \textbf{\bibinfo{volume}{7}}, \bibinfo{pages}{90} (\bibinfo{year}{2021}).

\end{thebibliography}


\begin{thebibliography}{21}%
\makeatletter
\providecommand \@ifxundefined [1]{%
 \@ifx{#1\undefined}
}%
\providecommand \@ifnum [1]{%
 \ifnum #1\expandafter \@firstoftwo
 \else \expandafter \@secondoftwo
 \fi
}%
\providecommand \@ifx [1]{%
 \ifx #1\expandafter \@firstoftwo
 \else \expandafter \@secondoftwo
 \fi
}%
\providecommand \natexlab [1]{#1}%
\providecommand \enquote  [1]{``#1''}%
\providecommand \bibnamefont  [1]{#1}%
\providecommand \bibfnamefont [1]{#1}%
\providecommand \citenamefont [1]{#1}%
\providecommand \href@noop [0]{\@secondoftwo}%
\providecommand \href [0]{\begingroup \@sanitize@url \@href}%
\providecommand \@href[1]{\@@startlink{#1}\@@href}%
\providecommand \@@href[1]{\endgroup#1\@@endlink}%
\providecommand \@sanitize@url [0]{\catcode `\\12\catcode `\$12\catcode
  `\&12\catcode `\#12\catcode `\^12\catcode `\_12\catcode `\%12\relax}%
\providecommand \@@startlink[1]{}%
\providecommand \@@endlink[0]{}%
\providecommand \url  [0]{\begingroup\@sanitize@url \@url }%
\providecommand \@url [1]{\endgroup\@href {#1}{\urlprefix }}%
\providecommand \urlprefix  [0]{URL }%
\providecommand \Eprint [0]{\href }%
\providecommand \doibase [0]{http://dx.doi.org/}%
\providecommand \selectlanguage [0]{\@gobble}%
\providecommand \bibinfo  [0]{\@secondoftwo}%
\providecommand \bibfield  [0]{\@secondoftwo}%
\providecommand \translation [1]{[#1]}%
\providecommand \BibitemOpen [0]{}%
\providecommand \bibitemStop [0]{}%
\providecommand \bibitemNoStop [0]{.\EOS\space}%
\providecommand \EOS [0]{\spacefactor3000\relax}%
\providecommand \BibitemShut  [1]{\csname bibitem#1\endcsname}%
\let\auto@bib@innerbib\@empty
\bibitem [{\citenamefont {Jin}\ \emph {et~al.}(2015)\citenamefont {Jin},
  \citenamefont {Kamal}, \citenamefont {Sears}, \citenamefont {Gudmundsen},
  \citenamefont {Hover}, \citenamefont {Miloshi}, \citenamefont {Slattery},
  \citenamefont {Yan}, \citenamefont {Yoder},\ and\ \citenamefont
  {Orlando}}]{Jin2015}%
  \BibitemOpen
  \bibfield  {author} {\bibinfo {author} {\bibfnamefont {X.~Y.}\ \bibnamefont
  {Jin}}, \bibinfo {author} {\bibfnamefont {A.}~\bibnamefont {Kamal}}, \bibinfo
  {author} {\bibfnamefont {A.~P.}\ \bibnamefont {Sears}}, \bibinfo {author}
  {\bibfnamefont {T.}~\bibnamefont {Gudmundsen}}, \bibinfo {author}
  {\bibfnamefont {D.}~\bibnamefont {Hover}}, \bibinfo {author} {\bibfnamefont
  {J.}~\bibnamefont {Miloshi}}, \bibinfo {author} {\bibfnamefont
  {R.}~\bibnamefont {Slattery}}, \bibinfo {author} {\bibfnamefont
  {F.}~\bibnamefont {Yan}}, \bibinfo {author} {\bibfnamefont {J.}~\bibnamefont
  {Yoder}}, \ and\ \bibinfo {author} {\bibfnamefont {T.~P.}\ \bibnamefont
  {Orlando}},\ }\href@noop {} {\bibfield  {journal} {\bibinfo  {journal}
  {Physical Review Letters}\ }\textbf {\bibinfo {volume} {114}},\ \bibinfo
  {pages} {565} (\bibinfo {year} {2015})}\BibitemShut {NoStop}%
\bibitem [{\citenamefont {Serniak}\ \emph {et~al.}(2019)\citenamefont
  {Serniak}, \citenamefont {Diamond}, \citenamefont {Hays}, \citenamefont
  {Fatemi}, \citenamefont {Shankar}, \citenamefont {Frunzio}, \citenamefont
  {Schoelkopf},\ and\ \citenamefont {Devoret}}]{Serniak2019}%
  \BibitemOpen
  \bibfield  {author} {\bibinfo {author} {\bibfnamefont {K.}~\bibnamefont
  {Serniak}}, \bibinfo {author} {\bibfnamefont {S.}~\bibnamefont {Diamond}},
  \bibinfo {author} {\bibfnamefont {M.}~\bibnamefont {Hays}}, \bibinfo {author}
  {\bibfnamefont {V.}~\bibnamefont {Fatemi}}, \bibinfo {author} {\bibfnamefont
  {S.}~\bibnamefont {Shankar}}, \bibinfo {author} {\bibfnamefont
  {L.}~\bibnamefont {Frunzio}}, \bibinfo {author} {\bibfnamefont {R.~J.}\
  \bibnamefont {Schoelkopf}}, \ and\ \bibinfo {author} {\bibfnamefont {M.~H.}\
  \bibnamefont {Devoret}},\ }\href {\doibase 10.1103/PhysRevApplied.12.014052}
  {\bibfield  {journal} {\bibinfo  {journal} {Phys. Rev. Applied}\ }\textbf
  {\bibinfo {volume} {12}},\ \bibinfo {pages} {014052} (\bibinfo {year}
  {2019})}\BibitemShut {NoStop}%
\bibitem [{\citenamefont {Gordon}\ \emph {et~al.}(2022)\citenamefont {Gordon},
  \citenamefont {Murray}, \citenamefont {Kurter}, \citenamefont {Sandberg},
  \citenamefont {Hall}, \citenamefont {Balakrishnan}, \citenamefont {Shelby},
  \citenamefont {Wacaser}, \citenamefont {Stabile}, \citenamefont {Sleight},
  \citenamefont {Brink}, \citenamefont {Rothwell}, \citenamefont {Rodbell},
  \citenamefont {Dial},\ and\ \citenamefont {Steffen}}]{Gordon2021}%
  \BibitemOpen
  \bibfield  {author} {\bibinfo {author} {\bibfnamefont {R.}~\bibnamefont
  {Gordon}}, \bibinfo {author} {\bibfnamefont {C.}~\bibnamefont {Murray}},
  \bibinfo {author} {\bibfnamefont {C.}~\bibnamefont {Kurter}}, \bibinfo
  {author} {\bibfnamefont {M.}~\bibnamefont {Sandberg}}, \bibinfo {author}
  {\bibfnamefont {S.}~\bibnamefont {Hall}}, \bibinfo {author} {\bibfnamefont
  {K.}~\bibnamefont {Balakrishnan}}, \bibinfo {author} {\bibfnamefont
  {R.}~\bibnamefont {Shelby}}, \bibinfo {author} {\bibfnamefont
  {B.}~\bibnamefont {Wacaser}}, \bibinfo {author} {\bibfnamefont
  {A.}~\bibnamefont {Stabile}}, \bibinfo {author} {\bibfnamefont
  {J.}~\bibnamefont {Sleight}}, \bibinfo {author} {\bibfnamefont
  {M.}~\bibnamefont {Brink}}, \bibinfo {author} {\bibfnamefont
  {M.}~\bibnamefont {Rothwell}}, \bibinfo {author} {\bibfnamefont
  {K.}~\bibnamefont {Rodbell}}, \bibinfo {author} {\bibfnamefont
  {O.}~\bibnamefont {Dial}}, \ and\ \bibinfo {author} {\bibfnamefont
  {M.}~\bibnamefont {Steffen}},\ }\href {\doibase 10.1063/5.0078785} {\bibfield
   {journal} {\bibinfo  {journal} {Appl. Phys. Lett.}\ }\textbf {\bibinfo
  {volume} {120}},\ \bibinfo {pages} {074002} (\bibinfo {year}
  {2022})}\BibitemShut {NoStop}%
\bibitem [{\citenamefont {Blais}\ \emph {et~al.}(2004)\citenamefont {Blais},
  \citenamefont {Huang}, \citenamefont {Wallraff}, \citenamefont {Girvin},\
  and\ \citenamefont {Schoelkopf}}]{Blais2004}%
  \BibitemOpen
  \bibfield  {author} {\bibinfo {author} {\bibfnamefont {A.}~\bibnamefont
  {Blais}}, \bibinfo {author} {\bibfnamefont {R.-S.}\ \bibnamefont {Huang}},
  \bibinfo {author} {\bibfnamefont {A.}~\bibnamefont {Wallraff}}, \bibinfo
  {author} {\bibfnamefont {S.~M.}\ \bibnamefont {Girvin}}, \ and\ \bibinfo
  {author} {\bibfnamefont {R.~J.}\ \bibnamefont {Schoelkopf}},\ }\href
  {\doibase 10.1103/PhysRevA.69.062320} {\bibfield  {journal} {\bibinfo
  {journal} {Phys. Rev. A}\ }\textbf {\bibinfo {volume} {69}},\ \bibinfo
  {pages} {062320} (\bibinfo {year} {2004})}\BibitemShut {NoStop}%
\bibitem [{\citenamefont {Rist{\`{e}}}\ \emph {et~al.}(2013)\citenamefont
  {Rist{\`{e}}}, \citenamefont {Bultink}, \citenamefont {Tiggelman},
  \citenamefont {Schouten}, \citenamefont {Lehnert},\ and\ \citenamefont
  {Dicarlo}}]{Riste2013}%
  \BibitemOpen
  \bibfield  {author} {\bibinfo {author} {\bibfnamefont {D.}~\bibnamefont
  {Rist{\`{e}}}}, \bibinfo {author} {\bibfnamefont {C.~C.}\ \bibnamefont
  {Bultink}}, \bibinfo {author} {\bibfnamefont {M.~J.}\ \bibnamefont
  {Tiggelman}}, \bibinfo {author} {\bibfnamefont {R.~N.}\ \bibnamefont
  {Schouten}}, \bibinfo {author} {\bibfnamefont {K.~W.}\ \bibnamefont
  {Lehnert}}, \ and\ \bibinfo {author} {\bibfnamefont {L.}~\bibnamefont
  {Dicarlo}},\ }\href {\doibase 10.1038/ncomms2936} {\bibfield  {journal}
  {\bibinfo  {journal} {Nat. Commun.}\ }\textbf {\bibinfo {volume} {4}},\
  \bibinfo {pages} {1913} (\bibinfo {year} {2013})}\BibitemShut {NoStop}%
\bibitem [{\citenamefont {Rafferty}\ \emph {et~al.}(2021)\citenamefont
  {Rafferty}, \citenamefont {Patel}, \citenamefont {Liu}, \citenamefont
  {Abdullah}, \citenamefont {Wilen}, \citenamefont {Harrison},\ and\
  \citenamefont {McDermott}}]{Rafferty2021}%
  \BibitemOpen
  \bibfield  {author} {\bibinfo {author} {\bibfnamefont {O.}~\bibnamefont
  {Rafferty}}, \bibinfo {author} {\bibfnamefont {S.}~\bibnamefont {Patel}},
  \bibinfo {author} {\bibfnamefont {C.~H.}\ \bibnamefont {Liu}}, \bibinfo
  {author} {\bibfnamefont {S.}~\bibnamefont {Abdullah}}, \bibinfo {author}
  {\bibfnamefont {C.~D.}\ \bibnamefont {Wilen}}, \bibinfo {author}
  {\bibfnamefont {D.~C.}\ \bibnamefont {Harrison}}, \ and\ \bibinfo {author}
  {\bibfnamefont {R.}~\bibnamefont {McDermott}},\ }\href
  {http://arxiv.org/abs/2103.06803} {\  (\bibinfo {year} {2021})},\ \Eprint
  {http://arxiv.org/abs/2103.06803} {arXiv:2103.06803} \BibitemShut {NoStop}%
\bibitem [{\citenamefont {Kraus}(2001)}]{Kraus1997}%
  \BibitemOpen
  \bibfield  {author} {\bibinfo {author} {\bibfnamefont {J.~D.}\ \bibnamefont
  {Kraus}},\ }\href@noop {} {\emph {\bibinfo {title} {Antennas for All
  Applications}}}\ (\bibinfo  {publisher} {McGraw-Hill Science},\ \bibinfo
  {year} {2001})\BibitemShut {NoStop}%
\bibitem [{\citenamefont {Pozar}(2011)}]{Pozar2011}%
  \BibitemOpen
  \bibfield  {author} {\bibinfo {author} {\bibfnamefont {D.~M.}\ \bibnamefont
  {Pozar}},\ }\href@noop {} {\emph {\bibinfo {title} {Microwave Engineering}}}\
  (\bibinfo  {publisher} {John Wiley},\ \bibinfo {year} {2011})\BibitemShut
  {NoStop}%
\bibitem [{\citenamefont {AnsysHFSS}()}]{AnsysHFSS}%
  \BibitemOpen
  \bibfield  {author} {\bibinfo {author} {\bibnamefont {AnsysHFSS}},\
  }\href@noop {} {}\bibinfo {howpublished}
  {\url{https://www.ansys.com/products/electronics/ansys-hfss}}\BibitemShut
  {NoStop}%
\bibitem [{\citenamefont {Orfanidis}(2008)}]{Sophocles2008}%
  \BibitemOpen
  \bibfield  {author} {\bibinfo {author} {\bibfnamefont {S.~J.}\ \bibnamefont
  {Orfanidis}},\ }\href@noop {} {\emph {\bibinfo {title} {Electromagnetic Waves
  and Antennas}}}\ (\bibinfo  {publisher}
  {http://eceweb1.rutgers.edu/orfanidi/ewa/},\ \bibinfo {year} {2008})\ pp.\
  \bibinfo {pages} {749--751}\BibitemShut {NoStop}%
\bibitem [{\citenamefont {Wilen}\ \emph {et~al.}(2021)\citenamefont {Wilen},
  \citenamefont {Abdullah}, \citenamefont {Kurinsky}, \citenamefont {Stanford},
  \citenamefont {Cardani}, \citenamefont {D'Imperio}, \citenamefont {Tomei},
  \citenamefont {Faoro}, \citenamefont {Ioffe}, \citenamefont {Liu},
  \citenamefont {Opremcak}, \citenamefont {Christensen}, \citenamefont
  {DuBois},\ and\ \citenamefont {McDermott}}]{Wilen2021}%
  \BibitemOpen
  \bibfield  {author} {\bibinfo {author} {\bibfnamefont {C.~D.}\ \bibnamefont
  {Wilen}}, \bibinfo {author} {\bibfnamefont {S.}~\bibnamefont {Abdullah}},
  \bibinfo {author} {\bibfnamefont {N.~A.}\ \bibnamefont {Kurinsky}}, \bibinfo
  {author} {\bibfnamefont {C.}~\bibnamefont {Stanford}}, \bibinfo {author}
  {\bibfnamefont {L.}~\bibnamefont {Cardani}}, \bibinfo {author} {\bibfnamefont
  {G.}~\bibnamefont {D'Imperio}}, \bibinfo {author} {\bibfnamefont
  {C.}~\bibnamefont {Tomei}}, \bibinfo {author} {\bibfnamefont
  {L.}~\bibnamefont {Faoro}}, \bibinfo {author} {\bibfnamefont {L.~B.}\
  \bibnamefont {Ioffe}}, \bibinfo {author} {\bibfnamefont {C.~H.}\ \bibnamefont
  {Liu}}, \bibinfo {author} {\bibfnamefont {A.}~\bibnamefont {Opremcak}},
  \bibinfo {author} {\bibfnamefont {B.~G.}\ \bibnamefont {Christensen}},
  \bibinfo {author} {\bibfnamefont {J.~L.}\ \bibnamefont {DuBois}}, \ and\
  \bibinfo {author} {\bibfnamefont {R.}~\bibnamefont {McDermott}},\ }\href
  {\doibase 10.1038/s41586-021-03557-5} {\bibfield  {journal} {\bibinfo
  {journal} {Nature}\ }\textbf {\bibinfo {volume} {594}},\ \bibinfo {pages}
  {369} (\bibinfo {year} {2021})},\ \Eprint {http://arxiv.org/abs/2012.06029}
  {2012.06029} \BibitemShut {NoStop}%
\bibitem [{\citenamefont {Balanis}(2012)}]{Balanis2012}%
  \BibitemOpen
  \bibfield  {author} {\bibinfo {author} {\bibfnamefont {C.~A.}\ \bibnamefont
  {Balanis}},\ }\href@noop {} {\emph {\bibinfo {title} {Advanced Engineering
  Electromagnetics}}}\ (\bibinfo  {publisher} {John Wiley},\ \bibinfo {year}
  {2012})\BibitemShut {NoStop}%
\bibitem [{\citenamefont {Catelani}(2014)}]{Catelani2014}%
  \BibitemOpen
  \bibfield  {author} {\bibinfo {author} {\bibfnamefont {G.}~\bibnamefont
  {Catelani}},\ }\href {\doibase 10.1103/PhysRevB.89.094522} {\bibfield
  {journal} {\bibinfo  {journal} {Phys. Rev. B}\ }\textbf {\bibinfo {volume}
  {89}},\ \bibinfo {pages} {094522} (\bibinfo {year} {2014})}\BibitemShut
  {NoStop}%
\bibitem [{\citenamefont {Houzet}\ \emph {et~al.}(2019)\citenamefont {Houzet},
  \citenamefont {Serniak}, \citenamefont {Catelani}, \citenamefont {Devoret},\
  and\ \citenamefont {Glazman}}]{Houzet2019}%
  \BibitemOpen
  \bibfield  {author} {\bibinfo {author} {\bibfnamefont {M.}~\bibnamefont
  {Houzet}}, \bibinfo {author} {\bibfnamefont {K.}~\bibnamefont {Serniak}},
  \bibinfo {author} {\bibfnamefont {G.}~\bibnamefont {Catelani}}, \bibinfo
  {author} {\bibfnamefont {M.}~\bibnamefont {Devoret}}, \ and\ \bibinfo
  {author} {\bibfnamefont {L.~I.}\ \bibnamefont {Glazman}},\ }\href {\doibase
  10.1103/PhysRevLett.123.107704} {\bibfield  {journal} {\bibinfo  {journal}
  {Phys. Rev. Lett.}\ }\textbf {\bibinfo {volume} {123}},\ \bibinfo {pages}
  {107704} (\bibinfo {year} {2019})}\BibitemShut {NoStop}%
\bibitem [{\citenamefont {Yan}\ \emph {et~al.}(2016)\citenamefont {Yan},
  \citenamefont {Gustavsson}, \citenamefont {Kamal}, \citenamefont {Birenbaum},
  \citenamefont {Sears}, \citenamefont {Hover}, \citenamefont {Gudmundsen},
  \citenamefont {Rosenberg}, \citenamefont {Samach}, \citenamefont {Weber},
  \citenamefont {Yoder}, \citenamefont {Orlando}, \citenamefont {Clarke},
  \citenamefont {Kerman},\ and\ \citenamefont {Oliver}}]{yan_flux_2016}%
  \BibitemOpen
  \bibfield  {author} {\bibinfo {author} {\bibfnamefont {F.}~\bibnamefont
  {Yan}}, \bibinfo {author} {\bibfnamefont {S.}~\bibnamefont {Gustavsson}},
  \bibinfo {author} {\bibfnamefont {A.}~\bibnamefont {Kamal}}, \bibinfo
  {author} {\bibfnamefont {J.}~\bibnamefont {Birenbaum}}, \bibinfo {author}
  {\bibfnamefont {A.~P.}\ \bibnamefont {Sears}}, \bibinfo {author}
  {\bibfnamefont {D.}~\bibnamefont {Hover}}, \bibinfo {author} {\bibfnamefont
  {T.~J.}\ \bibnamefont {Gudmundsen}}, \bibinfo {author} {\bibfnamefont
  {D.}~\bibnamefont {Rosenberg}}, \bibinfo {author} {\bibfnamefont
  {G.}~\bibnamefont {Samach}}, \bibinfo {author} {\bibfnamefont
  {S.}~\bibnamefont {Weber}}, \bibinfo {author} {\bibfnamefont {J.~L.}\
  \bibnamefont {Yoder}}, \bibinfo {author} {\bibfnamefont {T.~P.}\ \bibnamefont
  {Orlando}}, \bibinfo {author} {\bibfnamefont {J.}~\bibnamefont {Clarke}},
  \bibinfo {author} {\bibfnamefont {A.~J.}\ \bibnamefont {Kerman}}, \ and\
  \bibinfo {author} {\bibfnamefont {W.~D.}\ \bibnamefont {Oliver}},\ }\href
  {\doibase 10.1038/ncomms12964} {\bibfield  {journal} {\bibinfo  {journal}
  {Nat. Commun.}\ }\textbf {\bibinfo {volume} {7}},\ \bibinfo {pages} {12964}
  (\bibinfo {year} {2016})}\BibitemShut {NoStop}%
\bibitem [{\citenamefont {Chubov}\ \emph {et~al.}(1969)\citenamefont {Chubov},
  \citenamefont {Eremenko},\ and\ \citenamefont {Pilipenko}}]{Chubov1969}%
  \BibitemOpen
  \bibfield  {author} {\bibinfo {author} {\bibfnamefont {P.~N.}\ \bibnamefont
  {Chubov}}, \bibinfo {author} {\bibfnamefont {V.~V.}\ \bibnamefont
  {Eremenko}}, \ and\ \bibinfo {author} {\bibfnamefont {Y.~A.}\ \bibnamefont
  {Pilipenko}},\ }\href@noop {} {\bibfield  {journal} {\bibinfo  {journal}
  {Sov. Phys. JETP}\ }\textbf {\bibinfo {volume} {28}},\ \bibinfo {pages} {389}
  (\bibinfo {year} {1969})}\BibitemShut {NoStop}%
\bibitem [{\citenamefont {Court}\ \emph {et~al.}(2007)\citenamefont {Court},
  \citenamefont {Ferguson},\ and\ \citenamefont {Clark}}]{Court2007}%
  \BibitemOpen
  \bibfield  {author} {\bibinfo {author} {\bibfnamefont {N.~A.}\ \bibnamefont
  {Court}}, \bibinfo {author} {\bibfnamefont {A.~J.}\ \bibnamefont {Ferguson}},
  \ and\ \bibinfo {author} {\bibfnamefont {R.~G.}\ \bibnamefont {Clark}},\
  }\href {\doibase 10.1088/0953-2048/21/01/015013} {\bibfield  {journal}
  {\bibinfo  {journal} {Supercond. Sci. Technol.}\ }\textbf {\bibinfo {volume}
  {21}},\ \bibinfo {pages} {015013} (\bibinfo {year} {2007})}\BibitemShut
  {NoStop}%
\bibitem [{\citenamefont {Riwar}\ and\ \citenamefont
  {Catelani}(2019)}]{Riwar2019}%
  \BibitemOpen
  \bibfield  {author} {\bibinfo {author} {\bibfnamefont {R.-P.}\ \bibnamefont
  {Riwar}}\ and\ \bibinfo {author} {\bibfnamefont {G.}~\bibnamefont
  {Catelani}},\ }\href {\doibase 10.1103/PhysRevB.100.144514} {\bibfield
  {journal} {\bibinfo  {journal} {Phys. Rev. B}\ }\textbf {\bibinfo {volume}
  {100}},\ \bibinfo {pages} {144514} (\bibinfo {year} {2019})}\BibitemShut
  {NoStop}%
\bibitem [{\citenamefont {Pop}\ \emph {et~al.}(2014)\citenamefont {Pop},
  \citenamefont {Geerlings}, \citenamefont {Catelani}, \citenamefont
  {Schoelkopf}, \citenamefont {Glazman},\ and\ \citenamefont
  {Devoret}}]{Pop2014}%
  \BibitemOpen
  \bibfield  {author} {\bibinfo {author} {\bibfnamefont {I.~M.}\ \bibnamefont
  {Pop}}, \bibinfo {author} {\bibfnamefont {K.}~\bibnamefont {Geerlings}},
  \bibinfo {author} {\bibfnamefont {G.}~\bibnamefont {Catelani}}, \bibinfo
  {author} {\bibfnamefont {R.~J.}\ \bibnamefont {Schoelkopf}}, \bibinfo
  {author} {\bibfnamefont {L.~I.}\ \bibnamefont {Glazman}}, \ and\ \bibinfo
  {author} {\bibfnamefont {M.~H.}\ \bibnamefont {Devoret}},\ }\href {\doibase
  10.1038/nature13017} {\bibfield  {journal} {\bibinfo  {journal} {Nature}\
  }\textbf {\bibinfo {volume} {508}},\ \bibinfo {pages} {369} (\bibinfo {year}
  {2014})}\BibitemShut {NoStop}%
\bibitem [{\citenamefont {Wang}\ \emph {et~al.}(2014)\citenamefont {Wang},
  \citenamefont {Gao}, \citenamefont {Pop}, \citenamefont {Vool}, \citenamefont
  {Axline}, \citenamefont {Brecht}, \citenamefont {Heeres}, \citenamefont
  {Frunzio}, \citenamefont {Devoret}, \citenamefont {Catelani}, \citenamefont
  {Glazman},\ and\ \citenamefont {Schoelkopf}}]{Wang2014}%
  \BibitemOpen
  \bibfield  {author} {\bibinfo {author} {\bibfnamefont {C.}~\bibnamefont
  {Wang}}, \bibinfo {author} {\bibfnamefont {Y.~Y.}\ \bibnamefont {Gao}},
  \bibinfo {author} {\bibfnamefont {I.~M.}\ \bibnamefont {Pop}}, \bibinfo
  {author} {\bibfnamefont {U.}~\bibnamefont {Vool}}, \bibinfo {author}
  {\bibfnamefont {C.}~\bibnamefont {Axline}}, \bibinfo {author} {\bibfnamefont
  {T.}~\bibnamefont {Brecht}}, \bibinfo {author} {\bibfnamefont {R.~W.}\
  \bibnamefont {Heeres}}, \bibinfo {author} {\bibfnamefont {L.}~\bibnamefont
  {Frunzio}}, \bibinfo {author} {\bibfnamefont {M.~H.}\ \bibnamefont
  {Devoret}}, \bibinfo {author} {\bibfnamefont {G.}~\bibnamefont {Catelani}},
  \bibinfo {author} {\bibfnamefont {L.~I.}\ \bibnamefont {Glazman}}, \ and\
  \bibinfo {author} {\bibfnamefont {R.~J.}\ \bibnamefont {Schoelkopf}},\ }\href
  {\doibase 10.1038/ncomms6836} {\bibfield  {journal} {\bibinfo  {journal}
  {Nat. Commun.}\ }\textbf {\bibinfo {volume} {5}},\ \bibinfo {pages} {5836}
  (\bibinfo {year} {2014})}\BibitemShut {NoStop}%
\bibitem [{\citenamefont {Cardani}\ \emph {et~al.}(2021)\citenamefont
  {Cardani}, \citenamefont {Valenti}, \citenamefont {Casali}, \citenamefont
  {Catelani}, \citenamefont {Charpentier}, \citenamefont {Clemenza},
  \citenamefont {Colantoni}, \citenamefont {Cruciani}, \citenamefont
  {D'Imperio}, \citenamefont {Gironi}, \citenamefont {Gr\"unhaupt},
  \citenamefont {Gusenkova}, \citenamefont {Henriques}, \citenamefont {Lagoin},
  \citenamefont {Martinez}, \citenamefont {Pettinari}, \citenamefont {Rusconi},
  \citenamefont {Sander}, \citenamefont {Tomei}, \citenamefont {Ustinov},
  \citenamefont {Weber}, \citenamefont {Wernsdorfer}, \citenamefont {Vignati},
  \citenamefont {Pirro},\ and\ \citenamefont {Pop}}]{Cardani2020}%
  \BibitemOpen
  \bibfield  {author} {\bibinfo {author} {\bibfnamefont {L.}~\bibnamefont
  {Cardani}}, \bibinfo {author} {\bibfnamefont {F.}~\bibnamefont {Valenti}},
  \bibinfo {author} {\bibfnamefont {N.}~\bibnamefont {Casali}}, \bibinfo
  {author} {\bibfnamefont {G.}~\bibnamefont {Catelani}}, \bibinfo {author}
  {\bibfnamefont {T.}~\bibnamefont {Charpentier}}, \bibinfo {author}
  {\bibfnamefont {M.}~\bibnamefont {Clemenza}}, \bibinfo {author}
  {\bibfnamefont {I.}~\bibnamefont {Colantoni}}, \bibinfo {author}
  {\bibfnamefont {A.}~\bibnamefont {Cruciani}}, \bibinfo {author}
  {\bibfnamefont {G.}~\bibnamefont {D'Imperio}}, \bibinfo {author}
  {\bibfnamefont {L.}~\bibnamefont {Gironi}}, \bibinfo {author} {\bibfnamefont
  {L.}~\bibnamefont {Gr\"unhaupt}}, \bibinfo {author} {\bibfnamefont
  {D.}~\bibnamefont {Gusenkova}}, \bibinfo {author} {\bibfnamefont
  {F.}~\bibnamefont {Henriques}}, \bibinfo {author} {\bibfnamefont
  {M.}~\bibnamefont {Lagoin}}, \bibinfo {author} {\bibfnamefont
  {M.}~\bibnamefont {Martinez}}, \bibinfo {author} {\bibfnamefont
  {G.}~\bibnamefont {Pettinari}}, \bibinfo {author} {\bibfnamefont
  {C.}~\bibnamefont {Rusconi}}, \bibinfo {author} {\bibfnamefont
  {O.}~\bibnamefont {Sander}}, \bibinfo {author} {\bibfnamefont
  {C.}~\bibnamefont {Tomei}}, \bibinfo {author} {\bibfnamefont {A.~V.}\
  \bibnamefont {Ustinov}}, \bibinfo {author} {\bibfnamefont {M.}~\bibnamefont
  {Weber}}, \bibinfo {author} {\bibfnamefont {W.}~\bibnamefont {Wernsdorfer}},
  \bibinfo {author} {\bibfnamefont {M.}~\bibnamefont {Vignati}}, \bibinfo
  {author} {\bibfnamefont {S.}~\bibnamefont {Pirro}}, \ and\ \bibinfo {author}
  {\bibfnamefont {I.~M.}\ \bibnamefont {Pop}},\ }\href {\doibase
  10.1038/s41467-021-23032-z} {\bibfield  {journal} {\bibinfo  {journal} {Nat.
  Commun.}\ }\textbf {\bibinfo {volume} {12}},\ \bibinfo {pages} {2733}
  (\bibinfo {year} {2021})}\BibitemShut {NoStop}%
\end{thebibliography}%

\vspace{0.2in}

\vbox{}

\noindent \textbf{Acknowledgments}\\
We would like to thank Robert McDermott for enlightening discussions.
This work was supported by 
the Key-Area Research and Development Program of Guang-Dong Province (Grant No. 2018B030326001), 
the National Natural Science Foundation (NSF) of China (U1801661), 
the Guangdong Innovative and Entrepreneurial Research Team Program (2016ZT06D348), 
the Guangdong Provincial Key Laboratory (Grant No.2019B121203002), 
the Natural Science Foundation of Guangdong Province (2017B030308003), 
the Science, Technology and Innovation Commission of Shenzhen Municipality (KYTDPT20181011104202253), 
the Shenzhen-Hong Kong Cooperation Zone for Technology and Innovation (HZQB-KCZYB-2020050),
and the NSF of Beijing (Grant No. Z190012).
L.H. acknowledges support from the National Natural Science Foundation of China (Grant No. 11905098) and the Strategic Priority Research Program of Chinese Academy of Sciences (Grant No. XDB28000000).
Z.H.J. acknowledges support from the National Natural Science Foundation of China (Grants No. 62122019) and the High Level Innovation and Entrepreneurial Research Team Program in Jiangsu Province.
J.L. acknowledges support from the National Natural Science Foundation of
China (Grant No. 11874065).
G.C. acknowledges support by the German Federal Ministry of Education and Research (BMBF), funding program ``Quantum technologies - from basic research to market'', project QSolid (Grant No. 13N16149).

\vbox{}

\noindent \textbf{Author contributions}\\
L.H. and F.Y. conceived and designed the experiment. 
X.P., H.Y. and Y.Z. designed the device and performed the simulation. 
Y.Z. and L.Z. performed sample fabrication.
X.P., J.L., L.H. and F.Y. conducted the measurements.
X.P., G.C. and L.H. analysed the data.
Z.J. and G.C. provided the theory. 
X.P., Z.J., G.C., L.H. and F.Y. wrote the manuscript. 
S.L., L.H. and F.Y. supervised the project. 
All authors discussed the results and contributed to revising the manuscript and the supplementary material. 
All authors contributed to the experimental and theoretical infrastructure to enable the experiment.

\vbox{}

\noindent \textbf{Additional information}\\
Supplementary material is available in the online version of the paper.

\vbox{}

\noindent \textbf{Competing interests}\\
The authors declare no competing interests.

\vbox{}

\clearpage{}

\newpage{}

\newpage{}

\onecolumngrid
\renewcommand{\thefigure}{S\arabic{figure}}
\setcounter{figure}{0}
\renewcommand{\thepage}{S\arabic{page}}
\setcounter{page}{1}
\renewcommand{\theequation}{S.\arabic{equation}}
\setcounter{equation}{0}
\setcounter{section}{0}

\end{document}


\title{Supplementary information:\\
Engineering superconducting qubits to reduce quasiparticles and charge noise}

\newcommand{\SIQSE}{\affiliation{1}{Shenzhen Institute for Quantum Science and Engineering, Southern University of Science and Technology, Shenzhen, Guangdong, China}}
\newcommand{\IQA}{\affiliation{2}{International Quantum Academy, Shenzhen, Guangdong, China}}
\newcommand{\GDKL}{\affiliation{3}{Guangdong Provincial Key Laboratory of Quantum Science and Engineering, Southern University of Science and Technology, Shenzhen, Guangdong, China}}
\newcommand{\DPHY}{\affiliation{4}{Department of Physics, Southern University of Science and Technology, Shenzhen, Guangdong, China}}
\newcommand{\FJ}{\affiliation{5}{JARA Institute for Quantum Information (PGI-11), Forschungszentrum J\"ulich, 52425 J\"ulich, Germany}}
\newcommand{\TII}{\affiliation{6}{Quantum Research Centre, Technology Innovation Institute, Abu Dhabi, UAE}}
\newcommand{\SEU}{\affiliation{7}{State Key Laboratory of Millimeter Waves, School of Information Science and Engineering, Southeast University, Nanjing, China}}

\author{Xianchuang Pan}
\thanks{These authors have contributed equally to this work.}
\affiliation{\SIQSE}
\affiliation{\IQA}
\affiliation{\GDKL}

\author{Yuxuan Zhou}
\thanks{These authors have contributed equally to this work.}
\affiliation{\SIQSE}
\affiliation{\IQA}
\affiliation{\GDKL}
\affiliation{\DPHY}

\author{Haolan Yuan}
\affiliation{\SIQSE}
\affiliation{\IQA}
\affiliation{\GDKL}
\affiliation{\DPHY}

\author{Lifu Nie}
\affiliation{\SIQSE}
\affiliation{\IQA}
\affiliation{\GDKL}

\author{Weiwei Wei}
\affiliation{\SIQSE}
\affiliation{\IQA}
\affiliation{\GDKL}

\author{Libo Zhang}
\affiliation{\SIQSE}
\affiliation{\IQA}
\affiliation{\GDKL}

\author{Jian Li}
\affiliation{\SIQSE}
\affiliation{\IQA}
\affiliation{\GDKL}

\author{Song Liu}
\affiliation{\SIQSE}
\affiliation{\IQA}
\affiliation{\GDKL}

\author{Zhi Hao Jiang}
\affiliation{\SEU}

\author{Gianluigi Catelani}
\email{g.catelani@fz-juelich.de}
\affiliation{\FJ}
\affiliation{\TII}

\author{Ling Hu}
\email{hul@sustech.edu.cn}
\affiliation{\SIQSE}
\affiliation{\IQA}
\affiliation{\GDKL}

\author{Fei Yan}
\email{yanf7@sustech.edu.cn}
\affiliation{\SIQSE}
\affiliation{\IQA}
\affiliation{\GDKL}

\author{Dapeng Yu}
\affiliation{\SIQSE}
\affiliation{\IQA}
\affiliation{\GDKL}
\affiliation{\DPHY}



\maketitle

\begin{table*}[!tbh]
\renewcommand\arraystretch{0.8}
\centering
\begin{tabular}{c|ccccccccccccccccc}
\hline
\hline
\makecell[c]{Device-\\Qubit} &
\makecell[c]{$L$\\ $(\rm \mu m)$} &
\makecell[c]{$W$\\ $(\rm \mu m)$} &
\makecell[c]{$d$\\ $(\rm \mu m)$} &
\makecell[c]{$E_{\rm J}/h$\\ (GHz)} &
\makecell[c]{$E_{\rm C}/h$\\ (GHz)} &
\makecell[c]{$\frac{E_{\rm J}}{E_{\rm C}}$} &
\makecell[c]{$\omega_{\rm ge}^{\rm max}/2\pi$\\ (GHz)} &
\makecell[c]{$\omega_{\rm ge}^{\rm min}/2\pi$\\ (GHz)} &
\makecell[c]{$g/2\pi$\\ (MHz)} &
\makecell[c]{$\omega_{\rm r}/2\pi$\\ (GHz)} &
\makecell[c]{$T_1$\\ ($\mu$s)} &
\makecell[c]{$T_{\phi}$\\ ($\mu$s)} &
\makecell[c]{$T_P$\\~(s)} &
\makecell[c]{holder\\material} &
\makecell[c]{cap\\~($\mu$m)} &
\makecell[c]{CR110} &
\makecell[c]{Fig.}\\
\hline
\makecell[c]{S1$-$Q1}& 80 & 35 &  5 & 4.67 & 1.40 & 3.34 & 6.833  & 4.473  & 24.3  & 5.556 & 24.4 & 22.8 & 1.918 &&&&2, 4\\
\makecell[c]{S1$-$Q2}& 80 & 35 & 10 & 4.27 & 1.48 & 2.89 & 6.954  & 4.135  & 20.4  & 5.607 & 29.4 & 13.4 & 1.235 &&&&3d$\sim$g\\
\makecell[c]{S1$-$Q3}& 80 & 35 & 15 & 4.45 & 1.52 & 2.93 & 7.191  & 4.289  & 22.5  & 5.712 & 16.7 & 14.1 & 1.159 & Al&no&yes&S5a\\
\makecell[c]{S1$-$Q4}& 80 & 35 & 20 & 4.63 & 1.53 & 3.03 & 7.268  & 4.469  & 23.3  & 5.753 & 25.8 & 20.9 & 0.987 &&&&  S7$\sim$9  \\
\makecell[c]{S1$-$Q5}& 80 & 35 & 25 & 4.52 & 1.54 & 2.94 & 7.255  & 4.364  & 22.6  & 5.830 & 22.0 & 22.3 & 1.172 &&&&S18\\
\makecell[c]{S1$-$Q6}& 80 & 35 & 30 & 4.59 & 1.54 & 2.98 & 7.325  & 4.414  & 22.7  & 5.883 & 23.8 &  3.0 & 0.761 &&&&S21$\sim$23\\
\hline
\makecell[c]{S2$-$Q1}& 80 & 35 & 35 &  --    &  --    &  --   &    --    &    --    &   --    & 5.552 &   --   &    --  & 0.691 & \\
\makecell[c]{S2$-$Q2}& 80 & 35 & 40 &  --    &   --   &   --   &    --    &     --   &    --   & 5.610 &  --    &   --   & 0.531 & \\
\makecell[c]{S2$-$Q3}& 80 & 35 & 45 &   --   &   --   &  --    &    --    &   --     &   --    & 5.713 &  --    &   --   & 0.541 & Cu&no&yes&3d\\
\makecell[c]{S2$-$Q4}& 80 & 35 & 50 &   --   &   --   &   --   &    --    &    --    &   --    & 5.756 &  --    &   --   & 0.431 &  \\
\makecell[c]{S2$-$Q5}& 80 & 35 & 55 &   --   &   --   &   --   &    --    &    --    &   --    & 5.831 &  --    &  --    & 0.320 & \\
\makecell[c]{S2$-$Q6}& 80 & 35 & 60 &   --   &   --   &   --   &    --    &    --    &   --    & 5.882 &  --    &  --    & 0.353 & \\
\hline
\makecell[c]{S3$-$Q1}& 80 & 35 & 20 & 3.23 & 1.49 & 2.17 & 6.585  & 3.126  & 15.7  & 5.554 & 54.2 & 76.9 & 1.123 & \\
\makecell[c]{S3$-$Q2}& 80 & 35 & 40 & 3.36 & 1.51 & 2.23 & 6.729  & 3.279  & 15.5  & 5.609 & 42   & 400.0& 0.673 & \\
\makecell[c]{S3$-$Q3}& 80 & 35 & 80 & 3.24 & 1.52 & 2.13 & 6.728  & 3.130  & 13.2  & 5.712 &   --   &   --   & 0.175 & Al&no&yes&3d\\
\makecell[c]{S3$-$Q4}& 80 & 35 &100 & 3.47 & 1.51 & 2.30 & 6.806  & 3.386  & 14.3  & 5.754 &  --    &   --   & 0.244 &&&&S4b  \\
\makecell[c]{S3$-$Q5}& 80 & 35 &200 &  --    &   --   &   --   &    --    &    --    &   --    & 5.833 & --     &   --   & 0.164 & \\
\makecell[c]{S3$-$Q6}& 80 & 35 &400 &  --    &   --   &    --  &    --    &   --     &    --   & 5.885 &  --    &   --   & 0.251 & \\
\hline
\makecell[c]{S4$-$Q1}& 80 & 35 &  5 & 7.61 & 1.30 & 5.87 & 7.845  & 6.754  & 46.5  & 5.521 &   --   &   --   & 2.268 & \\
\makecell[c]{S4$-$Q2}&220 & 35 &  5 & 6.92 & 0.78 & 8.89 & 5.774  & 5.505  & 4.7   & 5.581 &  --    &   --   & 0.074 & \\
\makecell[c]{S4$-$Q3}&240 & 60 &  5 &12.25 & 0.44 &28.09 & 6.068  & 6.067  &   --    & 5.634 & 4.4  & 7.2  & 0.030 & Al&no&yes&3c\\
\makecell[c]{S4$-$Q4}&300 & 60 &  5 &15.31 & 0.40 &38.66 & 6.5402 & 6.5401 &   --    & 5.700 & 12.3 & 14.0 & 0.016 &&&&S5b,d  \\
\makecell[c]{S4$-$Q5}&350 & 90 &  5 &  --    &  --    &   --   &    --    &    --    &    --   & 5.756 &  --    &   --   &    --   &&&&S23  \\
\makecell[c]{S4$-$Q6}&420 &100 &  5 &   --   &   --   &   --   & 6.2184 & 6.2184 &   --    & 5.800 &  --    &   --   &   --    & &&&\\
\hline
\makecell[c]{S5$-$Q1}& 80 & 35 &  5 & 6.25 & 1.31 & 4.77 & 7.246  & 5.740  & 50.0  & 5.515 &  --    &   --   & 0.610 & \\
\makecell[c]{S5$-$Q2}&220 & 35 &  5 & 5.12 & 0.76 & 6.73 & 4.907  & 4.400  & 119.1 & 5.596 &  --   &   --   & 0.046 & & & & 3c \\
\makecell[c]{S5$-$Q3}&240 & 60 &  5 & 14.66& 0.37 &39.62 & 6.2013 & 6.2012 &   --    & 5.628 &  --    &  --    & 0.022 & Al&no&yes& 4\\
\makecell[c]{S5$-$Q4}&300 & 60 &  5 & 11.99& 0.43 &27.88 & 5.9853 & 5.984  &   --    & 5.695 & 8.6  & 45.7 & 0.011 & &&&S5d \\
\makecell[c]{S5$-$Q5}&350 & 90 &  5 &   --   &  --    &   --   & 6.061  & 6.061  &   --    & 5.752 & 13.1 & 28.3 &  --     &&&&S23 \\
\makecell[c]{S5$-$Q6}&420 &100 &  5 &   --   &  --    &  --    & 5.8345 & 5.8345 &    --   & 5.799 & 16.1 & 22.6 &   --    & &&&\\
\hline
\makecell[c]{S6$-$\\Q1$\sim$Q6}&80&35&20&3.50& 1.40 & 2.50 & 6.558  & 3.257  & 16.1  & 5.703 & 5    & 9    & 0.374 & Al&no&\makecell[c]{no}& \makecell[c]{S4a\\S20} \\
\hline
\makecell[c]{S7$-$Q1}& 80 & 35 & 20 & 2.79 & 1.43 & 1.95 & 6.214  & 2.774  & 11.25 & 5.556 &   --   &   --   & 0.940 & \\
\makecell[c]{S7$-$Q2}& 80 & 35 & 40 & -- & -- & -- & --  & --  & --  & -- &  --    &   --   & 0.610 & \\
\makecell[c]{S7$-$Q3}& 80 & 35 & 80 & -- & -- & --  & --  &  -- &  -- & -- &   --   &  --    & 0.072 & Al&no&\makecell[c]{no}&S4b\\
\makecell[c]{S7$-$Q4}& 80 & 35 &100 & -- & -- & -- & --  &  -- & --  & -- &  --    &  --    & 0.074 &  \\
\makecell[c]{S7$-$Q5}& 80 & 35 &200 &   --   &    --  &   --   &    --    &    --    &   --    & -- &   --   &  --    & 0.089 & \\
\makecell[c]{S7$-$Q6}& 80 & 35 &400 &  --    &    --  &   --   &   --     &    --    &   --    & -- &   --   &   --   & 0.140 & \\
\hline
\makecell[c]{S8$-$Q2}& 80 & 35 &  5 & 2.35 & 1.29 & 1.82 & 5.566  & 2.281  & 16.9  & 4.611 & 55.3 & 3.1  & 2.006 & &300\\
\makecell[c]{S8$-$Q3}&120 & 35 &  5 & 2.23 & 0.95 & 2.35 & 4.260  & 2.230  & 12.9  & 4.673 &   --   &   --   & 1.600 & &300\\
\makecell[c]{S8$-$Q4}&160 & 35 &  5 & 2.43 & 0.75 & 3.20 & 3.650  & 2.320  & 5.2   & 4.702 &   --   &   --   & 1.006 & &300 & & 3c\\
\makecell[c]{S8$-$Q5}&180 & 35 &  5 & 4.45 & 0.66 & 6.74 & 4.254  & 3.858  & 19.2  & 4.733 & 39.2 &   --   & 0.904 & Al&300&yes& S19\\
\makecell[c]{S8$-$Q6}&210 & 35 &  5 & 4.71 & 0.59 & 7.97 & 4.104  & 3.899  & 18.6  & 4.770 &  --    &    --  & 0.615 & & 300&& S23 \\
\makecell[c]{S8$-$Q7}&260 & 35 &  5 & 4.84 & 0.49 & 9.92 & 3.832  & 3.712  & 18.0  & 4.805 & 13.8 & 25.1 & 0.603 & & 500\\
\makecell[c]{S8$-$Q8}&260 & 35 &  5 & 5.79 & 0.46 & 12.58& 4.116  & 4.059  & 23.0  & 4.835 & 35.3 & 19.3 & 0.655 & &500\\
\hline
\makecell[c]{S9$-$Q1}&80 & 35 &  20 & -- & -- & -- & --  &  -- & --& 5.850 &   --   &   --   & 0.511 & \\
\makecell[c]{S9$-$Q2}&80 & 35 &  50 & -- & -- & -- & --  & --  & -- & 5.800 &   --   &   --   & 0.249 & \\
\makecell[c]{S9$-$Q3}&80 & 35 &  80 & -- & -- & -- &  -- &  -- &-- & 5.751 &   --   &   --   & 0.146 & Al&no&\makecell[c]{no}&S5c\\
\makecell[c]{S9$-$Q4}&80 & 35 &  110 & -- & -- & -- & --  &  -- & --& 5.750 &   --   &   --   & 0.196 & \\
\makecell[c]{S9$-$Q5}&80 & 35 &  140 & -- & -- & -- &  -- & --  & -- & 5.700 &   --   &   --   & 0.157 & \\
\makecell[c]{S9$-$Q6}&80 & 35 &  170 & -- & -- & -- & --  & --  & -- & 5.650 &   --   &   --   & 0.152 & \\
\hline
\makecell[c]{S10$-$Q1}&80 & 35 &  20 & -- & -- & -- & --  &  -- & --& 5.850 &   --   &   --   & 0.652 & \\
\makecell[c]{S10$-$Q2}&80 & 35 &  50 & -- & -- & -- & --  & --  & -- & 5.800 &   --   &   --   & 0.292 & \\
\makecell[c]{S10$-$Q3}&80 & 35 &  80 & -- & -- & -- &  -- &  -- &-- & 5.753 &   --   &   --   & 0.158 & Al&no&\makecell[c]{no}&S5c\\
\makecell[c]{S10$-$Q4}&80 & 35 &  110 & -- & -- & -- & --  &  -- & --& 5.750 &   --   &   --   & 0.181 & \\
\makecell[c]{S10$-$Q6}&80 & 35 &  170 & -- & -- & -- & --  & --  & -- & 5.650 &   --   &   --   & 0.149 & \\
\hline
\end{tabular}
\captionsetup{labelformat=empty,justification=raggedright}
\caption{\textbf{Supplementary Table 1: Device parameters and setup.}
$L, W, d$ are the pad length, pad width, and pad-to-
ground distances.
The listed $T_1$, $T_{\phi}$ and $T_P$ are measured at the minimum qubit frequency at $n_{\rm g}=0$. 
We also list the information about the sample holder material, the metallic cap size, the use of CR110 filter after the bias tee, and the figures the device is related to.
Note that the six qubits in S6 have all the same design; the typical data listed here comes from S6-Q1. Sample S7 has the same design as S3. 
Sample S8 is the flip-chip device. 
The wiring or shielding setup of sample S6, S7, S9, S10 is different from the standard (optimal) one shown in Supplementary Fig.~\ref{fig:Wirings}, so we did not include related data in the main text figure.
For samples S2, S7, S9, S10, we only measured $T_{\rm P}$ data.  The S10-Q5 qubit is broken. All the data in Figs.~S7$\sim$S9, S21$\sim$S22 are from S1-Q1. Unmeasured parameters are left blank. Here, Fig.~S\# means Supplementary Fig.~\#.}
\label{tab:tabel1}\vspace{-6pt}
\end{table*}

\section*{Supplementary Note 1: Device design and fabrication}

\begin{figure}[htbp]
\centering
\includegraphics{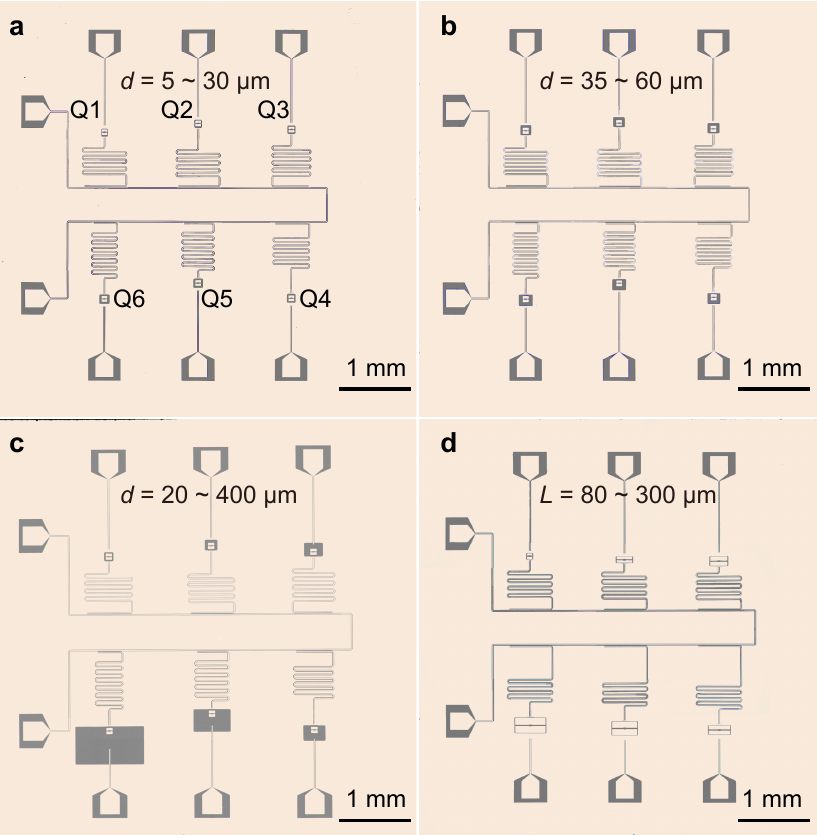}
\captionsetup{labelformat=empty,justification=raggedright}
\caption{\textbf{Supplementary Figure 1: Optical micrograph of planar samples (S1$\sim$S4).}
Dark: exposed sapphire substrate; light: base aluminium layer.
\textbf{a}, Sample S1, with variation of pad-to-ground distance $d$ from 5 to 30~$\mu$m.
\textbf{b}, Sample S2, with $d$ from 35 to 60~$\mu$m.
\textbf{c}, Sample S3, with $d$ from 20 to 400~$\mu$m.
\textbf{d}, Sample S4, with the pad length $L$ from 80 to 300~$\mu$m.
}
\label{fig:AllSample}\vspace{-8pt}
\end{figure}

\begin{figure}[htbp]
\centering
\includegraphics{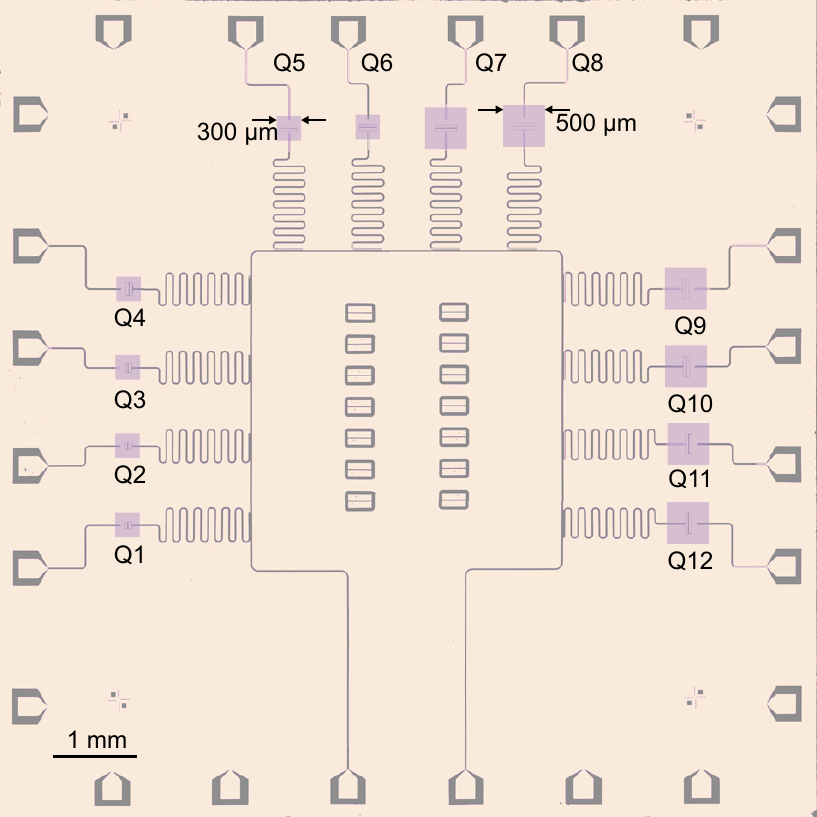}
\captionsetup{labelformat=empty,justification=raggedright}
\caption{\textbf{Supplementary Figure 2: Optical micrograph of the flip-chip sample S8 (bottom chip).}
The 12 qubits differ by the pad length $L$ and $E_{\rm J}$ (Supplement-
ary Table \ref{tab:tabel1}). 
Purple squares indicate the metallic caps on the top chip covering each of the 12 qubits.
The caps have two different size, $300~\mu$m and $500~\mu$m.
All qubits share the same pad width $W=35~\mu$m and pad-to-grou-
nd distance $d=5~\mu$m.}
\label{fig:FlipSample}\vspace{-6pt}
\end{figure}

Data presented in the main text are taken from five planar sample chips and one flip-chip sample. Each planar sample hosts 6 qubits with varying geometries as shown in Supplementary Fig.~\ref{fig:AllSample}. The device parameters are listed in Supplementary Table \ref{tab:tabel1}.
Data used to investigate the dependence of the parity switching rate on pad-to-ground distance $d$ (Fig.~3d in the main text) are from sample S1$\sim$S3; data used to investigate the dependence on pad length $L$ (Fig.~3c in the main text) are from sample S4$\sim$S5 (same design). Data shown in Fig.~2 of the main text are from the qubit S1$-$Q1.
Data shown in Fig.~4 of the main text are from sample S1, S4, and S5. There are two more planar samples (S6 and S7) considered in this supplement, see Supplementary Table~\ref{tab:tabel1} and Supplementary Note 2.
The chips are either packaged in an aluminium or copper sample holder for testing.
We do not observe appreciable differences in qubit performance between different holder materials and between different cooldowns with the same setup and wiring.

The devices are made in a two-step process on c-plane sapphire wafers. The first step is to pattern the base circuit. We deposited 100-nm-thick aluminum on c-plane sapphire substrate in a PLASSYS system at a growth rate of 1~nm/s with a base pressure of $10^{-10}$ Torr. After photolithography, we use $\rm BCl_3/Cl_2$ etch in an inductively coupled plasma (ICP) dry etcher. In the second step, the junctions are fabricated in the bridge-free Manhattan style. We use double-angle evaporation to make the $\rm Al/AlO_x/Al$ stack. The first aluminium film is about 30~nm thick and the second one is about 40~nm. Then the excess aluminium is lifted off with a stripper. An ion-milling step is performed before we add a final layer of aluminium film (200~nm thick), connecting the junction leads and capacitor pads in order to make direct galvanic contact.

For the flip-chip sample, the bottom chip, which host 12 qubits with varying geometries as shown in Supplementary Fig.~\ref{fig:FlipSample}, shares similar design and fabrication processes as the planar samples. 
Each qubit is covered by a metallic cap, a square-shaped aluminium pad, on the opposing die.
The two single-sided sapphire dies are bonded together using four spacers (2~mm$\times$2~mm) at the corners. The spacers made of SU-8 photoresist extend a vertical spacing of about $10~\mu$m.

\section*{Supplementary Note 2: Experimental setup and wiring}\label{sec:exp_wir}

\begin{figure}[!thb]
\includegraphics{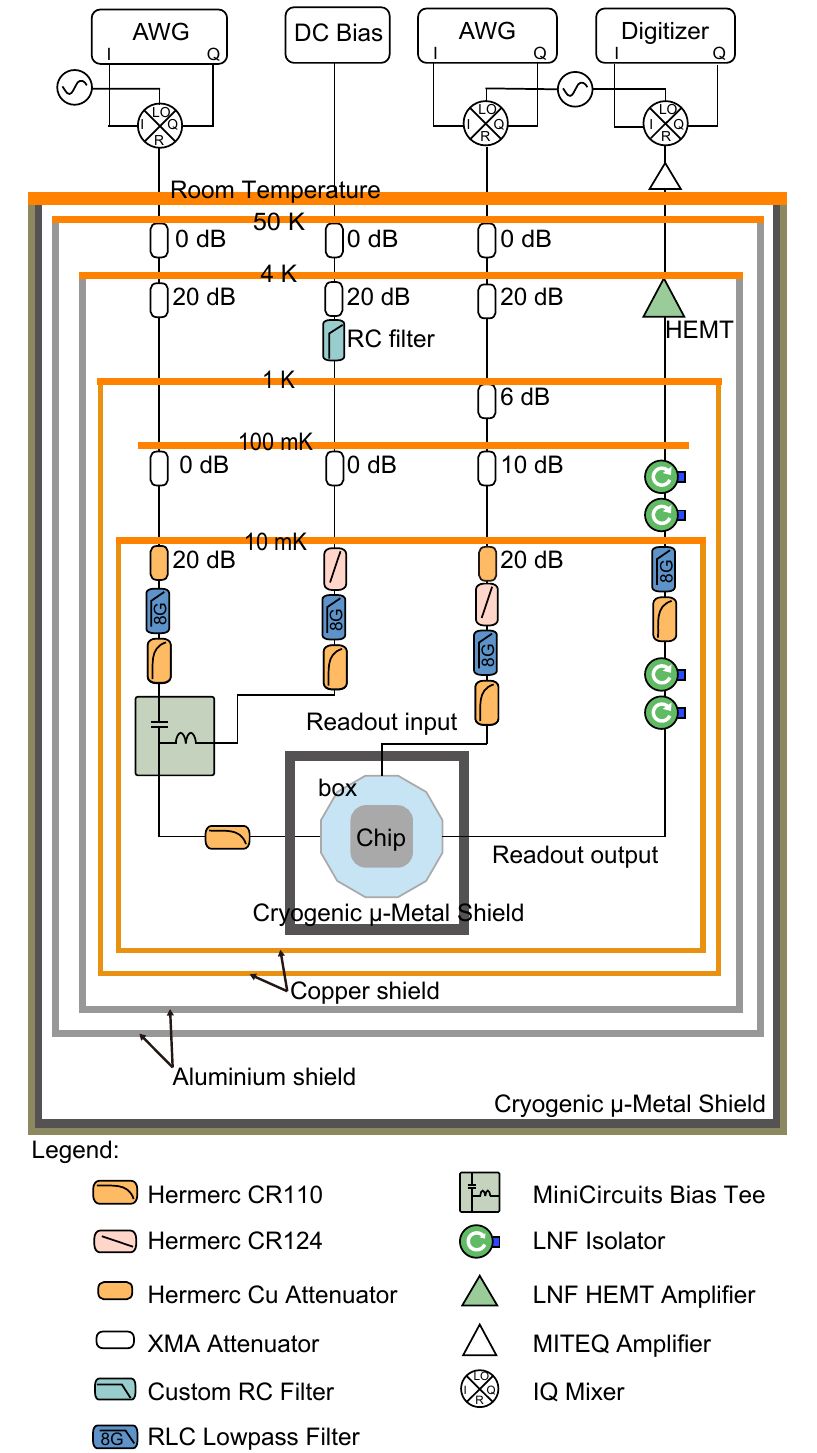}
\captionsetup{labelformat=empty,justification=raggedright}
\caption{\textbf{Supplementary Figure 3: Cryogenic wiring and signal synthesis at room temperature.}
We combine DC bias and RF signals at base temperature using a bias tee before sending them to the chip. All data shown in the main text is taken using this standard setup. A cryogenic $\mu$-metal shield is used for shielding.}
\label{fig:Wirings}\vspace{-8pt}
\end{figure}

\begin{figure}[!bht]
\centering
\includegraphics{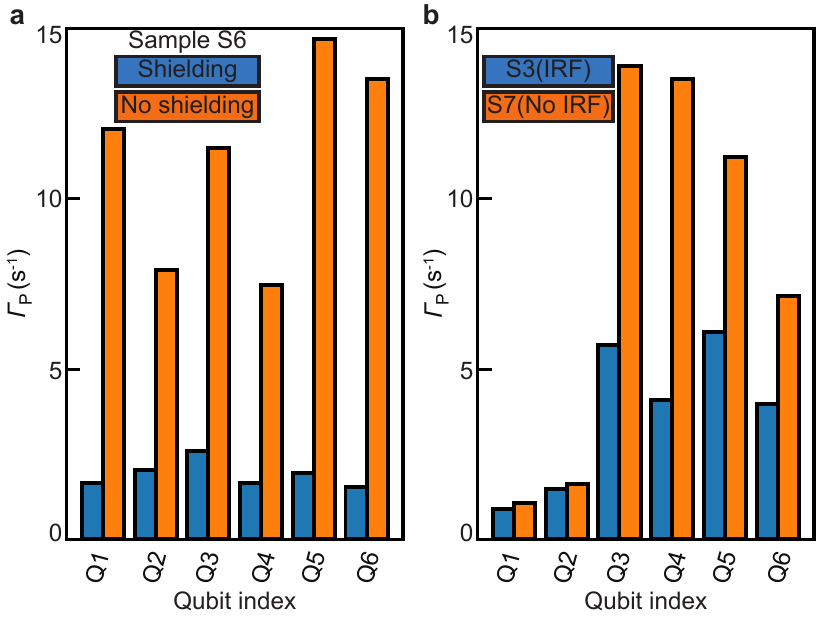}
\captionsetup{labelformat=empty,justification=raggedright}
\caption{\textbf{Supplementary Figure 4: Shielding and filtering.}
\textbf{a}, Comparison of $\Gamma_P$ with or without the shield. The data is taken in different cooldowns using sample S6 with six identical qubits. There are no CR110 filters after the bias tee in both cases.
\textbf{b}, Comparison of $\Gamma_P$ with or without CR110 IR filter after the bias tee. The data is taken in the same cooldown from sample S3 (blue) and S7 (orange); the samples share the same design and, except for the IR filter, are measured with identical setups. There are $\mu$-metal shields in both cases in \textbf{b}.}
\label{fig:TPvswiring}\vspace{-6pt}
\end{figure}

\begin{figure}[!thb]
\includegraphics{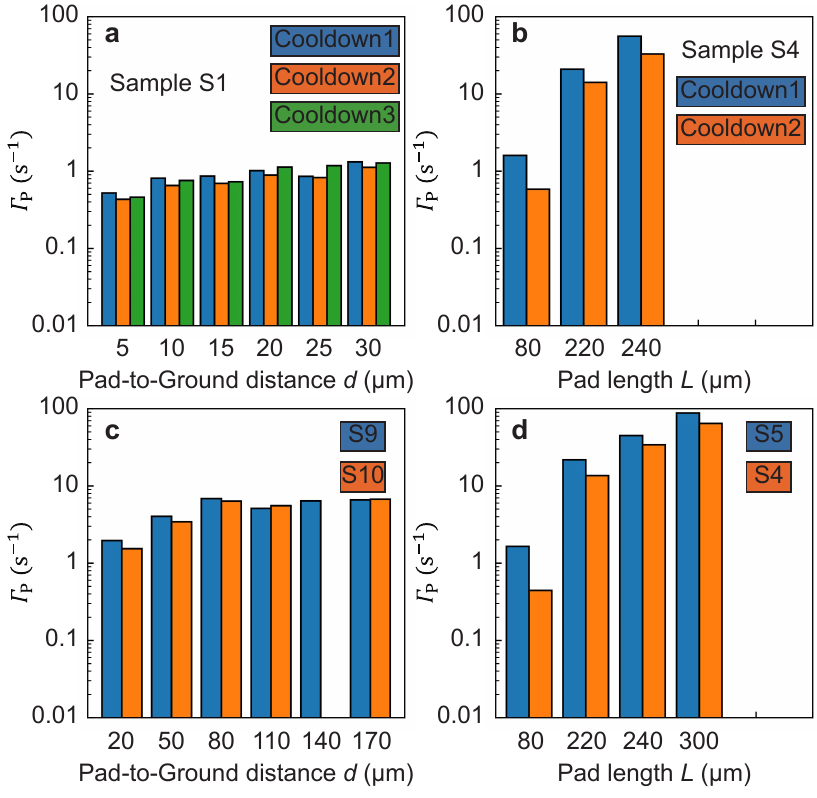}
\captionsetup{labelformat=empty,justification=raggedright}
\caption{\textbf{Supplementary Figure 5: $\Gamma_P$ stability.}
\textbf{a-b}, Comparison of measured parity switching rates between different cooldowns for a given sample.
\textbf{c-d}, Comparison of measured parity switching rates between samples of identical design during a given cooldown. The measurement setup is the same. Missing data is due to broken qubit or undetectable with our method (too small a dispersive shift).
}
\label{fig:compare}\vspace{-8pt}
\end{figure}

The samples are mounted inside a BlueFors LD400 dilution refrigerator at a nominal base temperature of less than 10~mK; however, the excited-state population measured with regular transmon qubits~\cite{Jin2015} in the same setup gives a typical device or electronic temperature of 50-60~mK. The experimental setup is depicted in Supplementary Fig.~\ref{fig:Wirings}. As shown in Supplementary Fig.~\ref{fig:Wirings}, from the inside out, the device is protected by a aluminum or copper holder box, a $\mu$-metal shield, a few layers of copper and aluminum shields, and an outer $\mu$-metal shield.
Control and readout signals are properly attenuated and filtered at multiple stages. At room temperature, we generally use a network analyzer for fast and cleaner data taking in the case of small parity switching rate $\Gamma_P<100$~Hz. However, for large $\Gamma_P$, one has to use pulsed measurement based on a Ramsey-like sequence (see Supplementary Note 6) using an AWG-digitizer-based setup.

Empirically, we find two aspects in the setup that make a difference in the measured parity switching time.
First, using a cryogenic $\mu$-mental shield can help suppressing $\Gamma_P$ by almost an order of magnitude, which suggests its effectiveness in radiation shielding (Supplementary Fig.~\ref{fig:TPvswiring}, left);
we also find it important to add an infrared (IR) filter to the common port of a bias tee (Supplementary Fig.~\ref{fig:TPvswiring}, right), which is consistent with observations reported by other groups~\cite{Serniak2019,Gordon2021}.
Based on the above findings, we believe that the shield and IR filter are effective in blocking stray photons of about 100~GHz (or higher) from reaching the sample via open space and cables, respectively.

The photon flux seen by the device may vary depending on the specific filtering and shielding setup. For a nominally identical setup, We assume that the photon flux stays constant between different cooldowns. 
In Supplementary Fig.~\ref{fig:compare}a$\sim$b, we compare the parity switching rates measured in different cooldowns for two devices, which show good agreement to our assumption. The measured switching rates for a given device have relatively small change between different cooldowns (about a factor of 2 in the worst case). In addition, it can be seen that, over different cooldowns, the switching rates have consistent geometric dependence which can vary by one to two orders of manginitude, much greater than the random fluctuations between cooldowns. We also compare the parity switching rates of samples with identical design, meausured with an identical setup during a same cooldown as shown in Supplementary Fig.~\ref{fig:compare}c$\sim$d. The data also show consistent behavior with only small fluctuations. Therefore, we can safely assume that the qubits see a same amount of photon flux for a given shielding and filtering setup.

\vspace{-9pt}
\section*{Supplementary Note 3: Qubit spectroscopy}

\begin{figure}[!bht]
\includegraphics{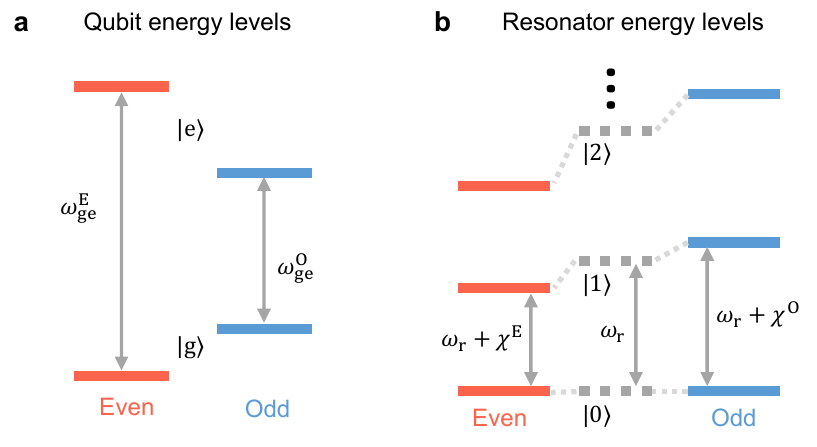}
\captionsetup{labelformat=empty,justification=raggedright}
\caption{\textbf{Supplementary Figure 6: Energy level diagram.}
\textbf{a}, Ground and excited states of a low-$E_{\rm J}/E_{\rm C}$ qubit for even and odd parities. Parity switching events, associated with quasiparticle tunnelling or pair-breaking at the junction, exchange the two parities. In the shown example, the transition frequency between ground and excited states is higher (lower) in the even (odd) parity.
\textbf{b}, Resonator levels in even and odd parity compared to the bare case (dashed grey). In the shown case, the resonator frequency is between the even-parity and odd-parity qubit frequency. The dressed resonator frequency in the even(odd) parity is red(blue)-shifted from its bare frequency (dashed).
}
\label{fig:EnergyLevelSketch}\vspace{-8pt}
\end{figure}

\begin{figure}[!hbt]
\includegraphics{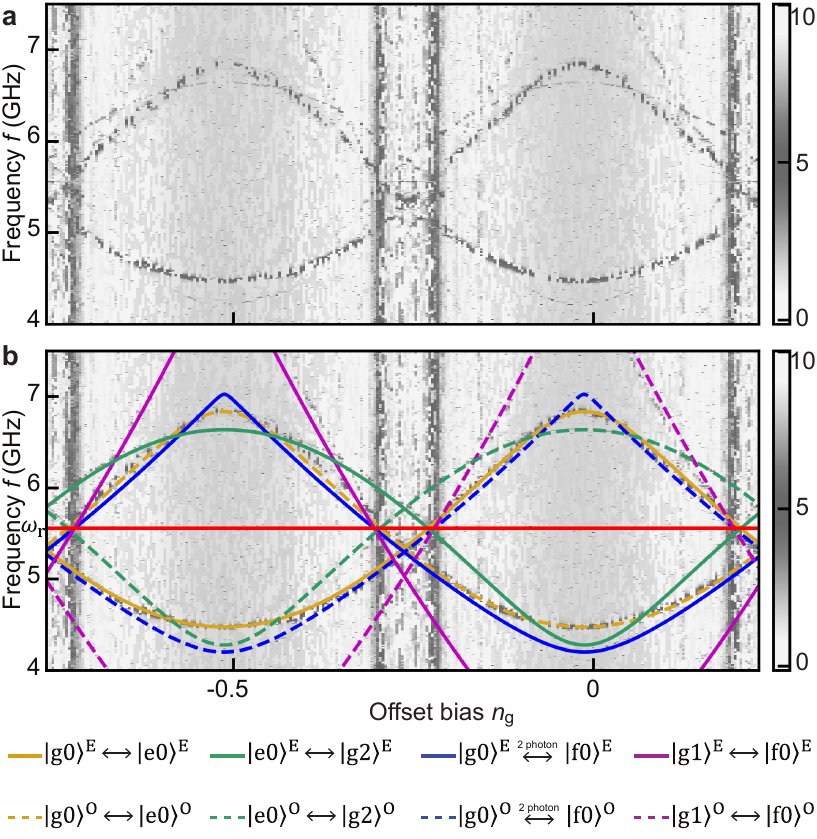}
\captionsetup{labelformat=empty,justification=raggedright}
\caption{\textbf{Supplementary Figure 7: Qubit spectroscopy with identified transitions.}
\textbf{a}, Qubit spectroscopy reproduced from Fig.~2b.
\textbf{b}, Same as \textbf{a} with added coloured lines denoting various transitions.
Solid (dashed) lines denote transitions in the even (odd) parity.
The spectroscopy pulse is about 9~$\mu$s long. The sequence is repeated every 100~$\mu$s.
Each datapoint presented is the average of 1000 repetitions.
}
\label{fig:EnergyLevel}\vspace{-8pt}
\end{figure}

For a typical qubit in the charge regime among our devices, Josephson and charging energies are in the range $E_{\rm J}/h=3.3$--4.6~GHz and $E_{\rm C}/h=1.4$--1.6~GHz ($E_{\rm J}/E_{\rm C}\sim3$).
As illustrated in Supplementary Fig.~\ref{fig:EnergyLevelSketch}a, in such a case, energy levels are considerably different between different parities, leading to strong discrepancy in qubit transition frequencies, $\omega_{\rm ge}^{\rm E}$ and $\omega_{\rm ge}^{\rm O}$, up to a few GHz depending on the offset bias.
In a circuit-QED architecture where the qubit is coupled to a resonator~\cite{Blais2004}, such large discrepancy can result in appreciable difference in the dressed resonator frequency because of level repulsion.
For example, Supplementary Fig.~\ref{fig:EnergyLevelSketch}b shows the level diagram when the resonator frequency $\omega_{\rm r}$ is between $\omega_{\rm ge}^{\rm E}$ and $\omega_{\rm ge}^{\rm O}$. The resonator frequency is pushed down (up) by $\chi^{\rm E(O)}$ ($\sim$1~MHz) when the qubit is in even (odd) parity.
Similar to the dispersive measurement of the qubit state, sending a probe tone near the resonator frequency allows us to distinguish between different qubit parities, provided that the acquisition time is short compared to the average latching time between parity switches.

In Supplementary Fig.~\ref{fig:EnergyLevel}, we reproduce the qubit spectroscopy shown in Fig.~2 of the main text and identify all the visible transitions which include, for both parities, the g-e transition of the qubit ($\ket{\mathrm{g}0} - \ket{\mathrm{e}0}$), the two-photon g-f transition ($\ket{\mathrm{g}0} - \ket{\mathrm{f}0}$), the qubit-resonator sideband transitions ($\ket{\mathrm{g}1} - \ket{\mathrm{f}0}$ and $\ket{\mathrm{e}0} - \ket{\mathrm{g}2}$), and the resonator mode at $\omega_{\rm r}$.

\vspace{-9pt}
\section*{Supplementary Note 4: Charge parity monitor}
\label{sec:ParityMonitor}

\begin{figure*}[!bht]
\includegraphics{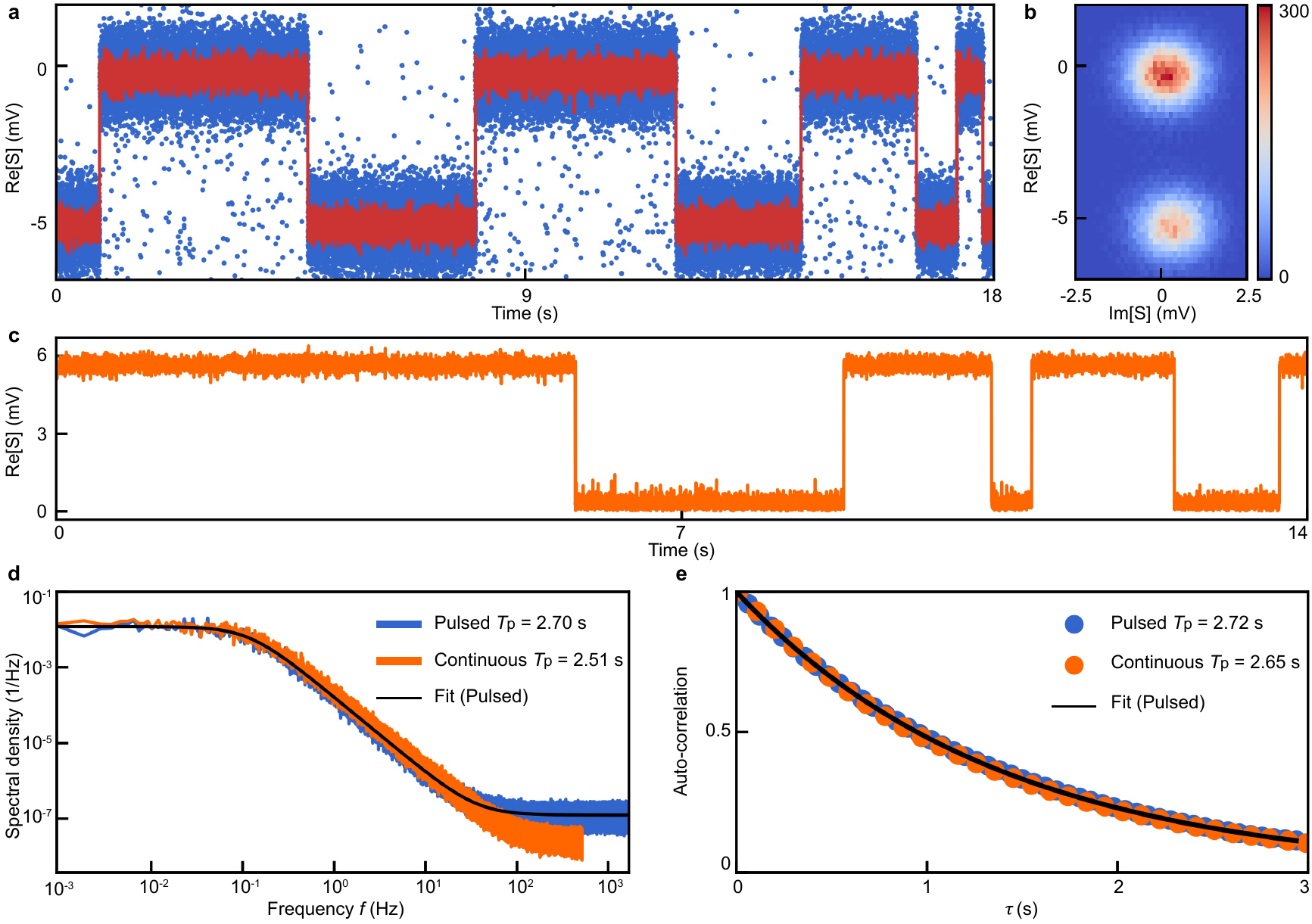}
\captionsetup{labelformat=empty,justification=raggedright}
\caption{\textbf{Supplementary Figure 8: Parity monitor with direct dispersive readout.}
\textbf{a}, Example of raw demodulated signal (blue dots) acquired in a pulsed measurement and its moving average (red line), calculated by taking the median of 10 consecutive points.
A typical trace taken with the pulsed method is 18~s long with a time interval of 0.3~ms between points.
\textbf{b}, Histogram of the raw complex data shown in \textbf{a}.
\textbf{c}, Example of raw data acquired in a continuous measurement with a network analyzer. A typical trace is 14~s long with the time interval of 1~ms between points.
\textbf{d}, Power spectrum of charge-parity fluctuations measured in pulsed (1200 repetitions) and continuous (1500 repetitions) measurement.
We concatenate the repeated traces in (a,c) to extend the lower limit of the spectrum to $10^{-3}$~Hz.
The background white noise is due to sampling noise which is stronger for the pulsed method, but does not influence the extracted $T_P$ times. The black line is a Lorentzian plus white noise fit to the pulsed data.
\textbf{e}, Normalized autocorrelation function $\langle P(0)P(\tau) \rangle$ of charge-parity fluctuations directly computed from the time-domain traces. The black line is an exponential fit to the pulsed data.}
\label{fig:TP}\vspace{-8pt}
\end{figure*}

\begin{figure*}[!bht]
\includegraphics{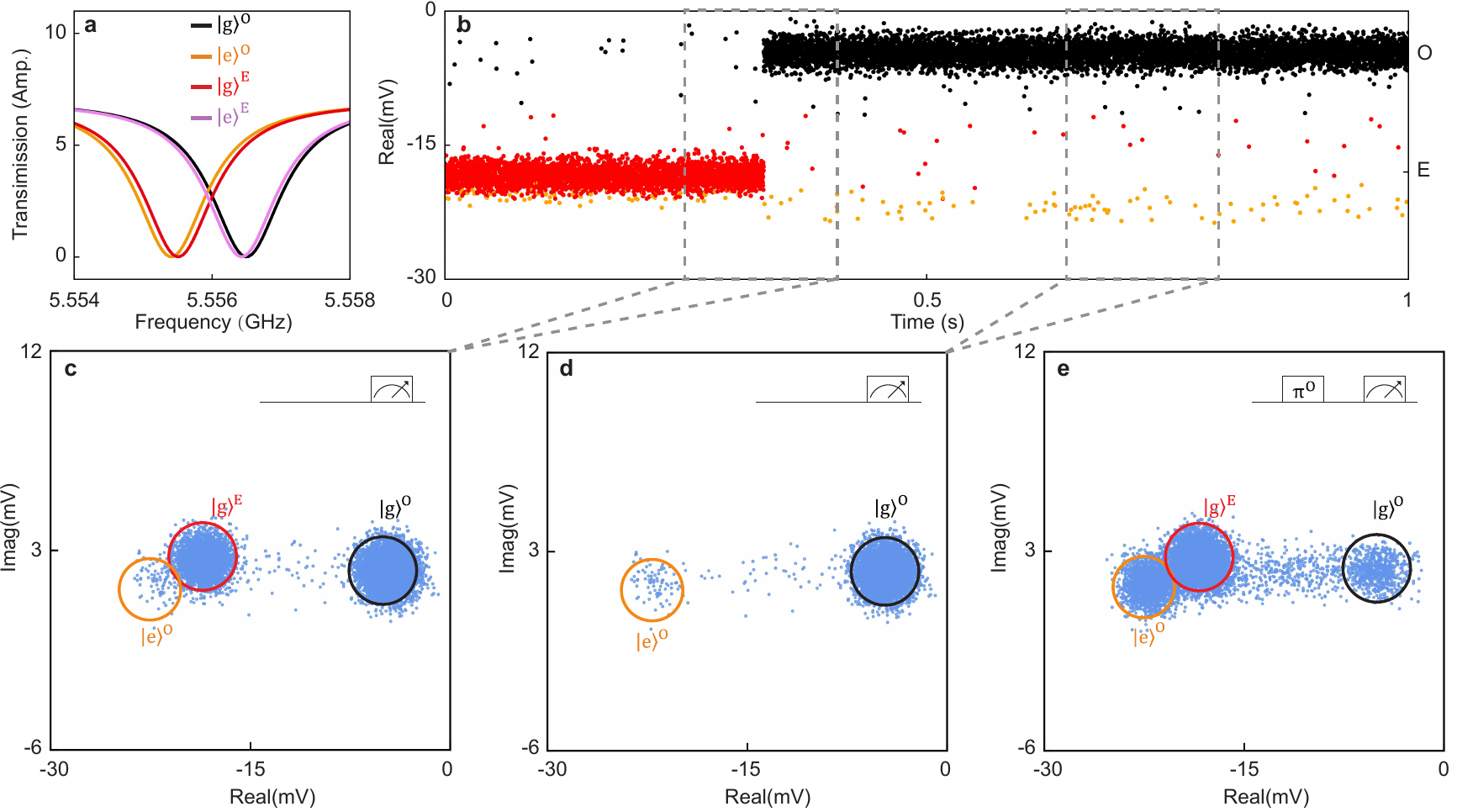}
\captionsetup{labelformat=empty,justification=raggedright}
\caption{\textbf{Supplementary Figure 9: Noise due to spurious qubit excitation induced by the probe signal.}
\textbf{a}, Readout resonator response for the four qubit states (ground/excited and even/odd) computed for sample S1-Q1. The resonant frequencies are $f_{\rm r} (\ket{\rm g}^{\rm O} )=5.55650$~GHz, $f_{\rm r} (\ket{\rm g}^{\rm E} )=5.55551$~GHz, $f_{\rm r} (\ket{\rm e}^{\rm O} )=5.55541$~GHz, $f_{\rm r} (\ket{\rm e}^{\rm E} )=5.55643$~GHz. \textbf{b}, An example of repeatedly measured single-shot time trace when the qubit is nominally prepared in the ground state. \textbf{c}, Demodulated signal amplitude plotted in the complex plane. Datapoints are taken from a section of time trace that samples both parities. \textbf{d}, Same as \textbf{c} but taken from a section that samples the odd parity only. \textbf{e}, Same as \textbf{c} but with a selective $\pi$ pulse for the odd parity. Colored circles indicate the distribution for corresponding states. Data in this figure is taken with a stronger probe signal than in Supplementary Fig.~\ref{fig:TP}.
}
\label{fig:SingleShotPopulation}\vspace{-8pt}
\end{figure*}

\subsection*{A. Parity detection with direct dispersive readout}

To monitor charge parity evolution, we implement two different methods, one based on direct dispersive readout with the qubit in the ground state and one based on a conditional bit flip.
For qubits with small $E_{\rm J}/E_{\rm C}$ ratio, the g-e transition frequencies at $n_{\rm g}=0$ for even and odd parity are drastically different, leading to dissimilar resonator frequency. We utilize such parity-dependent resonator response and use direct dispersive readout for distinguishing parity.
The measurement can be done either with pulsed probe signals generated with an AWG and collected by a digitizer, or with a network analyzer which probes continuously.
In the pulsed case, the probe pulse is typically $10~\mu$s long and repeated every 0.3~ms. The single-shot result -- $99.14\%$ fidelity for parity classification -- is relatively noisy, but it can be smoothed by taking a moving average, see Supplementary Fig.~\ref{fig:TP}a.
The noise has two parts: small sampling noise scattered around one of the telegraph state and less frequent strong jumps to the excited state about which we shall discuss later.

In our data processing, we rotate the raw demodulated data in the complex plane and use the real part for subsequent analysis, Supplementary Fig.~\ref{fig:TP}b.
In the continuous case, since the probe pulse is effectively always-on, the sampling noise becomes much reduced and a clean telegraph signal can be obtained, Supplementary Fig.~\ref{fig:TP}c.
The power spectral density (Supplementary Fig.~\ref{fig:TP}d) and the autocorrelation function (Supplementary Fig.~\ref{fig:TP}e) -- Fourier transform of power spectral density -- measured with the two setups show good agreement, validating our measurement and analysis protocols.

Now we discuss about the origin of the random jump events shown in Supplementary Fig.~\ref{fig:TP}a. In Supplementary Fig.~\ref{fig:SingleShotPopulation}a we plot the resonator response for qubit in four different states, $\ket{\rm g}^{\rm O}$, $\ket{\rm e}^{\rm O}$, $\ket{\rm g}^{\rm E}$, $\ket{\rm e}^{\rm E}$. It can be seen that $f_{\rm r} (\ket{\rm g}^{\rm O})$ is close to $f_{\rm r} (\ket{\rm e}^{\rm E})$, and $f_{\rm r} (\ket{\rm g}^{\rm E})$ is close to $f_{\rm r} (\ket{\rm e}^{\rm O})$. We can anticipate ambiguity between these pairs of states. We use a stronger probing amplitude for better pointer state separation in measurement associated with Supplementary Fig.~\ref{fig:SingleShotPopulation}b-e. Supplementary Fig.~\ref{fig:SingleShotPopulation}b shows an example time trace of repeated single-shot measurement where a parity switch is observed. From data collected from different sections of the trace, we can identify the corresponding state in the complex plane of the readout signal. First of all, $\ket{\rm g}^{\rm O}$ and $\ket{\rm g}^{\rm E}$ can be distinguished from each other by comparing Supplementary Fig.~\ref{fig:SingleShotPopulation}c and Supplementary Fig.~\ref{fig:SingleShotPopulation}d. Note that there is another small cluster adjacent to $\ket{\rm g}^{\rm E}$ which corresponds to $\ket{\rm e}^{\rm O}$, as can be confirmed by adding a $\pi$ pulse at the odd-parity qubit frequency, which selectively prepares the $\ket{\rm e}^{\rm O}$ state. The result is shown in Supplementary Fig.~\ref{fig:SingleShotPopulation}e, where significantly more datapoints are present in the $\ket{\rm e}^{\rm O}$ cluster marked by the orange circle. Therefore, those orange points (identified from being inside the orange circle) in Supplementary Fig.~\ref{fig:SingleShotPopulation}b correspond to the case of the qubit being in the excited state and in the odd parity. Since $f_{\rm r} (\ket{\rm g}^{\rm O} )$ and $f_{\rm r} (\ket{\rm e}^{\rm E}$) are even closer, the two clusters almost overlap with each other. We did not try to separate these two clusters. 
In a pulsed measurement setup, the data is taken with a digitizer; the acquisition time is usually a few microseconds (here, $3\,\mu$s) which is shorter than the typical $T_{\rm 1}$ time. Since we repeat every 100$\,\mu$s, any residual excitation is most likely to be observed as a single jump event in the repeated time trace. On the other hand, in the continuous measurement done with network analyzer, the effective acquisition time (bandwidth $\sim$10~kHz) is usually much longer than $T_{\rm 1}$ so that the analyzer signal -- being the averaged result -- cannot resolve those excitation events.

There are two major reasons for the spurious excitation. First, there is thermal excitation due to finite device temperature. With low enough readout power, we typically observe an excitation level of $0.5\%-1\%$, equivalent to 50-60~mK temperature, consistent with other regular transmon qubits measured in the same setup. Then, there is also measurement-induced excitation. When we increase the amplitude of the measurement pulse as in Supplementary Fig.~\ref{fig:SingleShotPopulation}, we tend to see increased contrast between the two parities, as well as more frequent jumps to the other state.

\begin{figure}[!htb]
\includegraphics{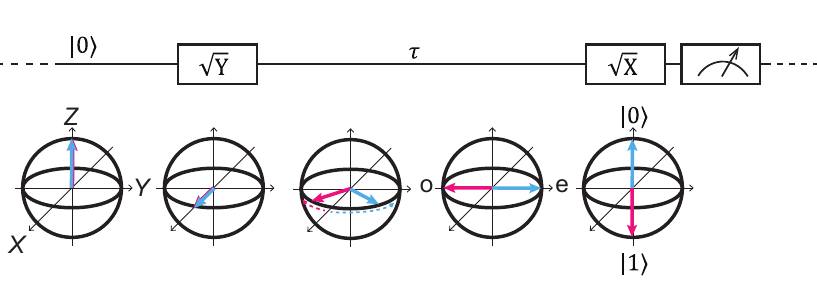}
\captionsetup{labelformat=empty,justification=raggedright}
\caption{\textbf{Supplementary Figure 10: Ramsey-based parity monitor.}
The sequence consists of two $\pi/2$ pulses separated by a waiting time $\tau$, which is chosen as $\tau=\pi/2(\omega_{\rm ge}^E-\omega_{\rm ge}^{\rm O})$, such that the odd and even states are mapped to ground and excited qubit states, respectively, at the end of the sequence. The time $\tau$ is pre-calibrated before repeated measurements.}
\label{fig:Ramseysequence}\vspace{-8pt}
\end{figure}

\subsection*{B. Parity detection with Ramsey sequence}\label{sec:RamseySequence}

For qubits with larger $E_{\rm J}/E_{\rm C}$ ratio (20$\sim$30), the frequency discrepancy between different parities (0.1$\sim$1~MHz) is exponentially suppressed in this ratio, and we can no longer use direct dispersive readout to distinguish between the parities. Instead, we use the Ramsey-type parity monitor as introduced in Ref.~\cite{Riste2013} and depicted in Supplementary Fig.~\ref{fig:Ramseysequence}.
In the Ramsey experiment, we set the microwave drive frequency at $\omega_{\rm drive}=(\omega_{\rm ge}^E+\omega_{\rm ge}^{\rm O})/2$ and the free-evolution time $\tau=\pi/2(\omega_{\rm ge}^E-\omega_{\rm ge}^{\rm O})$. Such a detuning setting transforms the qubit to the excited (ground) state for even (odd) parity, which is a conditional bit flip and enables us differentiate parity state from qubit state measurements.
The sequence is typically repeated every 0.1~ms.

\section*{Supplementary Note 5: Antenna mode of the transmon qubit}
The transmon qubit structure, which has a typical size of a few hundred microns, can be an efficient receiving antenna, transferring photons of a few hundred GHz to the junction \cite{Rafferty2021}. The absorbed photons at the junction can break Cooper pairs that tunnel through the barrier, giving rise to quasiparticle poisoning which is detectable as parity switching. Aided by finite-element electromagnetic simulations, we validate the photon absorption model and provide a semi-quantitative explanation for the experimental observations.

\begin{figure}[tbhp]
\includegraphics{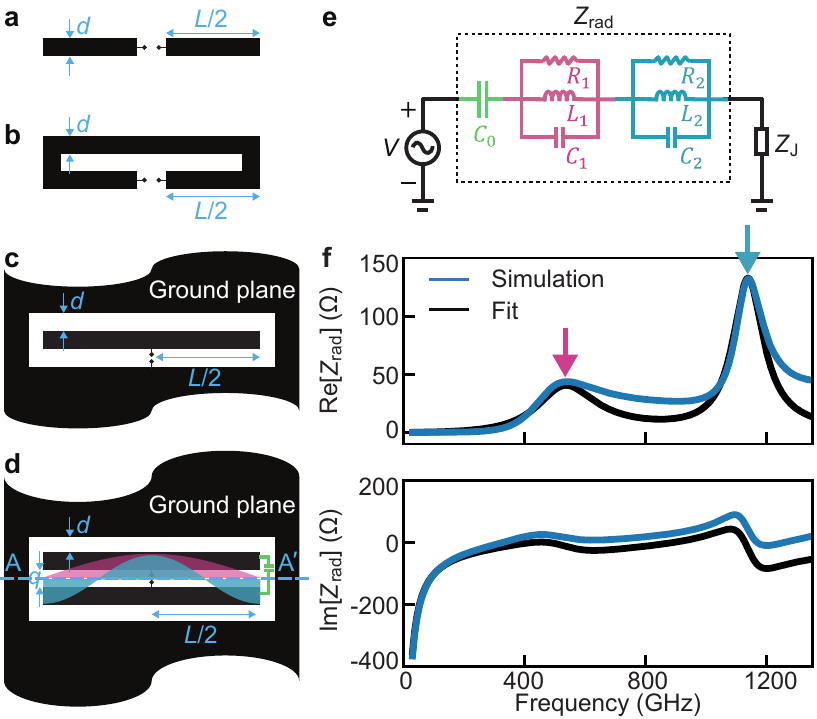}
\captionsetup{labelformat=empty,justification=raggedright}
\caption{\textbf{Supplementary Figure 11: Antenna mode of the qubit and equivalent circuit model.}
\textbf{a}, A simple half-wave dipole radiator.
\textbf{b}, A folded dipole radiator obtained by folding and connecting the two dipole ends in a.
\textbf{c}, A folded slot, which is the dual structure of b.
\textbf{d}, A paired folded slot radiator, obtained by mirroring c about the line AA', which shares a same structure as a symmetric floating transmon qubit.
\textbf{e}, Equivalent circuit of the qubit structure. $Z_{\rm rad}$ is the radiation impedance and $Z_{\rm J}$ is the impedance of Josephson junction.
\textbf{f}, Real and imaginary part of $Z_{\rm rad}$ from finite-element simulation and from calculating the equivalent circuit in \textbf{e}. Arrows indicate the corresponding resonance modes. In this example, the qubit parameters are $L=80\,\mu$m, $W=35\,\mu$m and $d=5\,\mu$m. In the equivalent circuit, $C_0=15$~fF, $R_1=40~\Omega$, $L_1=5.3$~pH, $C_1=17$~fF, $R_2=130~\Omega$, $L_2=2$~pH, $C_2=10$~fF.
}
\label{fig:AntennaModelFitting}\vspace{-8pt}
\end{figure}

\subsection*{A. Antenna mode of the qubit and equivalent circuit}

From the perspective of antenna theory, the floating qubit structure used in our work can be thought of as evolved from a simple dipole radiator. As shown in Supplementary Fig.~\ref{fig:AntennaModelFitting}a, we start with a conventional half-wave dipole containing two metallic arms, each of length $L/2$, and a feed at the center. The fundamental mode frequency is found equating $L$ to a half wavelength, while the higher-order resonances occur at frequencies such that $L$ corresponds to an integer multiple of a half-wavelength. By folding the two ends of the dipole back around and electrically connecting them together to form a loop, a folded dipole radiator can be obtained (Supplementary Fig.~\ref{fig:AntennaModelFitting}b), where the transverse length $L$ is the same. The input impedance of the folded dipole can be expressed as
\begin{equation}\label{equ:z_fd}
Z_{\rm fd} = 4Z_{\rm t}Z_{\rm d}/(Z_{\rm t}+2Z_{\rm d}) \;,
\end{equation}
where $Z_{\rm d}$ is the input impedance of a conventional dipole and $Z_{\rm t}$ is the input impedance of the transmission line formed by each folded arm with a short circuit loading~\cite{Kraus1997}. At the fundamental mode resonant frequency, $Z_{\rm fd} = 4Z_{\rm d}$ since $Z_{\rm t}$ approaches infinity. In order to adapt such radiating structures to superconducting qubits based on coplanar waveguide circuits, the Babinet's principle is applied to the folded dipole to obtain its dual structure, i.e. a folded slot (Supplementary Fig.~\ref{fig:AntennaModelFitting}c). The input impedance of this dual radiating structure can be related to that of the folded dipole radiator as $Z_{\rm fs} = Z_{0}^2/Z_{\rm fd}$, where $Z_0$ is the impedance of free space. Since the electric field within the folded slot is primarily perpendicular to the long edges of the slot, by utilizing image theory, another folded slot can be created by mirror reflection of the original one about the line AA', resulting in the paired folded slot radiator, which is structurally equivalent to a symmetric floating transmon (Supplementary Fig.~\ref{fig:AntennaModelFitting}d). It should be noted that the line AA' can be regarded as an electrical wall, such that the introduction of the additional folded slot below the line AA' will not affect the field distribution but increase the input impedance by a factor of two, i.e.\ $Z_{\rm rad} = 2Z_{\rm fs}$.

The frequency-dependent response of the radiation impedance ($Z_{\rm rad}$) over an ultra-wide frequency range, e.g.\ from $\sim$DC (10 GHz) to 1.2 THz, of the paired folded slot radiator can be modeled by an equivalent circuit. In order to ensure the passivity and causality of impedance response of the radiator, the equivalent circuit is made of capacitors, inductors, and resistors connected in series and/or parallel~\cite{Pozar2011}. Based on the structural characteristics of the pair folded slot radiator, it can be deduced that the fundamental resonance (mode 1) occurs at frequencies such that $L$ equals to about a half wavelength and the first higher-order resonant mode (mode 2) is found at frequencies such that $L$ is around a full wavelength. For slot radiators with the electric field polarized perpendicular to the long edges, the gap provides the capacitance while the metallic ground offers the inductance. Thus, both of the two slot modes can be modeled as a parallel RLC circuit with different circuit element values of $C_i$, $L_i$, and $R_i$, where $i = 1, 2$. In addition, there exists a coplanar capacitance, denoted as $C_0$, between the edges of the two metallic pads in the top and bottom folded slots. At frequencies away from the resonant modes, the two parallel RLC circuits possess a low impedance, behaving like a short circuit, indicating that these two RLC resonant circuits and the inter-pad capacitor should be connected in series.
Hence, the equivalent circuit can be modeled as shown in Supplementary Fig.~\ref{fig:AntennaModelFitting}e and the corresponding $Z_{\rm rad}$ can be experessed as
\begin{equation}
\begin{split}
Z_{\rm rad} = \frac{1}{j\omega C_0} &+\frac{1}{1/R_1+j\omega C_1+1/j\omega L_1} \\
&+\frac{1}{1/R_2+j\omega C_2+1/j\omega L_2},
\end{split}
\label{equ:Zradtheory}\vspace{-6pt}
\end{equation}

In order to validate this circuit model, we simulate the entire qubit structure using a finite-element electromagnetic field solver~\cite{AnsysHFSS} to obtain $Z_{\rm rad}$. We use a similar method as presented in Ref.~\cite{Rafferty2021}, which embeds the qubit layer in an uniform dielectric medium with effective permittivity $\epsilon_{\rm eff} = (1+\epsilon_{\rm r})/2 $ ($\epsilon_{\rm r}=11$ for sapphire).
The real and imaginary part of the obtained frequency responses of $Z_{\rm rad}$ are shown in Supplementary Fig.~\ref{fig:AntennaModelFitting}f for the example of a qubit with $L = 80\,\mu$m, $W = 35\,\mu$m, and $d = 5\,\mu$m,
and are fitted using the equivalent circuit model from Eq.~(\ref{equ:Zradtheory}); good agreement can be observed over a frequency range from near DC (10 GHz) to over 1.35~THz, thus validating the proposed circuit model for this physical transmon structure.
At low frequencies well below any resonance of the radiator, $Z_{\rm rad}$ should exhibit a purely capacitive response, as shown in the imaginary part of $Z_{\rm rad}$. The value of $C_0$ extracted from the simulation corresponds to a charging energy $E_{\rm C}/h \simeq 1.36\,$GHz, comparable to that estimated from qubit spectroscopy, see Supplementary Table~\ref{tab:tabel1}.
As frequency increases to the fundamental resonance, because the input impedance of the transmission line $Z_{\rm t} \to \infty$ and the dipole impedance $Z_{\rm d}$ has a large value right at resonance frequency, the folded dipole, $Z_{\rm fd} \approx 4Z_{\rm d}$ according to Eq.~(\ref{equ:z_fd}), has a large impedance. Therefore, the corresponding folded slot, $Z_{\rm fs} =  2Z_0^2/Z_{\rm fd}$, has a low impedance, as can be observed from the lower peak of the fundamental mode in the real part of $Z_{\rm rad}$ in Supplementary Fig.~\ref{fig:AntennaModelFitting}f.
At the frequency of the first higher-order resonant mode, a large $Z_{\rm rad}$ can be found since $Z_{\rm t}$ approaches zero when the value of $L$ is about a full wavelength. 

Next, we use the radiation impedance to estimate how efficiently pair-breaking photons are absorbed at the junction through the qubit antenna.
In the equivalent circuit, we model the external radiation as a Th\'{e}venin-equivalent generator with a voltage $V$~\cite{Sophocles2008}. The qubit structure apart from the junction can be seen as the internal impedance of the generator, which is $Z_{\rm rad}=R_{\rm rad}+jX_{\rm rad}$. The junction is the load with an impedance $Z_{\rm J}=R_{\rm J}+jX_{\rm J}$. The current is then $I=V/(Z_{\rm rad}+Z_{\rm J})$, and the power delivered to the antenna terminals $P_{\rm J}$ is:
\begin{equation}
\begin{split}
P_{\rm J}&=\frac{1}{2}|I_{\rm in}|^2R_{\rm J}\\
&=\frac{4R_{\rm J}R_{\rm rad}}{|Z_{\rm rad}+Z_{\rm J}|^2} \frac{|V|^2}{8R_{\rm rad}}\\
&=(1-|\Gamma_{\rm gen}|^2) P_{\rm J,max}\\
&=e_{\rm c} P_{\rm J,max} ,
\end{split}
\label{equ:PTer}\vspace{-6pt}
\end{equation}
where $P_{\rm J,max}$ is the maximum power that can be delivered to the junction when the conjugate matching condition $Z_{\rm rad}=Z_{\rm J}^*$ is satisfied or when the reflection coefficient vanishes, $\Gamma_{\rm gen}=\frac{Z_{\rm rad}-Z_{\rm J}^*}{Z_{\rm rad}+Z_{\rm J}}=0$. The coefficient $e_{\rm c}$ is a transfer function of frequency,
\begin{equation}
\begin{split}
e_{\rm c}=1-|\Gamma_{\rm gen}|^2=\frac{4R_{\rm J}R_{\rm rad}}{|Z_{\rm rad}+Z_{\rm J}|^2} ,
\end{split}
\label{equ:ec}\vspace{-6pt}
\end{equation}
valued between 0 and 1. It estimates the fraction of power transferred to the junction, the so called absorption or coupling efficiency~\cite{Rafferty2021}; $e_{\rm c}$ is proportional to the real parts of the impedances $Z_{\rm J}$ and $Z_{\rm rad}$, which are derived below.
Note that in the frequency response, we focus on photons with frequency $f^*$ twice the superconducting gap near the junction, $f^*=2\Delta/h=105$~GHz for a 30~nm thick aluminium film, because only photons above this frequency can break Cooper pairs and directly contribute to the observed parity switching. Moreover, since the emission spectrum of a radiation source at frequencies large compared to temperature, $hf \gg k_\mathrm{B}T$, generally takes an exponential distribution with frequency, $e^{-hf/k_{\rm B}T}$, only photons at $f^*$ make a significant contribution.
Therefore, in our comparison of power transfer efficiency and parity-switching rate, we use $e_{\rm c}^*=e_{\rm c}(f^*)$ instead of an integrated one.

\begin{figure}[htbp]
    \includegraphics{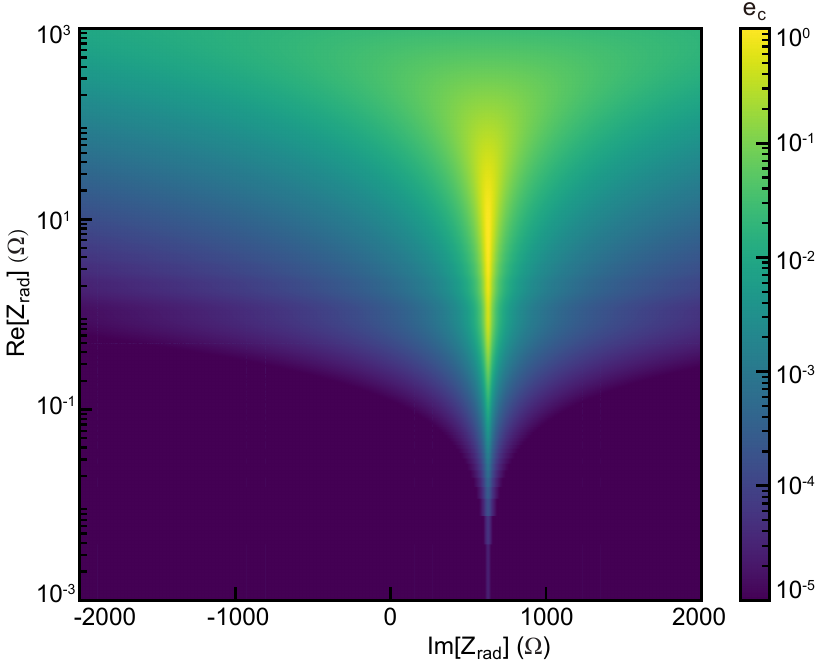}
\captionsetup{labelformat=empty,justification=raggedright}
\caption{\textbf{Supplementary Figure 12: Absorption efficiency versus $Z_{\rm rad}$.} The brightest point is the $Z_{\rm J}^*$ value. An effective way to decrease $e_{\rm c}^*$ is to decrease $R_{\rm rad}$, while for a given $R_{\rm rad}$, $X_{\rm rad}$ has little influence on $e_{\rm c}^*$. 
    }
    \label{fig:ecvsZrad}\vspace{-8pt}
\end{figure}

\begin{figure}[htbp]
\includegraphics{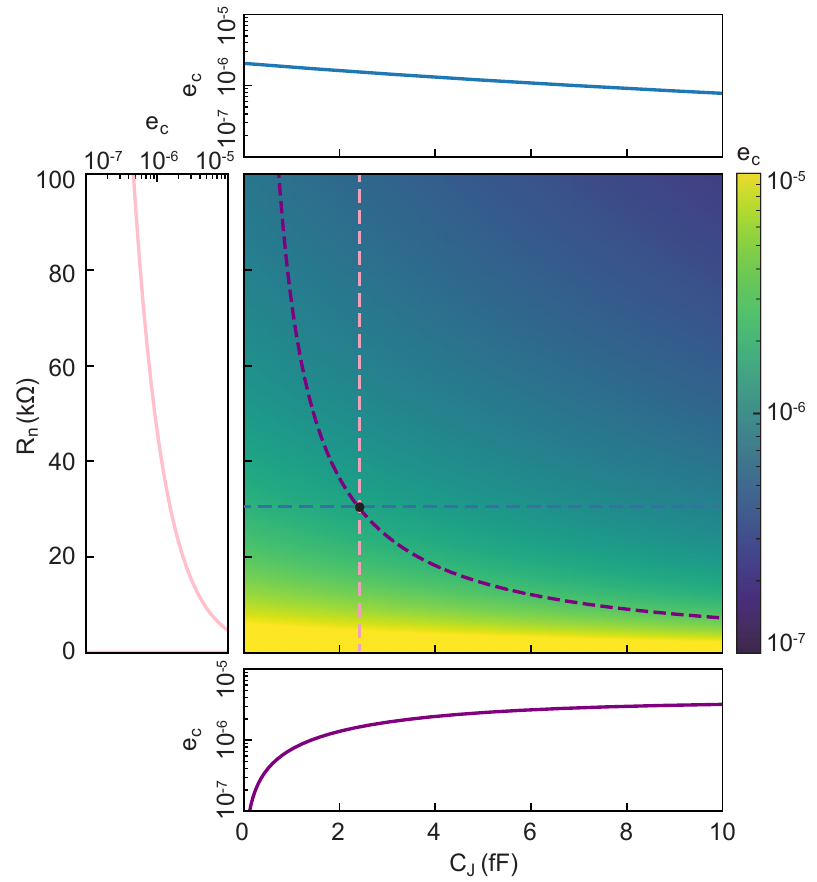}
\captionsetup{labelformat=empty,justification=raggedright}
\caption{\textbf{Supplementary Figure 13: Energy transfer efficiency of the qubit antenna $e_{\rm c}$ as a function of the junction capacitance $C_{\rm J}$ and the junction resistance $R_{\rm n}$.} Different dashed linecuts are shown in the surrounding panels in the same color coding.
}
\label{fig:junctionvary}\vspace{-8pt}
\end{figure}

For radiation at a frequency above twice the superconducting gap, the junction can be seen as a resistor, with resistance equal to its normal resistance $R_{\rm n}$, shunted by the junction capacitance $C_{\rm J}$. In our design, the qubit normal resistance $R_{\rm n}\simeq30~{\rm k\Omega}$, and the junction capacitance $C_{\rm J}\simeq2.4$~fF. The junction impedance is then:
\begin{equation}
\begin{split}
Z_{\rm J}=\frac{1-j\omega \tau}{1+\omega^2\tau^2}R_{\rm n} ,
\end{split}
\label{equ:ZJ}\vspace{-6pt}
\end{equation}
where $\tau = R_{\rm n}C_{\rm J}$.
For a given junction impedance ($R_{\rm J}=13~\Omega$, $X_{\rm J}=-624~\Omega$ evaluated at $f^*$), we plot in Supplementary Fig.~\ref{fig:ecvsZrad} the absorption efficiency $e_{\rm c}^*$ as a function of the real and imaginary parts of $Z_{\rm rad}$ using Eq.~(\ref{equ:ec}).
The maximum absorption occurs at $Z_{\rm rad}=Z_{\rm J}^*$, as expected from the conjugate matching condition.
According to Eq.~(\ref{equ:ec}), reducing $R_{\rm rad}$ is effective in suppressing $e_{\rm c}^*$. Therefore, in order to mitigate quasiparticle poisoning events, 
one should aim to
as small as possible $R_{\rm rad}$ when designing the device.
Besides, since the imaginary part of $Z_{\rm rad}$ is dominating in the denominator of Eq.~(\ref{equ:ec}), the more negative the imaginary part of $Z_{\rm rad}$, the smaller the absorption efficiency $e_{\rm c}^*$. This may be achieved by reducing the inter-pad capacitance $C_0$.

In Supplementary Fig.~\ref{fig:junctionvary}, we plot how the transfer efficiency $e_{\rm c}$ is affected by these factors according to the model described by the Eq.~(\ref{equ:ec}). The junction capacitance $C_{\rm J}$ is calculated by multiplying a specific capacitance of 75~fF/$\mu$m$^2$~\cite{Rafferty2021} with the design value of the junction area; both junction area and specific capacitance may vary from sample to sample due to fabrication variation. 
It can be seen that $e_{\rm c}$ is only weakly dependent on $C_{\rm J}$ (blue dashed line and top panel); it varies less than a factor of 2 over the whole range of $C_{\rm J}$ (0$\sim$10 fF). Because the junction resistance $R_{\rm n}$ scales inversely to the junction area, we have considered the case of covariation of $R_{\rm n}$ and $C_{\rm J}$ when the junction size fluctuates. The purple dashed line indicates where the product of $R_{\rm n}$ and $C_{\rm J}$ stays constant. In this case, $e_{\rm c}$ shows stronger dependence on $R_{\rm n}$; this is consistent with Eq.~(\ref{equ:ec}) where $e_{\rm c}$ has a near-linear scaling with $R_{\rm J}$ ($\propto R_{\rm n}$) and $R_{\rm rad}$. Considering typical fabrication variation, the variation in $e_{\rm c}$ is small around the typical working point ($R_{\rm n}$ = 30~k$\Omega$ and $C_{\rm J}$ = 2.43~fF) marked by the dark dot. In the paper, we deduce the value of $R_{\rm n}$ from the measured qubit spectrum.

\begin{figure*}[htbp]
    \includegraphics[width=0.95\textwidth]{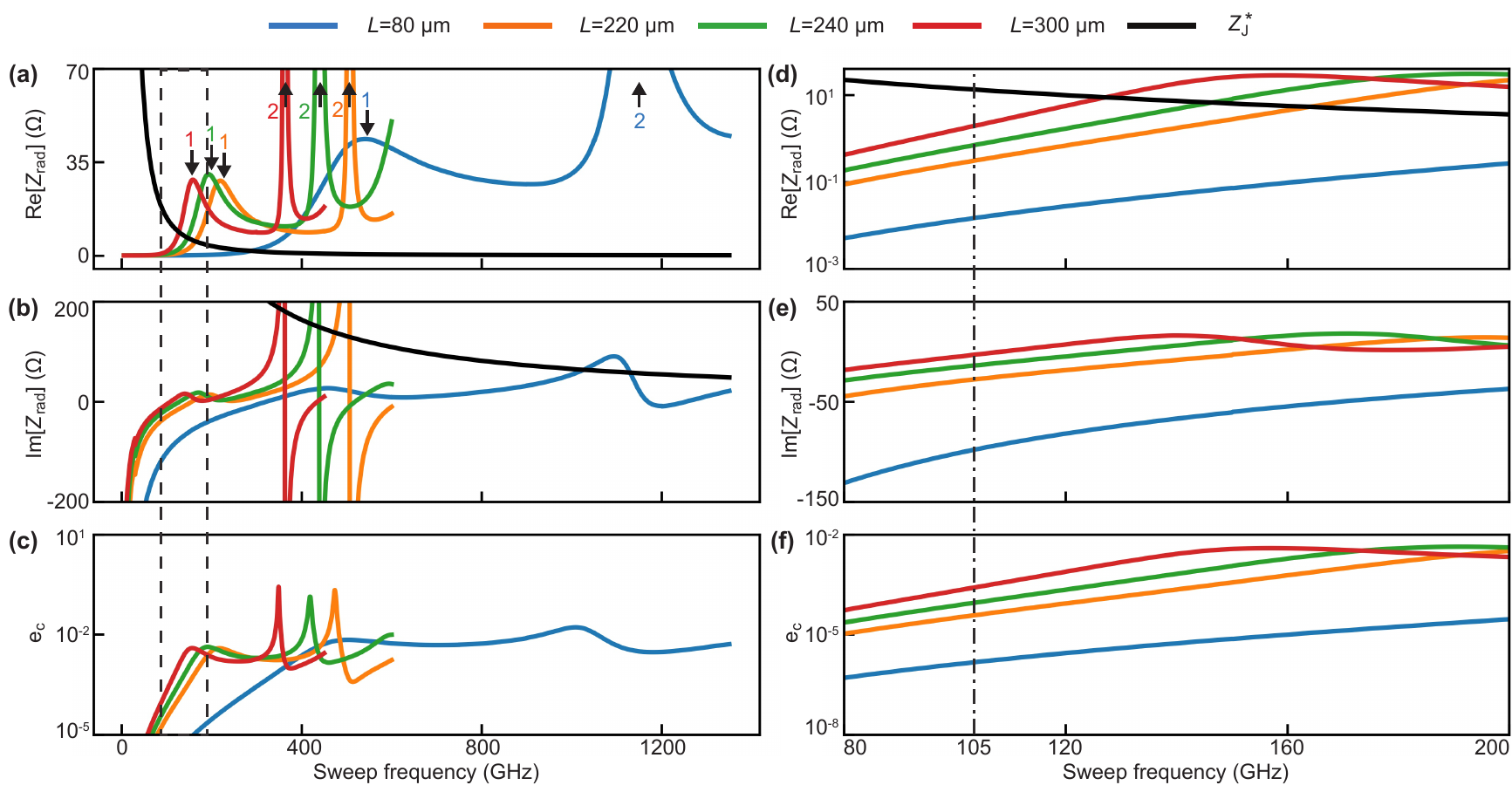}
\captionsetup{labelformat=empty,justification=raggedright}
\caption{\textbf{Supplementary Figure 14: Antenna impedance for different pad sizes.}
    \textbf{a}, The real part of $Z_{\rm rad}$. \textbf{b}, The imaginary part of $Z_{\rm rad}$.
    Arrows indicate the corresponding first and second resonance modes.
    \textbf{c}, The absorption efficiency $e_{\rm c}$.
    \textbf{d-f}, Zoom-in view of \textbf{a-c} between 80$\sim$200~GHz.
    For the junction impedance $\rm Z_{\rm J}$, $R_\mathrm{n}$=30~k$\Omega$ and $C_\mathrm{J}$=2.43~fF. }
    \label{fig:AntennaimpedanceL}\vspace{-8pt}
\end{figure*}

\begin{figure*}[htbp]
    \includegraphics[width=0.95\textwidth]{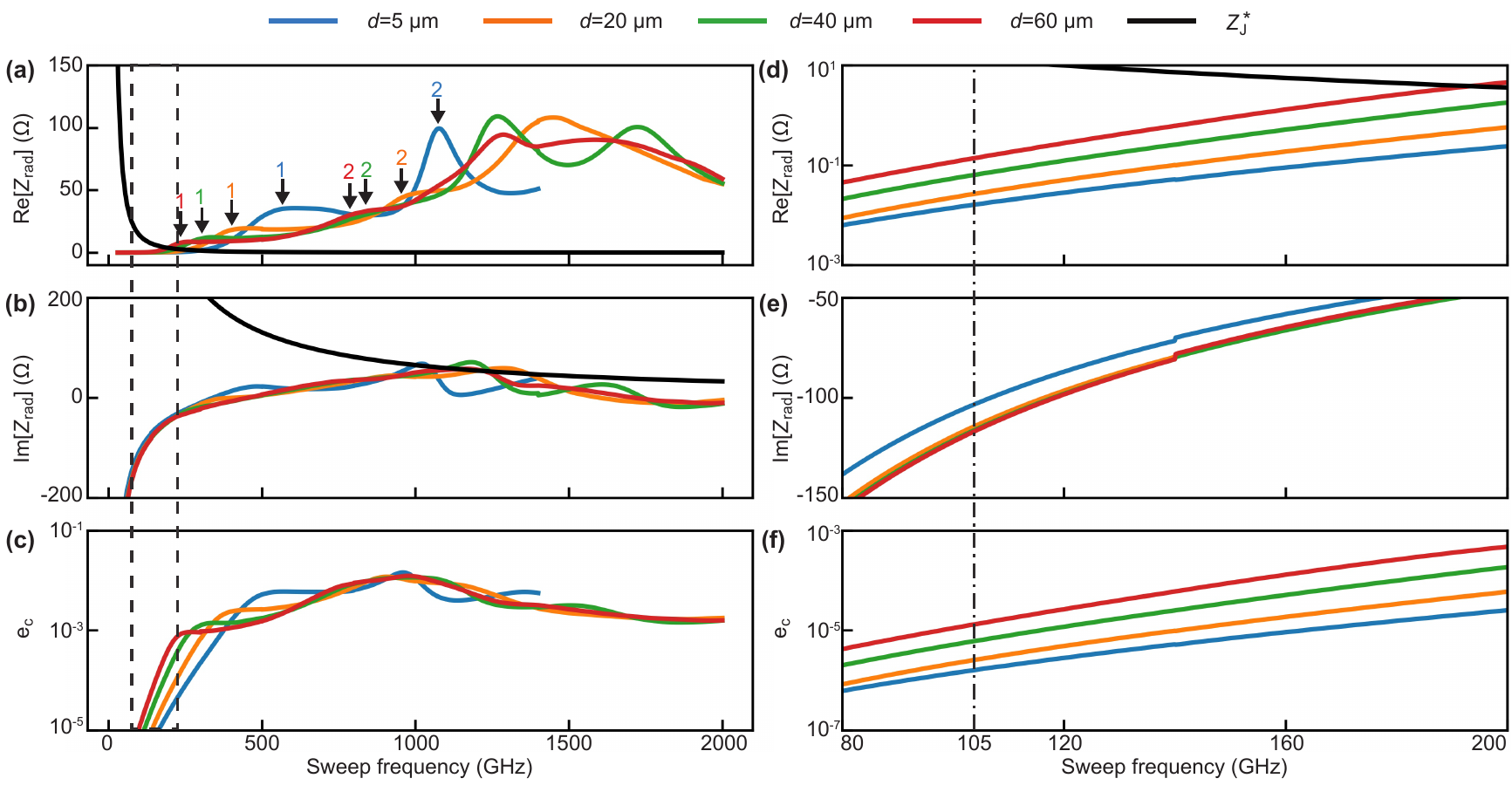}
\captionsetup{labelformat=empty,justification=raggedright}
\caption{\textbf{Supplementary Figure 15: Antenna impedance vs pad-to-ground distance.}
    \textbf{a}, The real part of $Z_{\rm rad}$. \textbf{b}, The imaginary part of $Z_{\rm rad}$.
    Arrows indicate the corresponding first and second resonance modes.
    \textbf{c}, The absorption efficiency $e_{\rm c}$.
    \textbf{d-f}, Zoom-in view of \textbf{a-c} between 80$\sim$200~GHz.
    For the junction impedance $\rm Z_{\rm J}$, $R_\mathrm{n}$=30k$\Omega$ and $C_\mathrm{J}$=2.43~fF. }
    \label{fig:Antennaimpedanced}\vspace{-8pt}
\end{figure*}

\begin{figure*}[htbp]
    \includegraphics[width=1\textwidth]{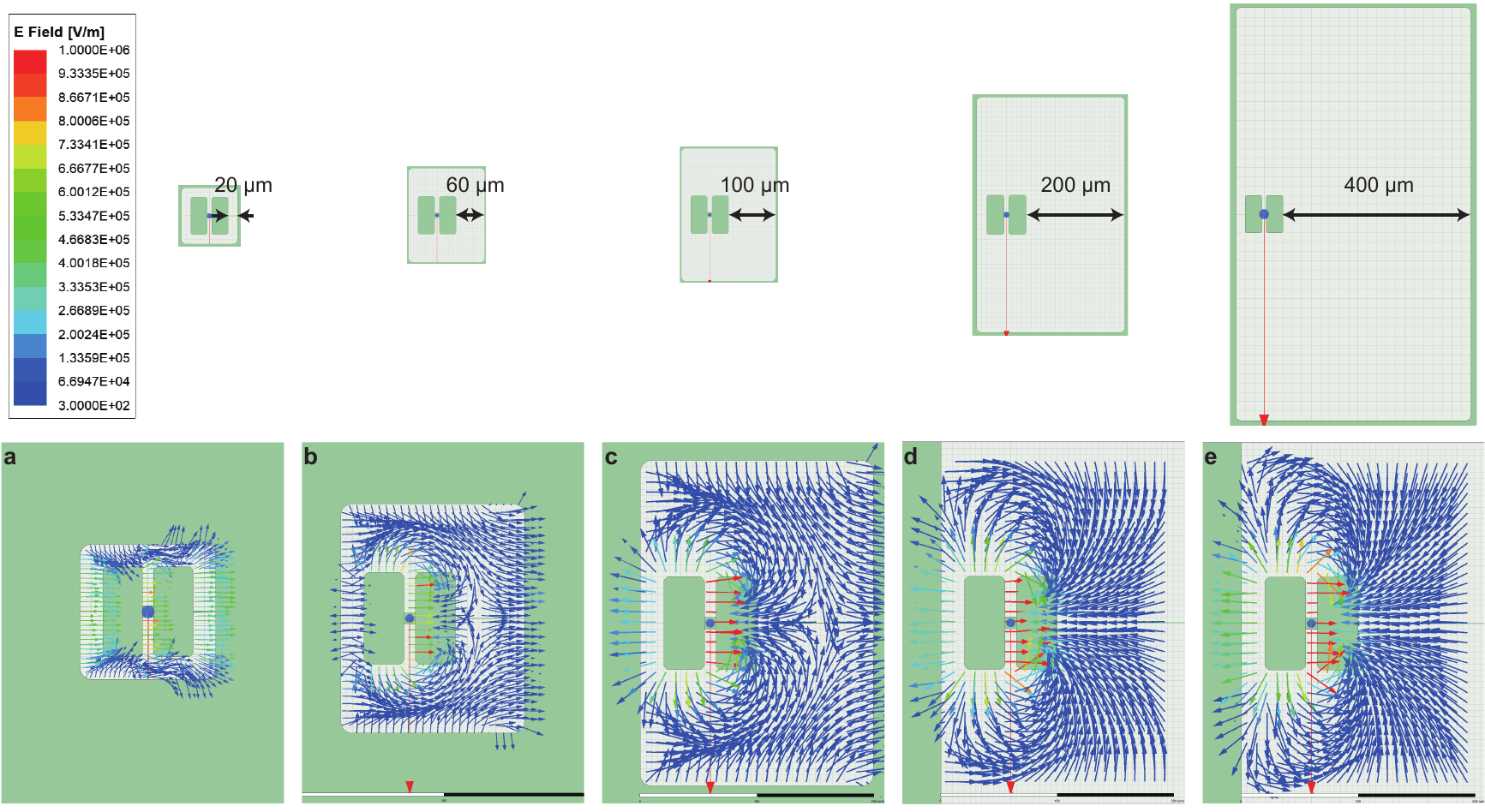}
\captionsetup{labelformat=empty,justification=raggedright}
\caption{\textbf{Supplementary Figure 16: E-field plot of the antenna mode for qubits with different pad-to-ground distance.}
    \textbf{a$\sim$e}, $d=20\sim400~\mu$m.
 All qubits have same pad length $L=80~\mu$m and pad width $W=35~\mu$m. As the pad-to-ground distance $d$ increases, the mode field spreads into the gap, and changes the effective wavelength. When $d$ is large enough ($d>L$), the effective wavelength tends to saturate at a fixed value. 
    }
    \label{fig:Efield}\vspace{-8pt}
\end{figure*}

\begin{figure}[htbp]
    \includegraphics{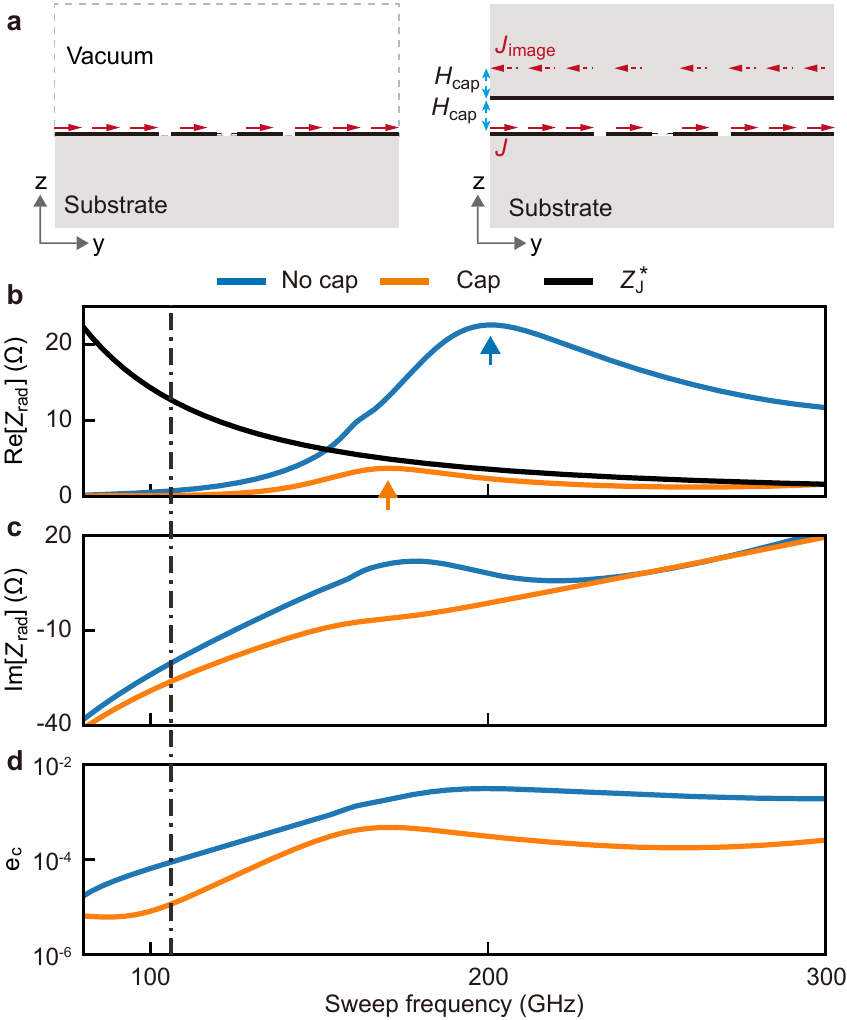}
\captionsetup{labelformat=empty,justification=raggedright}
\caption{\textbf{Supplementary Figure 17: Antenna impedance with and without cap.}
    \textbf{a}, The current distribution  with and without cap. The distance between chips is about $10~\mu\rm m$.
    \textbf{b}, The real part of $Z_{\rm rad}$. 
    \textbf{c}, The imaginary part of $Z_{\rm rad}$.
    \textbf{d}, The absorption efficiency $e_{\rm c}$.
    For the junction impedance $\rm Z_{\rm J}$, $R_\mathrm{n}$=30~k$\Omega$ and $C_\mathrm{J}$=2.43~fF.
     }
    \label{fig:3DCap}\vspace{-8pt}
\end{figure}

\subsection*{B. Geometric effect on antenna impedance}

To understand the impact of the key geometrical parameters to the radiation impedance of this radiating structure, we perform a parametric study.
We first simulate qubit geometries with varying pad size as used in the experiment. The results are shown in Supplementary Fig.~\ref{fig:AntennaimpedanceL}.
It can be seen that the first two resonances shift towards lower frequencies when $L$ increases. Since the fundamental mode is always higher than the characteristic frequency $f^*=105$~GHz in this range, the closer these two frequencies become, the more energy the qubit absorbs. Accordingly, the real part of $Z_{\rm rad}$ becomes also larger, leading to higher absorption efficiency $e_{\rm c}^*$, in agreement with theory.
We extend the simulation with $L$ densely sampled from $60~\mu$m to $320~\mu$m and with $W=35~\mu$m and $60~\mu$m. The results are shown in Fig.~3c in the main text.

We also simulate geometries with varying pad-to-ground distance $d$ as shown in Supplementary Fig.~\ref{fig:Antennaimpedanced}. As the gap between pad and ground widen, the resonances shift towards lower frequency. In a similar argument as above, such a change leads to an increased real part of $Z_{\rm rad}$ and greater $e_{\rm c}^*$.
We extend the simulation with $d$ densely sampled from $5~\mu$m to $400~\mu$m. The results are shown in Fig.~3d in the main text.
The phenomenon can be understood by taking a closer look at the field distribution of the fundamental resonance mode.
As shown in Supplementary Fig.~\ref{fig:Efield}, when $d$ is increased from $20~\mu$m to $400~\mu$m, the fundamental mode is no longer confined in a narrow space right next to the pad, but expands outwards from the pad edge, leading to a longer perimeter and hence a greater effective wavelength. However, when $d$ is large enough ($d>L$), the mode field tends to stabilize, not changing further. This qualitatively agrees with the saturation behavior when $d>100~\mu$m, but we do not know the reason for the quantitative discrepancy between experimental data and simulation result.

\begin{figure*}[htbp]
\includegraphics{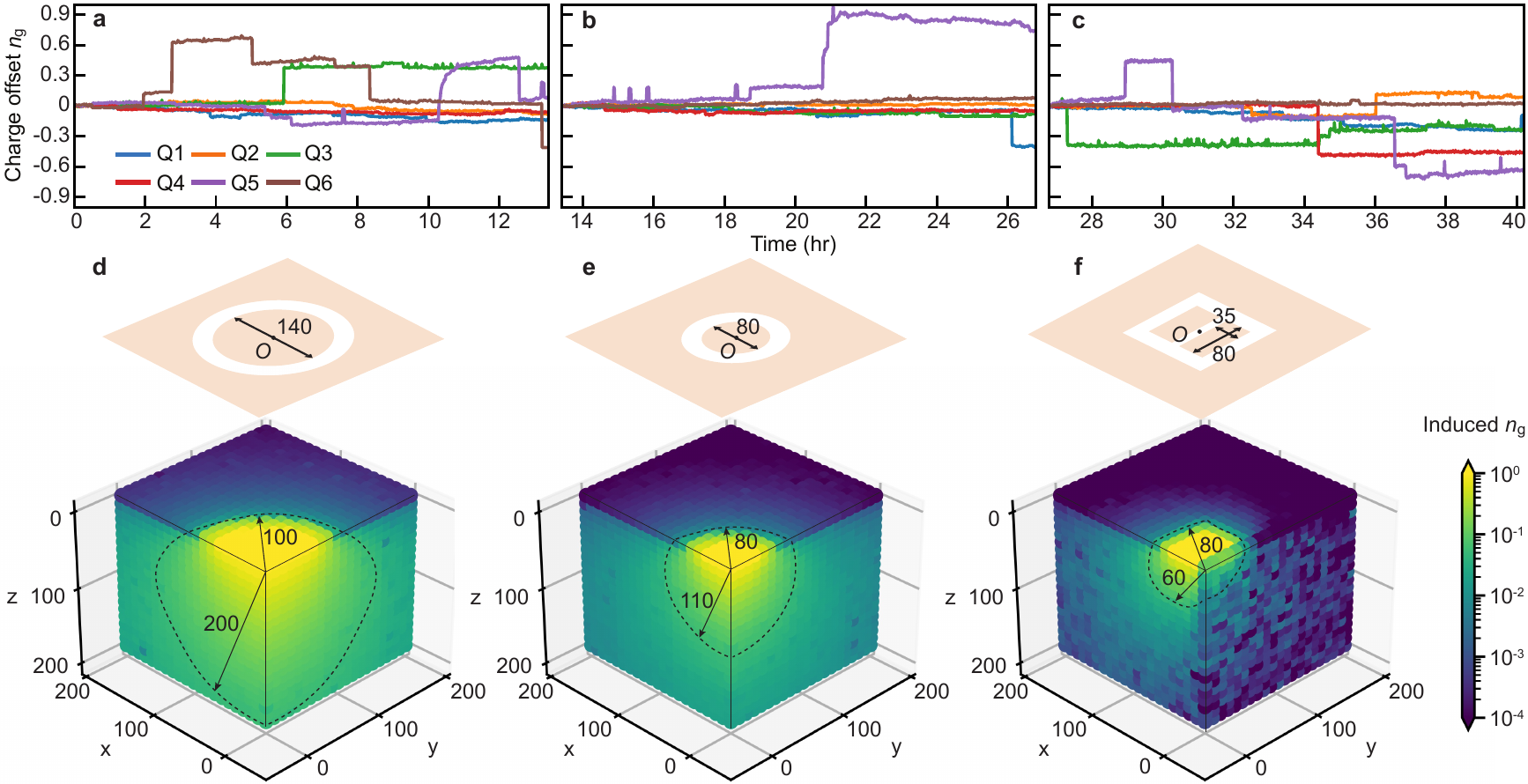}
\captionsetup{labelformat=empty,justification=raggedright}
\caption{\textbf{Supplementary Figure 18: Offset charge stability.}
\textbf{a-c}, Temporal trajectories of offset charge drift recorded during three separate sections over a total of 40 hours. The offset is reset at the beginning of each section.
\textbf{d-f}, Calculated offset charge induced by a test point charge in the substrate with \textbf{d}, a grounded transmon with a single circular pad (diameter $140~\mu$m, as used in Ref.~\cite{Wilen2021}); \textbf{e}, a grounded transmon with a smaller circular pad (diameter $80~\mu$m); and \textbf{f}, a floating transmon with two rectangular pads (each $80~\mu$m$\times35~\mu$m) on the top surface. $O$ is the original point (0,0,0) in the Cartesian space in which the coordinates are the location of the test charge. We show cross-sectional planes for $x=0$ and $y=0$, so a quarter of the substrate is displayed. The dashed line denotes the contour of $0.1\,e$ induced charge.
Note that in the floating transmon case, the induced offset charge vanishes on the $x=0$ plane because the induced charges are symmetric on the two pads.
}
\label{fig:ChargeOffsetJump}\vspace{-8pt}
\end{figure*}

\begin{figure*}[htbp]
\includegraphics{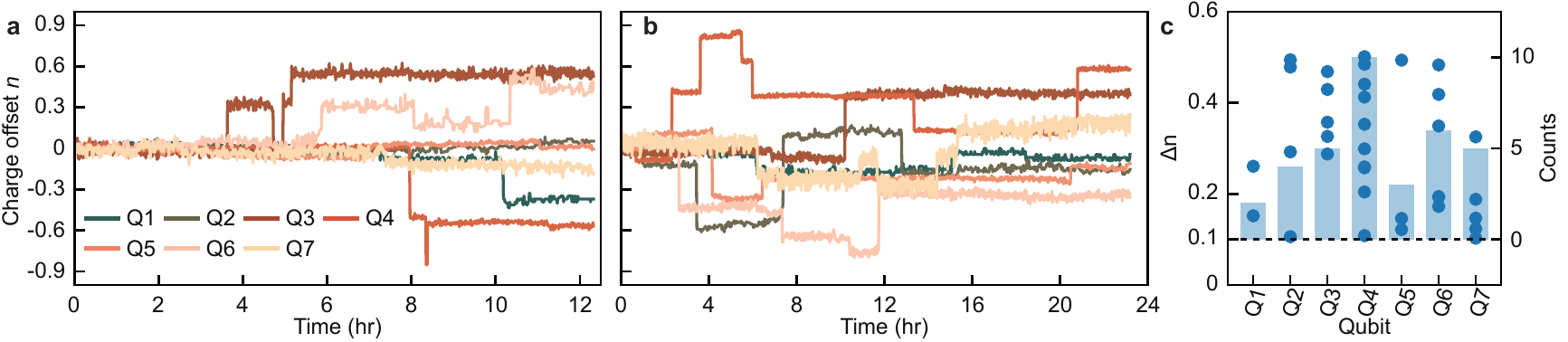}
\captionsetup{labelformat=empty,justification=raggedright}
\caption{\textbf{Supplementary Figure 19: Offset charge stability of capped floating qubits.}
\textbf{a-b}, Temporal trajectories of offset charge drift recorded during two separate sections over a total of 36 hours. The offset is reset at the beginning of each section.
\textbf{c}, Amplitudes (left axis) and total counts (right axis) of all offset charge jumps ($|\Delta q|$) that are greater than $0.1\,e$, identified from the data in \textbf{a-b}.
}
\label{fig:3DChargeOffsetJump}\vspace{-8pt}
\end{figure*}

\begin{figure}[htbp]
\includegraphics{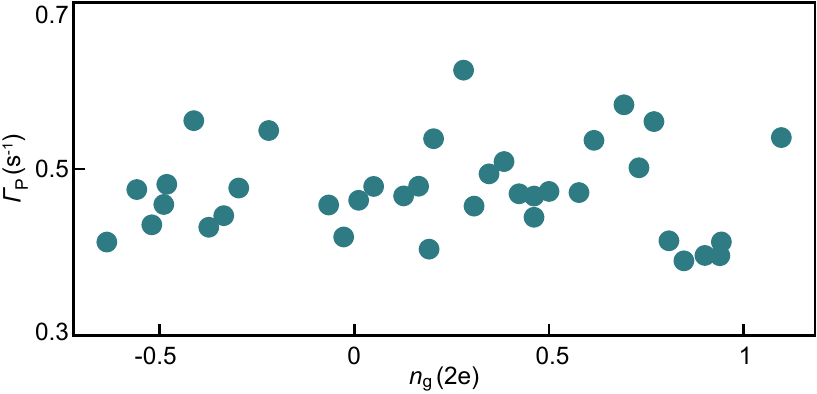}
\captionsetup{labelformat=empty,justification=raggedright}
\caption{\textbf{Supplementary Figure 20: Measured $\Gamma_P$ at different charge offset bias.}}
\label{fig:TpBias}\vspace{-8pt}
\end{figure}

In addition, capping the qubit helps reduce $e_{\rm c}$ without causing significant change to the qubit parameters such as $E_{\rm J}/E_{\rm C}$. 
When a square-shaped metallic cap with an edge length of 300--500~$\mu$m is placed on top of the qubit structure at a distance of $10~\mu$m, as shown in Supplementary Fig.~\ref{fig:3DCap}a (cf. Fig.~3a in the main text), the radiation properties of the qubit can be greatly affected. Specifically, the cap behaves as another floating ground plane located in close proximity of the paired folded slot radiator, on which currents are induced that would also contribute to the radiated field. Based on the image theory, virtual currents located on the other side of the cap flowing in the horizontal directions can be used to evaluate the impact of the added floating ground. Since the virtual image currents are pointing into directions opposite to those flowing on the qubit, i.e. they are out of phase to each other, and the vertical spacing is greatly smaller than the wavelength of interest, the radiated fields from the two kinds of currents will cancel each other, thereby resulting in a significantly reduced radiation impedance~\cite{Balanis2012} and $e_{\rm c}$, as shown in Supplementary Fig.~\ref{fig:3DCap}b-d. In other words, the radiator can be considered to be shorted out by the added metallic cap at a deep-subwavelength distance. 
We extend the simulation with $L$ densely sampled from $80~\mu$m to $260~\mu$m and with a cap sufficiently large ($500\times500~\mu {\rm m}^2$) to cover the qubit. The results are shown in Fig.~3c of the main text.

\section*{Supplementary Note 6: Offset charge stability}

In the experiment monitoring long-term offset charge stability, we simultaneously sweep the bias of all six qubits from sample S1 and probe their readout resonators.
Every scan takes 24~s, later fitted to identify zero bias. The scan is repeated for about 40 hours and divided into 3 sections. Trajectories of all qubits are shown in Supplementary Fig.~\ref{fig:ChargeOffsetJump}(a-c), which records bias charge fluctuations over time.

We find that in our devices, both the frequency and the amplitude of charge jumps are considerably less than the ones in Ref.~\cite{Wilen2021}.
We attribute the improvement to the difference in qubit geometries: it can be understood from the perspective of the effective volumes of the charge sensitive regions.
Take the example of the circular transmon from Ref.~\cite{Wilen2021}, which is a round-shaped pad with diameter of $140~\mu$m. The calculated induced charge on the pad given one electron charge at a certain location in the substrate is shown in Supplementary Fig.~\ref{fig:ChargeOffsetJump}d. The dashed contour line encircles the region where the induced charge is greater than $0.1~e$. The total volume of this charge-sensitive region is about $1.1\times10^7~\mu {\rm m}^3$.
Such a volume can be reduced by simply making the qubit smaller. For example, Supplementary Fig.~\ref{fig:ChargeOffsetJump}e shows the case in which the diameter of the circular transmon is reduced to $80~\mu$m, giving a charge-sensitive volume of $1.5\times10^6~\mu {\rm m}^3$.
Though our floating transmon is similar in size as the small circular transmon, the charge-sensitive volume is nevertheless smaller, being only $5.0\times10^5~\mu {\rm m}^3$ (Supplementary Fig.~\ref{fig:ChargeOffsetJump}f). The floating design is only sensitive to the differential charge induced on the two pads, so charges beneath the gap between the pads (the $x=0$ plane) induce equal amount of charges on the pads due to symmetry and thus do not contribute to the offset bias.
The effective volume can be considered similarly to a scattering cross section, which is proportional to the occurrence rate of observed events.
The effective volume in our design is almost two orders of magnitude smaller than the circular transmon of Ref.~\cite{Wilen2021}; this agrees with the ratio between the observed rates of charge jumps ($>0.1\,e$), 0.007-0.063~mHz in this work versus 1.35~mHz in Ref.~\cite{Wilen2021}.

For capped floating qubits with different pad length $L$ but same pad-to-ground distance $d$, the simultaneous charge stability monitoring is shown in Supplementary Fig.~\ref{fig:3DChargeOffsetJump}(a-b). Supplementary Fig.~\ref{fig:3DChargeOffsetJump}c shows the amplitudes and total counts of all offset charge jumps ($|\Delta q|$) that are greater than $0.1\,e$ extracted from Supplementary Fig.~\ref{fig:3DChargeOffsetJump}(a-b). The charge jump rates are about 0.015-0.077~mHz.

To exclude the possibility of the measured charge-parity switching rate being affected by charge jumps, we also measured $\Gamma_P$ at different offset charge $n_\mathrm{g}$ spanning more than a full period. A typical set of data is shown in Supplementary Fig.~\ref{fig:TpBias}. We do not observe significant bias dependence in our devices; this is in agreement with theoretical expectations: at leading order the matrix element determining the parity switching rate is independent of $n_\mathrm{g}$~\cite{Catelani2014,Houzet2019}, and while the spectral density depends on $n_\mathrm{g}$ through the energy difference $\epsilon_0$ between the state of opposite parities, this dependence is small in the parameter of order $\epsilon_0/2\Delta$~\cite{Houzet2019}.

\begin{figure}[!bth]
\includegraphics[scale=0.95]{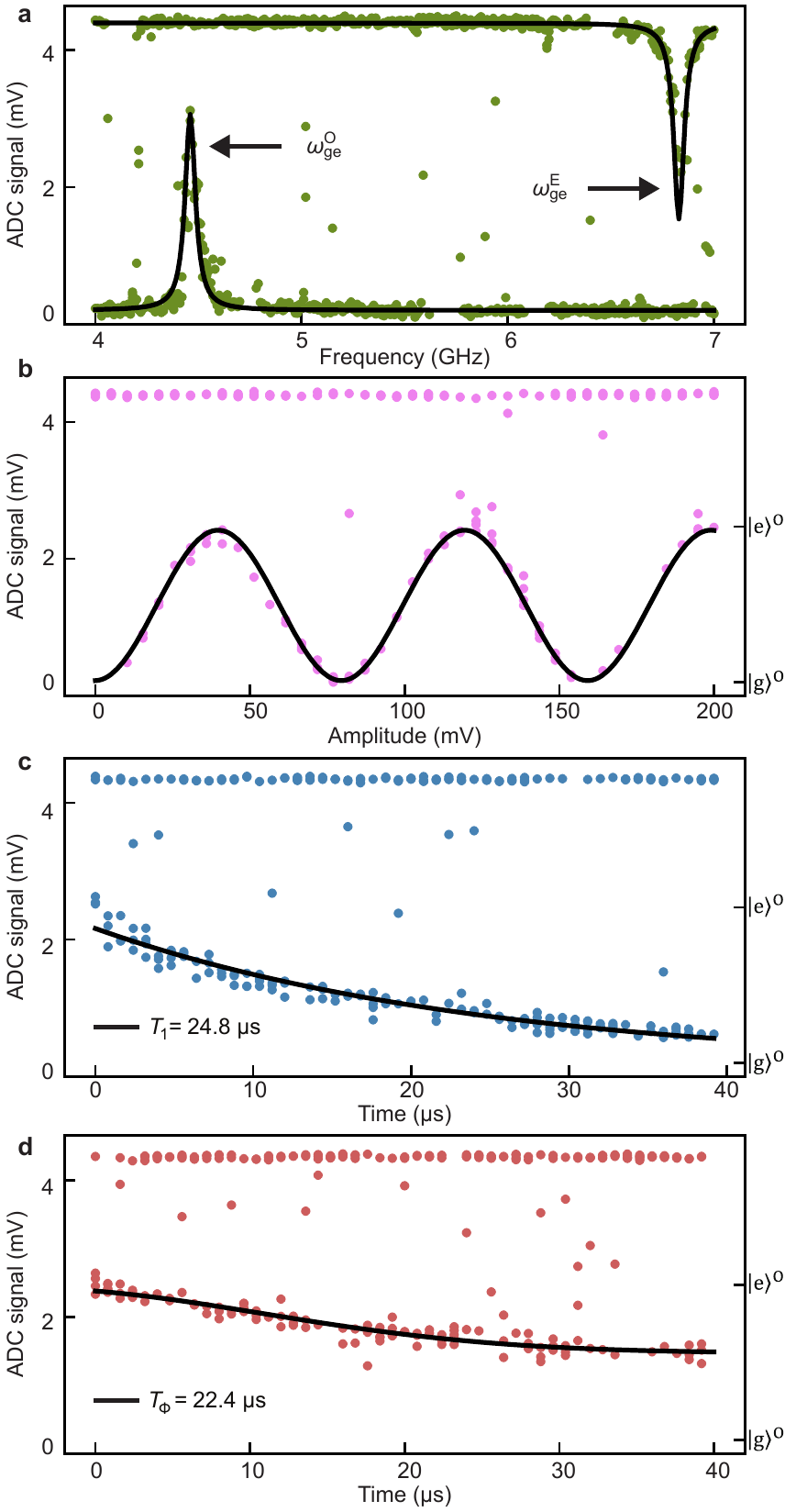}
\captionsetup{labelformat=empty,justification=raggedright}
\caption{\textbf{Supplementary Figure 21: Qubit characterization.}
\textbf{a}, Qubit spectrum measured at $n_{\rm g}=0$. Data points shown are from three repeated scans.
The outliers are caused by parity switching which happens during the time taking that data point which, therefore, takes a value in between those of the two parities.
\textbf{b}, Rabi oscillations obtained by driving the qubit at its odd-parity frequency $\omega_{\rm ge}^{\rm O}$.
\textbf{c}, Energy relaxation measured with a $\pi$ pulse calibrated from the odd-parity Rabi oscillations. Solid line is the exponential fit to $f(t)=A e^{-t/T_1} + B$.
\textbf{d}, Spin echo decay. Solid line is the fit to $G(t) = A e^{-t/2T_1-(t/T_{\phi})^2} + B$, which accounts for a Gaussian dephasing -- $1/f$ noise -- model and where $T_1$ is fixed to the energy relaxation value previously measured.}
\label{fig:T1T2E}\vspace{-8pt}
\end{figure}

\vspace{-10pt}
\section*{Supplementary Note 7: Qubit calibration and characterization}\label{sec:QubitCalibration}

\begin{figure}[!th]
\includegraphics{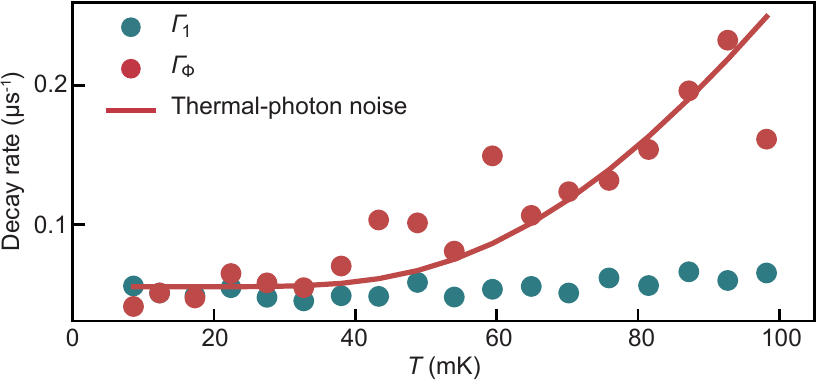}
\captionsetup{labelformat=empty,justification=raggedright}
\caption{\textbf{Supplementary Figure 22: Coherence versus temperature.}
Points: Temperature dependence of energy relaxation rate $\Gamma_1$ and pure dephasing rate $\Gamma_\phi$ measured at $n_{\rm g}=0$. Line: fit to Eq.~(\ref{eq:th_ph_noise}) with the resonator decay
rate $\kappa=0.8$~MHz and the dispersive shift $\chi=0.4$~MHz.}
\label{fig:CoherenceTemperature}\vspace{-6pt}
\end{figure}

Due to random parity switching, to calibrate and characterize qubits of small $E_{\rm J}/E_{\rm C}$ ratio requires repeated measurements.
As shown in the example of qubit spectrum measurement in Supplementary Fig.~\ref{fig:T1T2E}a, the qubit spectral peaks at $n_{\rm g}=0$ of both parities become visible by plotting together repeated scans, even though the signal may switch between parities during a single scan.
After finding the qubit frequency, we perform Rabi oscillations, to calibrate single-qubit gates, and then perform $T_1$ and $T_2$ measurements, as shown in Supplementary Fig.~\ref{fig:T1T2E}b-d.

We also measured the temperature dependence of coherence properties. As shown in Supplementary Fig.~\ref{fig:CoherenceTemperature}, the energy relaxation rate $\Gamma_{1}=1/T_1$ is largely temperature independent below 100~mK. This is consistent with the fact that in our model for the temperature dependence of $\Gamma_P$ (see Supplementary Note 8) up to this temperature only a small fraction (few percent) of the pads' quasiparticle density $x_\mathrm{qp}$ can be thermally excited, so that even in the worst case the quasiparticles would limit the relaxation time to several hundred microseconds.
Meanwhile the increase of the pure-dephasing rate $\Gamma_{\rm \phi}$ with temperature can be well explained by dephasing induced by thermal photons in the readout resonator according to~\cite{yan_flux_2016}
\begin{equation}\label{eq:th_ph_noise}
    \Gamma_{\rm \phi} = \frac{\kappa^2}{\kappa^2+4\chi^2} \frac{4\chi^2}{\kappa} \Bar{n} \, , 
\end{equation}
where $\Bar{n}=1/(e^{\hbar\omega_{\rm r}/k_{\rm B}T}-1)$ is the thermal photon number in the resonator, $\kappa=0.8\,$MHz is the resonator decay rate, and $\chi=0.4\,$MHz the dispersive shift due to the qubit state (not its parity).

\vspace{-10pt}
\section*{Supplementary Note 8: Temperature dependence of the parity switching rate}
\label{sec:GPvsT}

The change of parity in the state of a superconducting qubit is a clear signature of a transition mediated by quasiparticles. Generically, the rate $\Gamma_P$ of a parity-changing transition can have two contributions: one, $\Gamma_{\rm qp}$, originating
from the tunnelling of a quasiparticle from one side of the junction to the other;
the second, $\Gamma_{\rm pb}$, caused by the absorption of a Cooper pair-breaking photon at
the junction accompanied by the generation of one quasiparticle on each side of
the junction (for low quasiparticle densities, the opposite process of quasiparticle recombination with photon emission can be neglected):
\begin{equation}\label{equ:gammap0}
      \Gamma_P=\Gamma_{\rm qp}+\Gamma_{\rm pb} \, .
\end{equation}
As the pair-breaking photon energy is larger than $2\Delta$, corresponding in Al to a
temperature of a few degerees Kelvin, it is reasonable to assume that the corresponding rate
is independent of the temperature $T$ at the coldest stage of the fridge, which in
our experiment range from few mK up to about 100~mK. Thus, we attribute
the temperature dependence to quasiparticle tunnelling.

In quasiequilibrium, the temperature dependence of $\Gamma_{\rm qp}$ can be obtained
using the results of Ref.~\cite{Catelani2014}:
\begin{equation}\label{equ:gammaqp}
\Gamma_{\rm qp}(T)=\frac{16E_{\rm J}}{\Delta}c_0^2\frac{\epsilon_0}{h}e^{-(\Delta-\mu)/k_{\rm B}T}F\left(\frac{\epsilon_0}{2k_{\rm B}T},\frac{k_{\rm B}T}{2\Delta}\right),
\end{equation}
where $\epsilon_0$ is the energy difference between the two (ground) states of different parity, $c_0$ is the matrix element between those two states of the operator $\cos(\hat\phi/2)$,
$\mu$ is an effective chemical potential that encapsulates the deviation from
thermal equilibrium, and the function $F$ is
\begin{equation}\label{equ:Fxy}
F(x,y)=\cosh(x)[K_1(x)-xyK_0(x)]\,,
\end{equation}
where $K_i$ are the modified Bessel function of the second kind. The above formula
is an approximate expression which is accurate so long as $y\lesssim0.1$ and $xy\lesssim0.08$.
We note that $E_{\rm J}$, $\epsilon_0$, and $c_0$ can be extracted from the spectroscopic data and
hence, along with temperature, they are not free parameters. Not
known are the values of the gap $\Delta$ at the junction and of the chemical potential, which we consider next.

As discussed in the main text, our samples consist of two pads of thickness 100~nm connected by two thinner superconducting strips (30 to 40~nm thickness), with a Josephson tunnel junction formed where the strips overlap. It is well known~\cite{Chubov1969} that the gap in Al films depends on their thickness; for thick
films, such as those of the pads, the gap is close to the bulk value $\Delta_0=$180~$\mu$eV ($\Delta_0/h =$43.52~GHz). The gap increases as thickness decreases, and for 30~nm
thick films values between 200 and 220~$\mu$eV (48.36 to 53.20~GHz) were reported
in Ref.~\cite{Court2007}. An energy difference of 20 (40)~$\mu$eV corresponds to a temperature of about 230 (460)~mK; therefore, at temperatures well below this value the pads act as
traps for quasiparticles in the manner studied in Ref.~\cite{Riwar2019}. This means that at
low temperatures the quasiparticles should be largely confined to the pads and
hence, in qualitative agreement with our experimental data, the parity switching
rate should be independent of temperature. As temperature increases, quasiparticles can be thermally excited from the pads to the strips and hence reach the junction (cf. Fig.~1e in the main text), which could explain the increase in $\Gamma_P$ with temperature.

\begin{table*}[!bth]
\centering
\begin{tabular}{c|c|c|c|c|c||c|c|c|c||c|c|c}
\hline
\hline
& \makecell[c]{Pad length\\ $L\,(\rm \mu m)$}& \makecell[c]{Pad width \\$W\,(\rm \mu m)$} & \makecell[c]{Distance \\ $d\,(\rm \mu m)$} & $\omega^{\rm max}_{\rm ge}/2\pi$&$\omega^{\rm min}_{\rm ge}/2\pi$&$E_{\rm J}/h$ &$ E_{\rm C}/h$ &$\epsilon_0/h$&$c_0^2$&$\Gamma_P(0)$&$\Delta/h$&$10^{7}x_{\rm qp}$\\
\hline
S1$-$Q1 & 80 & 35 & 5  &6.833&4.473  & 4.67  &1.40 &0.238      &0.775&0.48  &50.9 &0.22 \\
S1$-$Q2 & 80 & 35 & 10 &6.954&4.135  & 4.27  &1.48 &0.319      &0.755&0.77 &50.9&1.94\\
S1$-$Q4 & 80 & 35 & 20 &7.268&4.469  & 4.63  &1.53 &0.305      &0.762&0.92 &50.9 &2.28 \\
\hline
S4$-$Q1 & 80 & 35 & 5  &7.845&6.754  &7.61  &1.30&0.0677     &0.838&0.57 &52.53&0.73\\
S4$-$Q2 & 220& 35 & 5  &5.774&5.505  & 6.92 &0.78&0.0117     &0.872&13.93 &52.53&3.51\\
S4$-$Q3 & 240& 60 & 5  &6.068&6.067  & 12.25&0.44&$2.18\times 10^{-5}$&0.931&34.06 &52.53&5.16\\
\hline
S5$-$Q1& 80  & 35  & 5 &7.246&5.740&6.25 &1.31&0.112&0.817&1.59&49.50&0.35  \\
S5$-$Q2& 220 & 35  & 5 &4.907&4.400&5.12 &0.76&0.027&0.850&19.11&49.50&3.90 \\
S5$-$Q3& 240 & 60  & 5 &6.2013&6.2012&14.66 &0.37&$1.6\times10^{-6}$&0.942&44.12&49.50&2.87\\
\hline
S8$-$Q2 & 80  & 35 & 5  &5.566&2.281  & 2.35  &1.29 &0.490      &0.691&0.42  &48.55 &0.96 \\
S8$-$Q3 & 120 & 35 & 5  &4.260&2.230  & 2.23  &0.95 &0.272      &0.726&0.63 &48.55&1.15\\
S8$-$Q5 & 180 & 35 & 5  &4.254&3.858  & 4.45  &0.66 &0.024      &0.846&1.10 &48.55 &0.42 \\
S8$-$Q6 & 210 & 35 & 5  &4.104&3.899  & 4.71  &0.59 &0.013      &0.858&1.60  &48.55 &0.28 \\
S8$-$Q7 & 260 & 35 & 5  &3.832&3.712  & 4.84  &0.49 &0.005      &0.871&1.60 &48.55&0.70\\
S8$-$Q8 & 260 & 35 & 5  &4.116&4.059  & 5.79  &0.46 &0.002      &0.881&1.60 &48.55 &0.44 \\
\hline
\end{tabular}
\captionsetup{labelformat=empty,justification=raggedright}
\caption{\textbf{Supplementary Table 2: Qubit parameters.} Qubit design parameters $L$, $W$, $d$, maximum and minimum frequencies $\omega^{\rm max/min}_{\rm ge}$, and Josephson and charging energies $E_{\rm J}$ and  $E_{\rm C}$ are as in Supplementary Table~\ref{tab:tabel1} and are reported here for convenience. 
The energy difference between g states of different parities $\epsilon_0$ and the squared matrix element $c_0^2$ are calculated numerically from $E_{\rm C}$ and $E_{\rm J}$ (although $c_0^2$ can be estimated using the analytical formula given in~\cite{Catelani2014} within 2\% percent in the worst case of S8-Q2, which has the smallest ratio $E_J/E_C$), while $\Gamma_P(0)$, $\Delta$, and $x_{\rm qp}$ are found by fitting $\Gamma_P$ versus temperature using Eq.~(\ref{equ:gammaP1}). Numbers $c_0^2$ and $x_{\rm qp}$ are dimensionless, $\Gamma_P(0)$ is given in Hz, and the remaining quantities in GHz.}
\label{tab:tabel2}\vspace{-6pt}
\end{table*}

To make the above consideration quantitative, we make the simplifying assumption that in the pads there is a fixed normalized density of quasiparticle $x_{\rm qp}\ll1$. This density can be related to the effective chemical potential $\mu$ through the formula:
\begin{equation}\label{equ:xqp}
x_{\rm qp}\simeq\sqrt{\frac{2\pi k_\mathrm{B} T}{\Delta_0}}e^{-(\Delta_0-\mu)/k_\mathrm{B}T}
\end{equation}
We can invert this expression to obtain the chemical potential as function of $T$, $\Delta_0$, and $x_{\rm qp}$. Substituting the result into Eq.~(\ref{equ:gammaqp}) we can finally write:
\begin{align}\label{equ:gammaP1}
\Gamma_P(T) & =\Gamma_P(0) \\ \nonumber + & \frac{16E_{\rm J}}{\Delta}c_0^2\frac{\epsilon_0}{h}e^{\frac{-(\Delta-\Delta_0)}{k_{\rm B}T}}x_{\rm qp}\sqrt{\frac{\Delta_0}{2\pi k_{\rm B}T}}F\left(\frac{\epsilon_0}{2k_{\rm B}T},\frac{k_{\rm B}T}{2\Delta}\right)
\end{align}
where $\Gamma_P(0)$ accounts for all possible temperature-independent contribution to $\Gamma_P$ (including $\Gamma_{\rm pb}$, but also other possible sources of parity switching). Thus, we have three free parameters: $\Gamma_P(0)$, $\Delta$ (the gap at the junctions), and $x_{\rm qp}$, that can be used in comparing theory to experiment. Note that of these parameters, only $\Delta$ and $x_{\rm qp}$ appear in the temperature-dependent term of Eq.~(\ref{equ:gammaP1}), while $\Gamma_{P}(0)$ is independent of temperature.

We perform such a comparison for fifteen qubits belonging to four samples, Sample S1, S4, S5, and S8; the corresponding parameters are given in Supplementary Table~\ref{tab:tabel2}. For the first three samples, the experimental data with fit (in log scale) are presented in Fig.~4 of the main text. When fitting the data, we require $\Delta$ to be the same for qubits of a given sample, since the thicknesses of their superconducting arms are expected to be the same; this leaves only $x_{\rm qp}$ as a parameter to fit the temperature dependence for a given qubit in a chip. The normalized quasiparticle densities are of order $10^{-7}$, comparable to those reported in the literature~\cite{Pop2014,Wang2014}. It should be noted that even
at 100~mK the expected thermal equilibrium value for $x_\mathrm{qp}$ is about $4.6\times 10^{-10}$
and hence always two orders of magnitude smaller than our extracted densities. This justifies assuming the temperature independence of the quasiparticle density in the pads, its value being determined by some non-equilibrium process. In fact, the data in Fig.~4b of the main text suggest that pair breaking at the junction contributes significantly to the non-equilibrium density $x_\mathrm{qp}$. In the steady-state, $x_\mathrm{qp}$ is determined by the balance between generation with rate $g$, recombination with rate $r$ (as mentioned in the main text), and possibly single-quasiparticle trapping with rate $s$~\cite{Wang2014},
\begin{equation}
    0 = g - s x_\mathrm{qp} - r x_\mathrm{qp}^2 \, .
\end{equation}
The recombination rate is material and geometry dependent; for thin films of aluminium it is estimated to be $r\approx 1/(120\,$ns). The trapping rate $s$ can be written as the sum of a background rate $s_0$ plus a term due to vortices $s_v$: $s=s_0+s_v$; assuming no vortices, we take $s=s_0\approx 0.05\,$ms$^{-1}=50\,$s$^{-1}$. (For both recombination and trapping rates, their values are taken from Ref.~\cite{Wang2014}.) Finally, the generation rate can in principle have several contributions, such as pair-breaking phonons being generated in the substrate by radiation~\cite{Cardani2020} or pair-breaking photons absorbed directly in the pads. For our purposes, we assume that the quasiparticles generated at the junction with rate $\Gamma_P(0)$ quickly diffuse to the pads, relax by phonon emission to an energy lower than $\Delta$ (the gap in the junction's arms), and are hence trapped in the pads; in this scenario, the generation rate is
\begin{equation}\label{eq:g_rate}
    g = \frac{\Gamma_P(0)}{2\nu_0 \Delta_0 V}\, ,
\end{equation}
where $\nu_0 = 0.73\times10^{47}\,($J m$^3)^{-1}$ is the single-spin density of states at the Fermi energy in Al, $\Delta_0$ is the gap in the pads, and $V$ the volume of one pad.

For the densities of interest ($x_\mathrm{qp} < 6\times 10^{-7}$), the recombination term is smaller than the trapping one, $r x_\mathrm{qp}< s$, so trapping should be dominant, giving $x_\mathrm{qp} \sim g/s$; however, even for the largest value of $g\sim 10^{-8}\,$s$^{-1}$ (corresponding to the largest $\Gamma_P(0)$ measured in the qubit with the biggest pads), the expected density would be of order $x_\mathrm{qp}\sim 10^{-10}$, much smaller than the measured densities. In the devices of Ref.~\cite{Wang2014}, the pads consisted of a bilayer where two aluminium films were separated by a thin insulating layer of aluminium oxide; this oxide, absent in our devices, could be the source of background trapping.
If trapping can be neglected, one instead finds $x_\mathrm{qp} = \sqrt{g/r} \sim 3\times10^{-8}$, which is still one order of magnitude smaller than the largest measured density. 

\begin{figure}[!bth]
    \centering
    \includegraphics[width=0.48\textwidth]{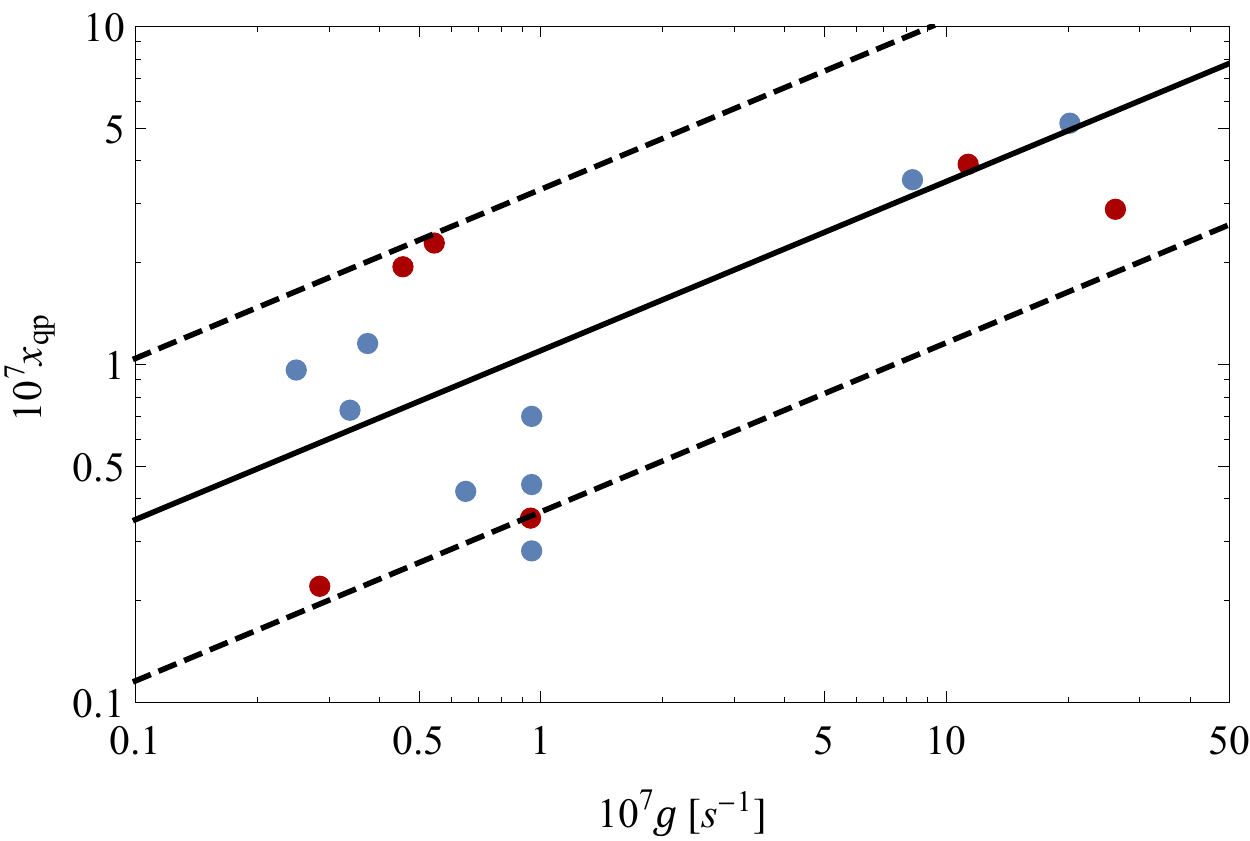}
\captionsetup{labelformat=empty,justification=raggedright}
\caption{\textbf{Supplementary Figure 23:} Points: experimental data, where the generation rate $g$ is calculated using Eq.~(\ref{eq:g_rate}) with $V=4\,\mu{\rm m}^3$ the volume of a bandage; red: data from qubits in samples S1 and S5 (same as in Fig.~4b of the main text), blue: data from qubits in S4 and S8. Solid line: theory estimate using the same recombination rate $r$ as in Ref.~\cite{Wang2014}. Dashed lines: theory estimates multiplied or divided by 3.}
    \label{fig:xg}\vspace{-8pt}
\end{figure}

A possible resolution of the discrepancy is as follows: while we have assumed that the quasiparticles are trapped in the pads, they could be trapped in the bandages instead. The bandages are twice as thick as the pads (200 vs 100~nm), so their gap is likely a few $\mu$eV smaller, equivalent to a temperature of a few tens of mK, hence lower than base temperature. Moreover, due to their small lateral dimensions ($10\times 2\,\mu$m$^2$), the bandages are unlikely to host vortices, explaining the absence of trapping. For the purpose of fitting the temperature dependence of the parity switching rate, Fig.~4a of the main text, it doesn't matter if the non-equilibrium quasiparticles are activated from the pads or the bandages; however, if the bandages' volume is used in Eq.~(\ref{eq:g_rate}) to calculate the generation rate $g$, we find that the data is reasonably well explained (within a factor of $\sim 3$) by the relation $x_\mathrm{qp} = \sqrt{g/r}$ with the same recombination rate $r$ of Ref.~\cite{Wang2014}, see Supplementary Fig.~\ref{fig:xg}. Interestingly, this result suggests that either no significant quasiparticle generation takes place in the pads, or that the quasiparticles generated there are quickly trapped, for instance due to one or a few vortices, so that they do not reach the bandages (for reference, the trapping rate $s_v$ by a single vortex in the largest pad is $s_v \approx 400\,$s$^{-1}$ and $s_v$ is larger for smaller pads~\cite{Wang2014}). 

\vspace{-10pt}
\bibliography{supp}